\documentclass[10pt,journal,compsoc]{IEEEtran}

\usepackage{url}

\ifCLASSINFOpdf
   \usepackage[pdftex]{graphicx}
\else
\fi
\usepackage{tikz}
\usepackage{booktabs}
\usepackage{tabularx}
\usetikzlibrary{arrows, arrows.meta}
\usepackage{amsthm,amsmath}
\makeatletter
\newtheorem*{rep@theorem}{\rep@title}
\newcommand{\newreptheorem}[2]{%
\newenvironment{rep#1}[1]{%
 \def\rep@title{#2 \ref{##1}}%
 \begin{rep@theorem}}%
 {\end{rep@theorem}}}
\makeatother

\newtheorem*{rep@corollary}{\rep@title}
\newcommand{\newrepcorollary}[2]{%
\newenvironment{rep#1}[1]{%
 \def\rep@title{#2 \ref{##1}}%
 \begin{rep@corollary}}%
 {\end{rep@corollary}}}

\makeatother
\newtheorem*{rep@lemma}{\rep@title}
\newcommand{\newreplemma}[2]{%
\newenvironment{rep#1}[1]{%
 \def\rep@title{#2 \ref{##1}}%
 \begin{rep@lemma}}%
 {\end{rep@lemma}}}
\makeatother

\newtheorem{theorem}{Theorem}
\newreptheorem{theorem}{Theorem}
\newtheorem{corollary}{Corollary}[theorem]
\newrepcorollary{corollary}{Corollary}

\newreplemma{lemma}{Lemma}

\newtheorem{subtheorem}{Theorem}

\usepackage{pgfplots}
\pgfplotsset{width=10cm,compat=1.9}
\usepackage{caption}
\usepackage{subcaption}
\usepackage[font=small,labelfont=bf]{caption}

\usepackage{xstring}
\newcommand{\figref}[1]{%
  \IfSubStr{#1}{,}{Fig.~\ref{#1}}{Fig.~\ref{#1}}%
}
\newcommand{\tabref}[1]{%
  \IfSubStr{#1}{,}{Tables~\ref{#1}}{Table~\ref{#1}}%
}
\newcommand{\eqnref}[1]{%
  \IfSubStr{#1}{,}{Equations~(\ref{#1})}{Equation~(\ref{#1})}%
}
\usepackage{xspace}
\newcommand{\etal}{\textit{et al.}\xspace}

\usepackage{amsmath}
\usepackage{amsfonts}
\DeclareMathAlphabet\mathbfcal{OMS}{cmsy}{b}{n}

\DeclareMathOperator*{\argmin}{arg\,min}
\newcommand{\equaltext}[1]{\ensuremath{\stackrel{\text{#1}}{=}}}

\newcommand{\largertext}[1]{\ensuremath{\stackrel{\text{#1}}{>}}}

\usepackage{xcolor}
\colorlet{mygreen}{green!50!black}

\begin{document}
%
\title{Rate-Distortion Theory in Coding for Machines and its Applications}
%
%
%
%

\author{Alon~Harell,~\IEEEmembership{Student Member,~IEEE,}
        Yalda Foroutan,~\IEEEmembership{Student Member,~IEEE,}
        Nilesh Ahuja,
        Parual Datta,
        Bhavya Kanzariya,
        V. Srinivasa Somayazulu,
        Omesh Tickoo,
        Anderson de Andrade,~\IEEEmembership{Student Member,~IEEE,}
        Ivan V. Baji\'{c},~\IEEEmembership{Senior Member,~IEEE}
\thanks{ Alon Harell, Yalda Foroutan, Anderson de Andrade, and Ivan V. Baji\'c are with Simon Fraser University, 8888 University Drive, Burnaby, BC, V5S1A6, Canada. Email communications to Alon Harell: aharell@sfu.ca. Nilesh Ahuja, Parual Datta, Bhavya Kanzariya, Srinivasa Somayazulu, and Omesh Tickoo are with Intel Labs. This work was funded by Intel Labs and NSERC.}
}

%
%

\markboth{Submitted to IEEE Transactions on Pattern Analysis and Machine Intelligence}%
{Shell \MakeLowercase{\textit{et al.}}: Bare Demo of IEEEtran.cls for Computer Society Journals}
%



\IEEEtitleabstractindextext{%
\begin{abstract}
Recent years have seen a tremendous growth in both the capability and popularity of automatic machine analysis of \textcolor{black} {media, especially images and video}. As a result, a growing need for efficient compression methods optimised for machine vision, rather than human vision, has emerged. To meet this growing demand, significant developments have been made in image and video coding for machines.  Unfortunately, while there is a substantial body of knowledge regarding rate-distortion theory for human vision, the same cannot be said of machine analysis. In this paper, we greatly extend the current rate-distortion theory for machines, providing insight into important design considerations of machine-vision codecs. We then utilise this newfound understanding to improve several methods for learned image coding for machines. Our proposed methods achieve state-of-the-art rate-distortion performance on several computer vision tasks -- classification, instance \textcolor{black}{and semantic} segmentation, and object detection. 
\end{abstract}

\begin{IEEEkeywords}
Rate-Distortion Theory, Collaborative Intelligence, Image Coding, Coding for Machines, Learned Compression, Compression for Machines, Split Computing, Neural Compression.
\end{IEEEkeywords}}

\maketitle

\IEEEdisplaynontitleabstractindextext

%
\IEEEpeerreviewmaketitle

\IEEEraisesectionheading{\section{Introduction}\label{sec:introduction}}

%
%
%
%

\IEEEPARstart{I}{n} machine analysis, an automated system processes an input to provide some semantically meaningful information. Common examples include object detection in images, action recognition in video, and speech-to-text conversion in audio. Recent advancements in deep neural networks (DNNs) have resulted in a rapid increase in the accuracy and reliability of automated machine analysis. As a result, DNN models (often known simply as deep models) have become ubiquitous in many tasks such as object detection~\cite{Redmon2018_yolov3,ren2015faster,Liu_2021_ICCV} and segmentation~\cite{Liu_2021_ICCV,he2017mask,kirillov2023segment}, machine translation~\cite{bert,gpt3}, and speech-recognition~\cite{whisper,wave2vec_u}.  
Unfortunately, alongside their tremendous improvements, deep models have very high computational costs both in terms of memory and floating point operations~\cite{growth}.  The resulting complexity, coupled with the growing diversity of deep models, is a significant limiting factor in the deployment of such models, especially to devices such as smart speakers or wearable devices with limited computing resources.  


Beyond the simple improvement of edge-device hardware capabilities, several different \textcolor{black}{approaches} are used to deploy resource-intensive deep models to end-users. One approach, known as edgification, is to directly reduce the complexity of the DNN models in the design process. This reduction can be the result of changes to the model architecture~\cite{Howard_2019_ICCV}, the use of lower numerical precision~\cite{dos2019impact}, or via the simplification of existing models to match the computational capabilities of edge devices~\cite{gou2021knowledge}. Although these methods and more hold great promise, they are not always adequate, and are in limited use today. In practice, the most common approach to date~\cite{Shlezinger_Bajic_IOTM_2022} is simply to avoid deployment of computationally demanding models on edge devices altogether.  Instead, the majority of the computation is offloaded to remote servers with tremendous computational capacity ("cloud"), which often utilise  graphics processing units (GPU) or tensor processing units (TPU). 


In order for the computation to be performed remotely, the edge device must transmit information to the cloud. The most common solution in use today is the straightforward approach, wherein the input itself is compressed and sent to the cloud using coding methods developed for humans (for example a video codec such as VVC\cite{vvc_std}, or an audio codec such as AAC\cite{aac_std}). In recent work, this naive approach has been shown to be sub-optimal both theoretically~\cite{hyomin} and empirically~\cite{hyomin,kang2017neurosurgeon}. Furthermore, in a growing number of applications such as traffic monitoring or home security, the majority of inputs are never observed by a human, and thus there is no reason to preserve them in their original, human-friendly form. As a result \emph{coding for machines}, also known as \emph{compression for machines} \textcolor{black}{(CfM)}, an umbrella term used for methods of transmitting information to facilitate automated analysis (rather than human consumption), has garnered increasing attention of late. 

Another important development is the emergence of learned compression (also known as neural compression), most commonly used for images~\cite{minnen2018joint, balle2017end, cheng2020image, he2022elic, jiang2022multi, ho2021anfic} and video, \cite{Rippel_2019_ICCV, ho2022canf, kim2022scalable}. In this setup, complex hand-engineered codecs (such as JPEG~\cite{wallace1992jpeg}, VVC~\cite{vvc_std}, and HEVC~\cite{hevc_std_2019}), are replaced with trainable models (often DNNs). 
\textcolor{black}{It is important to distinguish between learned compression designed around human perception, which can be thought of as \emph{compression for humans by machines}; and \emph{compression for machines}, learned or otherwise, which is focused on automated analysis models. This difference also exists in standardisation efforts where JPEG-AI~\cite{jpeg_ai} focuses more on the former, while MPEG-VCM~\cite{mpeg_vcm} on the latter.}

The most common objective ($L$) used to train learned codecs is derived from the information bottleneck~\cite{tishby_deep_2015}, and is of the following Lagrange multiplier form:
\begin{equation} \label{eq:learned}
    L = R + \lambda D,
\end{equation}
where $R$ is an estimate of the size required for encoding the input (the \emph{bitrate} or simply \emph{rate}), $D$ is some measure of distortion for the resulting reconstructed output, and $\lambda$ is the Lagrange multiplier that controls the trade-off between rate and distortion. Using this approach, a codec for machines can be created by simply choosing a distortion metric corresponding to the performance of some desired task model as was done, for example, in the work of~\cite{entropic_student}. 

Generally, information theory~\cite{shannon1948mathematical} allows us to discuss 
ultimate bounds on the amount of bits needed to describe a random variable (RV). As such, its understanding is crucial in the development of efficient compression methods for complex signals such as audio, images, or video. More specifically within information theory, rate-distortion~\cite{Cover_Thomas_2006,Berger_1971} describes the inherent trade-off between the rate needed to describe an RV and the fidelity, or conversely the distortion, with which said RV can be reconstructed. In the case of coding for machines, however, the traditional, well-developed rate-distortion (RD) theory does not apply directly, because we are no longer required to reconstruct the RV in its original form. Thus, extending RD theory to cases where input reconstruction is no longer needed is key to understanding the limits of codecs for machines.

\begin{figure}[ht]
    \centering
    \begin{subfigure}[b]{\linewidth}
        \centering
        \includegraphics[width=0.8\textwidth]{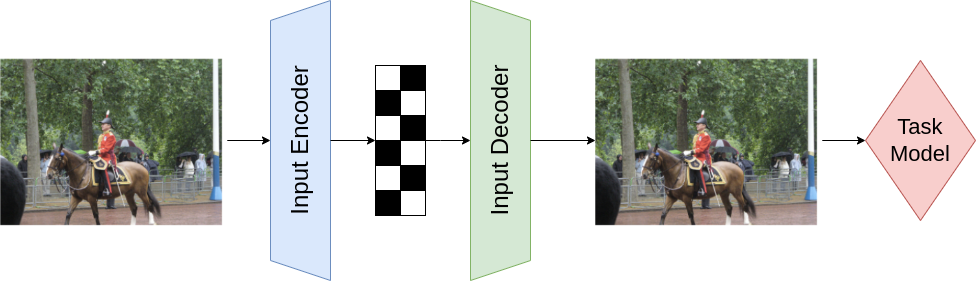}
        \caption{Full-input coding for machines}
        \label{fig:approach-full}
    \end{subfigure}

    \begin{subfigure}[b]{\linewidth}
        \vspace{5pt}
        \centering
        \includegraphics[width=0.65\textwidth]{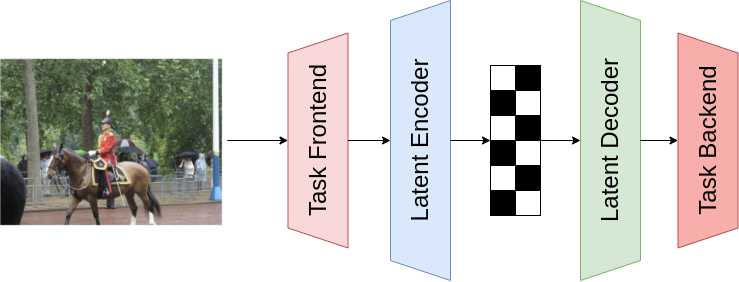}
        \caption{Model-Splitting coding for machines}
        \label{fig:approach-split}
    \end{subfigure}

    \begin{subfigure}[b]{\linewidth}
        \vspace{7pt}
        \centering
        \includegraphics[width=0.55\textwidth]{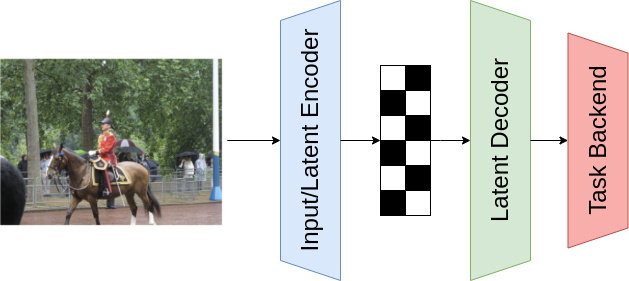}
        \caption{Direct coding for machines}
        \label{fig:approach-direct}
    \end{subfigure}

    \caption{Three common approaches for coding for machines.}
    \label{fig:approaches}
\end{figure}

We focus on three main approaches for coding for machines that 
have been presented in the literature so far: the \emph{full-input} approach \textcolor{black}{\cite{Choi_2018_ICASSP}}, the \emph{model-splitting} approach \textcolor{black}{\cite{choi2018deep,datta2022low}}, and the \emph{direct} approach \textcolor{black}{\cite{torfason2018towards,entropic_student,duan2022pcs}}. In the full-input approach, an input signal is encoded and reconstructed before being passed to the machine task model for analysis. Note that this approach can differ from the naive approach presented above, because the codec used to encode and reconstruct the input may be specifically optimised for machine analysis. In the model-splitting approach, sometimes also known as \emph{collaborative intelligence}~\cite{kang2017neurosurgeon} or \emph{collaborative inference}~\cite{Shlezinger_Bajic_IOTM_2022}, the 
initial part (\emph{frontend}) of the task model is run on the edge device, yielding a latent representation of the input specific to a given task. This representation is then encoded, whether by using traditional codecs~\cite{kang2017neurosurgeon,choi2018deep}, or by purpose-built learned methods~\cite{datta2022low,Choi2018NearLosslessDF} and transmitted to the cloud. There, the 
remainder (\emph{backend}) of the task model is used to perform the desired analysis. Finally, in direct CfM, an asymmetric approach is used. The input signal is encoded and transmitted to the cloud, but instead of reconstructing the input itself, the decoder recreates the desired latent representation directly. Then, as in the model-splitting approach, the reconstructed latent representation is passed on to the 
backend of the task model, which performs the task. \figref{fig:approaches} provides a visual representation of these three approaches. \textcolor{black}{Note also that there has been recent work on scalable human-machine coding, where the goal is to support the machine task(s) with partial decoding and human viewing with full decoding of the compressed bitstream. This has been done for images~\cite{face_scalable,hyomin,semantics_scalable},  video~\cite{Choi2022MMSP,HMFVC,hadi2023scalable_video}, and point clouds~\cite{ulhaq2024scalable_pc}. However, our focus here is on coding for machines only, which corresponds to the base layer of these scalable codecs.}

In this paper, we 
develop and utilise rate-distortion theory for machines, with specific attention to coding for deep models. The main contributions of our work are:
\textcolor{black}{
\begin{itemize}
    \item We present a framework that allows for theoretical analysis of  coding for machines, including all three of the commonly used approaches in the field. 
    \item Using our new formulation, we greatly extend the current theoretical understanding of rate-distortion theory in coding for machines. We prove that under reasonable conditions, all 3 approaches can achieve the same optimal performance in the asymptotic ideal case. Furthermore we prove that supervised optimisation yields better optimal rate-distortion performance in coding for machines.
    \item Where our theory does not provide concrete proofs, we offer guidance for design considerations through 
    empirically demonstrated 
    hypotheses. Specifically we analyze the effect of different optimisation targets when optimising a CfM codec without task-labels.
    \item Using 
    the new theoretic understanding, we better explain our previously published work in image coding for machines. Additionally, we utilise our insights into the design of learned image codecs for machines to present new or improved results for additional tasks and task models. The resulting codecs all achieve, to the best of our knowledge, state-of-the-art (SOTA) rate-distortion performance in their corresponding setting. 
\end{itemize}
}
We present our work in the following order: Section~\ref{sec:RD_theory} presents background on rate-distortion, including prior theory in coding for machines, followed by our formulation and theoretic results; Section~\ref{sec:design} discusses design considerations in cases beyond the scope of our theory, including 
evidence-based motivating 
hypotheses; In Section~\ref{sec:applications}, we demonstrate the benefits of our theory in practice, achieving state-of-the-art rate-distortion performance on several computer-vision tasks; \textcolor{black}{Finally, Section~\ref{sec:conclusion} 
provides a summary and cites several examples of CfM in other settings which corroborate our conclusions.} Throughout the paper, upper case variables $X,Y, T$ represent random variables; their lower case counterparts $x,y,t$ represent single observations; sets and spaces are represented by script letter such as $\mathbfcal{P},\mathbfcal{R}$; and other notation is explained when introduced. For ease of reading, we present all theorems and corollaries in the main body of the paper, and all proofs in Appendix~\ref{app:proofs} of the supplemental material.

\vspace{-0.3cm}
\section{Rate-Distortion Theory for Machines} \label{sec:RD_theory}
\vspace{-0.7cm}
\textcolor{black}{\subsection{Introduction}}
Due to its emerging nature, theory governing rate-distortion in \emph{coding for machines} is still evolving. Recently, \cite{Xu2020A} proposed "$\mathcal{V}$-Information", a theoretic structure for measuring information in terms of its usability by a downstream family of models. While this framework allows for better understanding of the inner workings of deep models, it does not relate directly to compression because the resulting measure cannot be used for encoding a desired RV. Meanwhile,~\cite{dubois2021lossy} present a theory of rate-distortion for machine tasks, based on the invariance of a task 
(or several) to certain changes in the input. Using their formulation, the authors in~\cite{dubois2021lossy} are able to show that significant degradation of the input is possible without affecting the task performance. Furthermore, 
\cite{dubois2021lossy} demonstrate a closed form for rate-distortion for certain tasks when using log-loss as their distortion metric. An important conclusion of this work is that utilising a measure of distortion directly related to the task is preferable to simply attempting to create a high-fidelity reconstruction of the input. 
In another example of a multi-task setting,~\cite{alvar2020pareto} examine bit allocation optimality in coding for machines. The proposed method achieves multi-objective optimality in settings where the codec and task model are fixed, only allowing for bit allocation via changing of quantisation coefficients for various tensors. 

More directly related to our formulation is the work of~\cite{hyomin} which proposes a scalable codec for both humans and machines. Importantly, in~\cite{hyomin}, 
the authors present a formulation which allows comparing the performance of the full-input method and the model-splitting method. Furthermore, the authors prove that the model-splitting approach achieves a rate no worse than the full-input approach for equivalent task performance (including the lossless case). Our own previous work~\cite{pcs}, extends the work of~\cite{hyomin}, discussing the importance of the choice of splitting point, as well as the chosen measure of distortion. 

The results proven in~\cite{hyomin, pcs} are directly referenced in this section, and form an important basis of our newfound formulation. Nonetheless, we will show that in many settings, those results are insufficient in explaining (and thus optimising) rate-distortion behaviour in coding for machines. We present our formulation step by step, beginning with introducing the reader to traditional rate-distortion theory. Next, we restate the formulation presented in~\cite{hyomin}, while adding important nuance, including the introduction of notation to support the direct coding for machines approach. Lastly, we prove several important results regarding the rate-distortion behaviour of the various methods, as well as define and discuss the difference between supervised and unsupervised coding for machines.


\subsection{Traditional Rate-Distortion Theory}
In traditional digital compression settings, 
an input $x\in\mathbfcal{X}$ (such as an image, video, or audio) is encoded, 
usually for the purpose of storage or transmission. 
It is common to separate between lossless compression, where the input is perfectly reconstructed after decoding; and lossy compression where some level of distortion is acceptable after decoding. In both settings, a coding scheme should minimise the size of the encoded representation, often measured in bits per sample. 
The minimum achievable bitrate for lossless coding of a discrete RV, $X$,  is 
given by its entropy~\cite{Cover_Thomas_2006}: 
$H(X) = -\sum_{x\in\mathbfcal{X}}p_X(x)\log_2\left(p_X(x)\right),$ where  $p_X(x)$ is the probability of occurrence of the input $x$.\footnote{The use of $\log_2$ here ensures that the entropy is measured in bits. Other logarithm bases, such as $e$ or $10$, are also acceptable, and lead to different units for entropy such as nats and bans respectively.} In lossy compression, the rate clearly depends on the acceptable level of distortion, leading to the following formulation: given an input  $X\in\mathbfcal{X}$, an approximation $\widehat{X}\sim p_{\widehat{X}|X}(\hat{x}|x)$, and a measure of distortion between two observations $d(x,\hat{x})$, the minimal achievable rate~\cite{Cover_Thomas_2006} 
that allows an expected distortion no more than $D$ is given by:
\begin{equation}
    R(D) = \min_{p_{\hat{x}|x}(\hat{x}|x)~:~\mathbb{E} \left[d(X,\widehat{X})\right]~\leq~D} I(X;\widehat{X}).
    \label{eq:RD_function} 
\end{equation}
Above, $I(X;\widehat{X})$ is the Mutual Information~\cite{Cover_Thomas_2006} between the reconstruction and the input, and $R(D)$ is commonly known as the rate-distortion function. In traditional rate-distortion analysis, the distortion function is chosen to reflect the perceptual degradation in the quality of \emph{input} reconstruction, for example the squared $\ell_2$ error: $\|x-\hat{x}\|_2^2$.

As explained in the introduction, in many modern settings there is no necessity to reconstruct the original input. Instead, on the decoder side, the only requirement is that some downstream task, such as voice recognition in audio, or object detection in images, be performed successfully. Thus, a CfM codec is required to create a minimal encoded representation of the input such that the downstream task performance output is compromised as little as possible (or not at all, resulting in lossless coding for machines). 

\subsection{Problem Formulation} \label{subsec:formulation}

We 
begin by defining a task $T = f(X),f:\mathbfcal{X}\rightarrow \mathbfcal{T}$, defined by the model $f(\cdot)$, and some corresponding measure of distortion $d_T(f(X),\widehat{T})$, where $\widehat{T}$ represents the reconstructed output of the task-model, accounting for compression. The reconstructed $\widehat{T}(\cdot)$ can be calculated in a variety of ways, corresponding to the different CfM approaches explained in Section~\ref{sec:introduction}, and detailed below. 
Next, we assume that the task model can be decomposed into two parts $f = h\circ g$, where we refer, as is commonly done, to $g:\mathbfcal{X}\rightarrow \mathbfcal{Y}$ as a \emph{feature extractor}, and to $h: \mathbfcal{Y}\rightarrow \mathbfcal{T}$ as the \emph{task backend}. \textcolor{black}{$\circ$ denotes function composition}. We then define $Y = g(X)\in\mathbfcal{Y}$ to be the resulting intermediate representation of the input, to which we refer to as a \emph{feature tensor},  as illustrated below:
\begin{equation*}
\begin{tikzpicture}[
  inner sep=0pt,
  outer sep=0pt,
  baseline=(x.base),
]
  \tikzstyle{arrow} = [thin,->,>=stealth]
  
  \path[every node/.append style={anchor=base west}]
    (0, 0)
    \foreach \name/\code in {
      x/ X~,
      tmp/\,\qquad,
      t/ ~Y~,
      tmp/\,\qquad,
      y/ ~T%
    } {
      node (\name) {$\code$}
      (\name.base east)
    }
  ;
  \path[
    every node/.append style={
      anchor=base,
      font=\scriptsize,
    },
  ]
    (x.base) -- node[above=2.0\baselineskip] (f) {$f$} (y)
    (x.base) -- node[below=0.2\baselineskip] (g) {$g$} (t)
    (t.base) -- node[below=0.2\baselineskip] (h) {$h$} (y)
  ;
  \draw [arrow] (x) -- (t);
  \draw [arrow] (t) -- (y);
    
  \begin{scope}[
    >={Stealth[length=5pt]},
    thin,
    rounded corners=2pt,
    shorten <=.3em,
    shorten >=.3em,
  ]
    
    \def\GebArrow#1#2#3{
      \draw[->]
        (#2.south) ++(0, -.3em) coordinate (tmp)
        (#1) |- (tmp) -| (#3)
      ;%
    }
    \GebArrow{x}{f}{y}
  \end{scope}
\end{tikzpicture}
\label{eq:mappings}
\end{equation*}
\textcolor{black}{Note that, other than the decomposition stated above, we make no assumptions as to the nature of the input signal or the task-model. Thus, any results presented in the following sections apply to any modality of input signal, any task, and any model implementing said task. }

In full-input compression (\figref{fig:approaches}(a)), which remains the most common approach used in practice as of today, the input $X$ is compressed and then decoded into $\widehat{X}$, yielding $\widehat{T}(\widehat{X}) = f(\widehat{X})$. 
This can be accomplished either using traditional codecs~\cite{Choi_2018_ICASSP} or learned ones~\cite{e2eicm_2021_DCC}. 
The model-splitting approach (\figref{fig:approaches}(b)) utilises the existing feature extractor to obtain the latent representation $Y$ and then proceed to encode it, whether by using traditional codecs~\cite{kang2017neurosurgeon, choi2018deep}, or by purpose-built learned methods~\cite{datta2022low,Cohen_quantcode_ICME2020, Choi2018NearLosslessDF}. Next, the intermediate
representation is decoded, giving $\widehat{Y}\sim p_{\widehat{Y}|Y}(\hat{y}|y)$ which is passed onto the task backend resulting in $\widehat{T}(\widehat{Y}) = h(\widehat{Y})$. 
Finally, in direct coding for machines~\cite{hyomin, entropic_student} (\figref{fig:approaches}(c)), the feature encoder is not used, but instead the input is encoded directly 
by a CfM codec. On the decoder side however, the input itself is not reconstructed but rather the latent representation of the task model $\widetilde{Y}\sim q_{\tilde{Y}|X}(\tilde{y}|x)$. This representation is subsequently passed on to the task backend, resulting in $\widehat{T}(\widetilde{Y}) = h(\tilde{Y})$. \textcolor{black}{Note that decoded representations for model splitting and direct coding are assigned two different symbols -- $\widehat{Y}$ and $\widetilde{Y}$ -- because, although they both approximate $Y$, they are not exactly the same.}

Next, we present and extend the notation introduced by~\cite{hyomin} for the rate-distortion function in coding for machines, which is used as a foundation for our notation. We begin with sets of conditional distributions (sometimes referred to as \emph{quantisers}), that achieve some desired average level of distortion $D>0$ measured at the task $T$:

\begin{equation}
\small
    \mathbfcal{P}_{{X}}(D;T)=\left\{p\left(\hat{{x}}|{x}\right)~:~\mathbb{E}\left[d_T\left(f({X}),f(\widehat{{X}})\right)\right]\leq D\right\},
    \label{eq:quant_X}
\end{equation}
\begin{equation}
\small
\mathbfcal{P}_{{Y}}(D;T)=\left\{p\left(\hat{{y}}|{y}\right)~:~\mathbb{E}\left[d_T\left(h({Y}),h(\widehat{{Y}})\right)\right]\leq D\right\}.
    \label{eq:quant_Y}
\end{equation}
The difference between~\eqref{eq:quant_X} and~\eqref{eq:quant_Y} lies in the RV that is quantised: the input $X$ or the feature tensor $Y$. Next, we define the rate-distortion function corresponding to each of the quantisers, analogously to the traditional approach~\cite{Cover_Thomas_2006}. That is, we minimise the mutual information 
between the compressed RV 
and the original RV, while 
preserving the distortion requirement:
\begin{equation}
\small
    R_{{X}}(D;T) = \min_{p(\hat{{x}}|{x})~\in~\mathbfcal{P}_{{X}}(D;T)} I({X};\widehat{{X}}).
    \label{eq:RD_function_X}
\end{equation}
\begin{equation}
\small
    R_{{Y}}(D;T) = \min_{p(\hat{{y}}|{y})~\in~\mathbfcal{P}_{{Y}}(D;T)} I({Y};\widehat{{Y}}).
    \label{eq:RD_function_Y}
\end{equation}
The definitions in~\eqref{eq:RD_function_X} and~\eqref{eq:RD_function_Y} result in the minimal achievable rate~\cite{Cover_Thomas_2006} for compressing $X,\,Y$, respectively, while maintaining average distortion no higher than $D$ (as measured at the task $T$). Note the notation $(D;T)$ for task distortion as opposed to 
distortion related to the input $X$. In~\cite{hyomin}, the authors have proven that $R_{{Y}}(D;T)\leq R_{{X}}(D;T)$, meaning that, for a given task distortion, the lowest achievable rate for the model splitting approach is 
no worse than the full-input compression approach.
We expand on this notation to include a rate-distortion function for direct coding for machines as follows:
\begin{equation}
    \mathbfcal{Q}_{X{Y}}(D;T)=\left\{q\left(\tilde{{y}}|{x}\right)~:~\mathbb{E}\left[d_T\left(f({X}),h(\widetilde{{Y}})\right)\right]\leq D 
    \right\},
    \label{eq:quant_XY}
\end{equation}
\begin{equation}
    R_{XY}(D;T) = \min_{q(\tilde{{x}}|{y})~\in~\mathbfcal{Q}_{{XY}}(D;T)} I(X;\widetilde{{Y}}).
    \label{eq:RD_function_XY}
\end{equation}
Using this formulation, we are able to prove several important results in the remainder of this section. As mentioned earlier, we only state the results in the main body of the paper; the proofs are provided in Appendix~\ref{app:proofs} of the supplemental material.

\begin{theorem}\label{thm:split-direct}
The minimal achievable rates for direct coding for machines and model splitting are identical, that is, $R_{XY}(D;T) = R_Y(D;T)$  
\end{theorem}
Importantly, in practical settings, there may nonetheless be a significant difference between the direct approach and the model splitting approach, because practical codecs are generally sub-optimal. An important benefit of Theorem~\ref{thm:split-direct} is that, moving forward, any theoretic discussion of rate-distortion for machines need not differentiate between model-splitting and the direct approach. Next, we show that, under reasonable conditions, the inequality proven in~\cite{hyomin} is in fact an equality. The only necessary constraints are that the output of the task backend for any given approximation of a feature layer,\footnote{In fact, it is sufficient to require this for any approximation $\hat{y}$ derived from a quantiser, $p(\hat{y}|y)\in\mathbfcal{P}_Y(D;T)$, that achieves the optimal rate $I(\widehat{Y};Y) = R_Y(D;T)$.} be in the image of the function $f$ representing the task model. 

\begin{theorem}\label{thm:distortion-only}
Let $\mathbfcal{I}_f(\mathbfcal{X})\subseteq \mathbfcal{T}$ be the image set of the function $f$ (the task model) on all possible inputs, and let $\widehat{\mathbfcal{Y}}\subseteq \mathbfcal{Y}$ be the set of all possible approximations of $Y$.
If $h(\widehat{\mathbfcal{Y}})\subseteq \mathbfcal{I}_f(\mathbfcal{X})$ then, for any given distortion $D>0$, the minimal achievable rate for model splitting is equal to the minimal achievable rate for 
full-input compression:
\begin{equation}\label{eq:equality}
    R_X(D;T) = R_Y(D;T).
\end{equation}
\end{theorem}
The constraint 
$h(\widehat{\mathbfcal{Y}})\subseteq \mathbfcal{I}_f(\mathbfcal{X})$ holds trivially for classification-like problems (such as object detection and semantic/object segmentation) since the output must be in the set of defined classes. In regression problems, where this condition does not trivially hold, the result above remains true if 
any approximation $\hat{y}$ whose task output is not in the image of $f$ (the task model) is sub-optimal in terms of distortion. For such cases we provide the following result. 

\setcounter{subtheorem}{0}
\begin{subtheorem}
\label{subthm:distortion_only}
    Let $\hat{y}: h(\hat{y})\notin \mathbfcal{I}_f(\mathbfcal{X})$ be an approximation of \textcolor{black}{a subset $\mathbfcal{Y}_{\hat{y}} \subseteq~\mathbfcal{Y}$ of values of $Y$}, for which the output of the task backend is not contained in the image set of $f$ (the task model). If, for any such $\hat{y}$, there exists an alternative approximation, $\tilde{y}$ such that $h(\tilde{y}) \in \mathbfcal{I}_f(\mathbfcal{X})$ and $d_T\left(h(y),h(\tilde{y})\right)\leq d_T\left((h(y),h(\hat{y})\right) \textcolor{black}{\forall y \in\mathbfcal{Y}_{\hat{y}}} $ then \eqnref{eq:equality} still holds. 
\end{subtheorem}
\color{black}
\subsubsection{Example of the Equivalence of the Three Approaches}
\label{sec:thought_experiment}
As explained above, the equalities proved in Theorems~\ref{thm:split-direct} and~\ref{thm:distortion-only} deal with the optimal case. This means, that as in many cases in traditional rate distortion theory, designing a practical codec to achieve the RD-function is challenging. Nonetheless, we present the following thought experiment to demonstrate the equivalence of the three approaches in an optimal case. 

Consider a task model $f:\mathbfcal{X} \rightarrow \mathbfcal{T}$ such that $f(X) = (h\circ g) = h(g(X)) = T$ (as in our previous formulation), where $\mathbfcal{T}$ is simply a set of $N$ equally probable classes. Furthermore, assume that we 
posses an encoder-decoder pair $e_{opt}:\mathbfcal{T}\rightarrow \mathbfcal{B}$, $d_{opt}:\mathbfcal{B}\rightarrow \mathbfcal{T}$, which 
achieves any one point on the RD bound for a discrete memoryless source (DMS) 
with $N$ possible outcomes. Here, $\mathbfcal{B}$ is the set of arbitrary-length binary representations - the bitstream. 
Since we are working in an optimal setting, we assume that the full task-model $f$ is known to encoders and decoders, and thus for each input $x$ (and for each latent representation $y$) the corresponding class label $t$ is also known.

Under this setting, we describe 
three codecs corresponding to the three approaches for coding for machines.  For a given input $x$, the full-input encoder, $e_{full}$, computes the class label $t=f(x)$ and then encodes it as $b=e_{opt}(t)$, where $b$ is the bitstream. The direct-coding encoder, $e_{direct}$, operates in the same way: it computes the class label $t=f(x)$ and encodes it as $b=e_{opt}(t)$. The model-splitting encoder, $e_{split}$, does not receive the input $x$, but rather the representation $y=g(x)$. From here, it finds the label $t=h(y)$ and encodes it as $b=e_{opt}(t)$. In all three cases, the optimal DMS encoder $e_{opt}$ receives the same label $t$ and therefore produces the same bitstream $b$. Hence, the three rates are the same.    
Let $\hat{t}=d_{opt}(b)$ be the decoded class label, which will be the same as $t$ in the lossless case (if the rate is equal to the entropy of the DMS), and might be different in the lossy case (with a lower rate).

Given $\hat{t}$, the full-input decoder, $d_{full}$, 
produces one of $N$ pre-selected inputs whose label matches $\hat{t}$. Formally, if the pre-selected inputs are $\{x^{(1)}, x^{(2)}, ..., x^{(N)}\}: f(x^{(i)}) = i$, the full-input decoder produces $d_{full}(b) = x^{(d_{opt}(b))}$. For example, if $\hat{t}=$``dog'', the full-input decoder would produce a pre-selected image of a dog (not necessarily the dog from the input image). Recall that by definition, $f\left(x^{(d_{opt}(b))}\right)=\hat{t}$. On the other hand, the direct-coding decoder, $d_{direct}$, would produce one of $N$ pre-selected latent representations whose label matches $\hat{t}$. Formally, if the pre-selected latent representations are $\{y^{(1)}, y^{(2)}, ... , y^{(N)}\}: h(y^{(i)}) = i$, we have $d_{direct}(b) = y^{(d_{opt}(b))}$. In line with the previous example, if $\hat{t}=$``dog'', the direct-coding decoder would produce a latent representation of a dog (not necessarily the dog from the input image). The model-splitting decoder, $d_{split}$, operates in the same way and produces the same latent representation as the direct-coding decoder.  
Hence, with both direct-coding and model splitting, $h\left(y^{(d_{opt}(b))}\right)=\hat{t}$.

As noted above, all three codecs achieve the same rate. Moreover, they all eventually produce the same decoded class label $\hat{t}$. Hence, all of them achieve the same rate-distortion point. By varying the rate from $0$ to the entropy of the DMS, which involves varying the optimal DMS encoder-decoder pair $(e_{opt},d_{opt})$, the three coding approaches trace out the same RD curve. Hence, in the optimal case, all three approaches are equivalent. 

\color{black}
\subsection{Multiple Splitting Options} \label{subsec:multiple}
In some cases, and especially 
when the task model is a neural network, 
there are multiple options for decomposing the model into a feature extractor and task backend. 
In such a case we 
can analyze two alternative decompositions of the task model by defining $f = h_1\circ g_1 = h_2 \circ g_2 \circ g_1$, and correspondingly $Y_1 = g_1(X), Y_2 = g_2(Y_1)$, as seen below:
\begin{equation*}
\begin{tikzpicture}[
  inner sep=0pt,
  outer sep=0pt,
  baseline=(x.base),
]
  \tikzstyle{arrow} = [thin,->,>=stealth]
  
  \path[every node/.append style={anchor=base west}]
    (0, 0)
    \foreach \name/\code in {
      x/ X~,
      tmp/\,\qquad,
      y1/ ~Y_1~,
      tmp/\,\qquad,
      y2/ ~Y_2~,
      tmp/\,\qquad,
      t/ ~T
    } {
      node (\name) {$\code$}
      (\name.base east)
    }
  ;
  \path[
    every node/.append style={
      anchor=base,
      font=\scriptsize,
    },
  ]
    (x.base) -- node[above=3.0\baselineskip] (f) {\small $f$} (t)
    (x.base) -- node[below=0.2\baselineskip] (g1) {\small $g_1$} (y1)
    (y1.base) -- node[above=1.8\baselineskip] (h1) {\small $h_1$} (t)
    (y1.base) -- node[below=0.2\baselineskip] (g2) {\small $g_2$} (y2)
    (y2.base) -- node[below=0.2\baselineskip] (h2) {\small $h_2$} (t)
  ;
  \draw [arrow] (x) -- (y1);
  \draw [arrow] (y1) -- (y2);
  \draw [arrow] (y2) -- (t);
    
  \begin{scope}[
    >={Stealth[length=5pt]},
    thin,
    rounded corners=2pt,
    shorten <=.3em,
    shorten >=.3em,
  ]
    
    \def\GebArrow#1#2#3{
      \draw[->]
        (#2.south) ++(0, -.3em) coordinate (tmp)
        (#1) |- (tmp) -| (#3)
      ;%
    }
    \GebArrow{x}{f}{t}
    \GebArrow{y1}{h1}{t}
  \end{scope}
\end{tikzpicture}
\label{eq:double_mappings}
\end{equation*}
Under this notation we refer to $Y_2$ as a \emph{deeper} feature tensor than $Y_1$ or, equivalently, $Y_2$ as a \emph{deeper} split point than $Y_1$. In our previous work~\cite{pcs}, we had proven that choosing 
a deeper split point yields no worse rate-distortion performance:  $R_{Y_2}(D;T)\leq R_{Y_1}(D;T)$. However, we can apply Theorem~\ref{thm:distortion-only} under equivalent conditions in $Y_2$ and $Y_1$ to show the following.
\begin{corollary}\label{cor:design_considerations}
    If $h(\widehat{\mathbfcal{Y}_1})\subseteq \mathbfcal{I}_f(\mathbfcal{X})$ and $h(\widehat{\mathbfcal{Y}_2})\subseteq \mathbfcal{I}_f(\mathbfcal{X})$, then for any given 
    task distortion $D>0$, there is no difference in the minimum achievable rate encoding either intermediate feature:
    $$R_{Y_2}(D;T) = R_{Y_1}(D;T) = R_X(D;T).$$ 
\end{corollary} 
Furthermore, applying Theorem~\ref{thm:split-direct} means that this is also true when using the direct coding for machine approach, resulting in the following:
\begin{corollary}\label{cor:direct_design_considerations}
Under the conditions of Corollary~\ref{cor:design_considerations}, and for any given task distortion $D>0$, encoding directly from the input 
has the same minimum achievable rate 
as either choice of intermediate layer, that is:
    $$R_{Y_2}(D;T) = R_{XY_2}(D;T) = R_{XY_1}(D;T) = R_{Y_1}(D;T).$$
\end{corollary}

The combination of the theorems and corollaries presented up to this point can be summarised into the following important conclusion: \textbf{\emph{In the theoretically optimal case, the only consideration that affects rate-distortion for machines is the task and its corresponding distortion metric}}. This important fact is a central point of both the rest of our theoretical discussion as well as the bulk of our experimental work.

\subsubsection{Alternative Distortion}
Up to this point, we have always considered distortion at the output of our task $T$. However, in general, we may also be interested in measuring distortion elsewhere in the model. To denote distortion measured some point other than $T$, we simply replace $T$ 
in $d_T$ and 
$(D;T)$ with a 
symbol indicating the point at which distortion is measured. We refer to the point at which we  measure distortion as the \emph{distortion target}. Consider, for example the case of measuring the distortion at some intermediate layer $Y$. In that case we denote:
\begin{equation}\label{eq:quant_X_Y}
    \small
    \mathbfcal{P}_{X}(D;Y)=\left\{p\left(\hat{{x}}|{x}\right):\mathbb{E}\left[d_{Y}\left(g(X),g(\widehat{X})\right)\right]\leq D\right\}
\end{equation}
\begin{equation}
    R_{{X}}(D;Y) = \min_{p(\hat{{x}}|{x})~\in~\mathbfcal{P}_{{X}}(D;Y)} I({X};\widehat{{X}}).
    \label{eq:RD_function_X_Y}
\end{equation}

\noindent Using this notation we can equivalently define $R_{X}(D;X)$, $R_{X}(D;Y_1)$, $R_{Y_1}(D;{Y_1})$, $R_{Y_1}(D;{Y_2})$, $R_{Y_2}(D;{Y_2})$, and more. In some cases we may 
be interested in two feature tensors. The first, which we will call the \emph{cut point} or \emph{split-point}, is used for model splitting or direct coding for machines. The other, deeper layer, which we call \emph{distillation point}, will be used to measure the distortion. Importantly, the same logic used to prove all theorems up to this point holds when changing the definition of distortion \textbf{\emph{so long as the same distortion is used whenever making comparisons}}. This leads to the following results:

\begin{corollary}\label{cor:split_direct}
The model-splitting and direct coding for machines approaches have equal minimal rates for achieving distortion at some intermediate layer $Y$. That is, for any given $D>0$: $$R_{XY}(D;Y) = R_{Y}(D;Y).$$
\end{corollary}
\begin{corollary} \label{cor:distillation}
Under equivalent conditions to Theorem~\ref{thm:distortion-only} or~\ref{subthm:distortion_only}, the model-splitting approach, the direct coding for machines, and the 
full-input compression 
all 
have equal minimal rate for achieving distortion $D>0$ at some intermediate layer $Y_2$ (the distillation point).  That is: $$ R_{XY_2}(D;Y_2) = R_{Y_2}(D;Y_2) = R_X(D;Y_2).$$ Furthermore, cases where the cut-point $Y_1$ is different from the distillation point are also equivalent:
$$R_{XY_1}(D;Y_2) = R_{Y_1}(D;Y_2) = R_{Y_2}(D;Y_2).$$
\end{corollary}

One notable conclusion from the corollaries above is that one may choose a cut point different from the distillation point without, in the theoretic limit, diminishing rate-distortion performance. In practice, as we will show in Section~\ref{sec:applications}, this result allows for the use of deeper distillation points without having to change the structure of the codec. 
\subsection{Unsupervised Setting}
An important aspect of measuring distortion at points other than the task output is that it allows us to design a coding scheme without access to the full model $f(X)$ or, equivalently, the task labels. In particular, when our coding scheme is learned from data, measuring the distortion at some intermediate representation $Y$ (in other words, using $Y$ as a \emph{distillation point}), allows unsupervised training of our compression model. Unfortunately, when performing coding for machines, the objective generally still remains to perform well on the entire task, rather than just match some intermediate representation well. This means that in order to truly discuss unsupervised training of learned coding for machines, we must account for the statistical relationship between the distortion at the input $X$: $d_X(X,\widehat{X})$, at an intermediate point $Y$: $d_Y(Y,\widehat{Y})$, and at the full task: $d_T(T,\widehat{T})$. 

We begin by noting that intuition suggests that an unsupervised approach cannot achieve better rate-distortion performance than the fully-supervised equivalent. To prove this formally, we first define the set of all possible approximations of $X$ that achieve some desired input-distortion $D>0$ with a rate equal to $R_X(D;X)$:
\begin{equation*}
\begin{aligned}
     \mathbfcal{R}_X(D;X) = \Big\{p(\hat{x}|x)\;\; :\;\; & p(\hat{x}|x)\in  \mathbfcal{P}_X(D;X), \\
     & I(\widehat{X};X) = R_X(D;X)\Big\}
\end{aligned}
\end{equation*}
We call the distributions in this set \emph{$X$-optimal} for 
the corresponding distortion $D$. When this set contains more than one distribution, we have multiple alternative approximations of the input, which are indistinguishable when evaluated by their rate-distortion at $X$. Next, we define the \emph{$X$-optimal induced distortion set of $T$} as the set of all possible  task distortion values corresponding to these aforementioned distributions:  
\begin{equation*}
    \resizebox{\hsize}{!}{$\mathbfcal{D}^*_{T}(D_X;X) = \bigg\{\mathbb{E}\Big[d_T\big(f(x),f(\hat{x})\big)\Big] : p(\hat{x}|x) \in \mathbfcal{R}_X(D_X;X)\bigg\}$}
\end{equation*}
Note the added subscript to $D_X$, denoting the difference in the numerical values of the average distortion at the input and in the set $\mathbfcal{D}^*_T$. These values are generally different due to $X$ and $T$ being different RVs and may be in completely different scales or units, depending on the RVs themselves and the distortion metrics $d_X$ and $d_T$. 

Similarly, we can define equivalent sets for some intermediate representation $Y$: $\mathbfcal{R}_T(D_Y;Y)$ and $\mathbfcal{D}^*_T(D_Y;Y)$\footnote{In some cases we may also consider the intermediate representation as our target giving us $\mathbfcal{D}^*_Y(D_X;X)$.}. 
We observe the $X$-optimal distortion set and note the following: if $|\mathbfcal{D}^*_{T}(D_X;X)|>1$, then two (or more) indistinguishably optimal solutions (in terms of input rate-distortion) will perform differently at the task output. In other words, even though one such distribution 
may perform better at our desired task, we have no way of preferring it over other solutions when looking only at the input distortion. Of course, this is true for the $Y$-optimal distortion set as well. Although this is a clear advantage of the supervised approach, we show next that there are further benefits compared with the unsupervised approach. 

Consider a situation where we have some way to pick the best alternative in terms of task distortion from our $X$-optimal (or $Y$-optimal) compression methods. We show next that even under these conditions, optimising directly at the task achieves no worse rate-distortion performance.
\begin{theorem}\label{thm:task-optimal-non-strict} 
For a given input distortion $D_X>0$, and the corresponding lowest possible task distortion achievable by an $X$-optimal approximation, $D_T^{min} = \min \mathbfcal{D}^*_{T}(D_X;X)$,  the minimal achievable rate of the supervised approach is upper-bounded by the input rate-distortion (for corresponding distortions). Formally:
$$R_X(D_T^{min};T)\leq R_X(D_X;X).$$
\end{theorem}
\noindent All of this also holds when using an intermediate representation for optimising our coding scheme, directly resulting in the following.
\begin{corollary}\label{cor:task_optimal_non_strict} 
    For a given distortion $D_Y>0$ measured at an intermediate representation $Y$, and the corresponding lowest possible task distortion achievable by a $Y$-optimal approximation, $D_T^{min} = \min \mathbfcal{D}^*_{T}(D_Y;Y)$,  the minimal achievable rate of the supervised approach is upper-bounded by the intermediate-representation rate-distortion (for the corresponding distortion values). Formally:
$$R_Y(D_T^{min};T)\leq R_X(D_Y;Y) = R_Y(D_Y;Y).$$
\end{corollary}
\noindent Equivalently, and of interest when choosing which layer to use for distillation, we can use the logic of Theorem~\ref{thm:task-optimal-non-strict}, but consider the latent representation $Y$ as our target:
\begin{corollary} \label{cor:unsupervised-mid-input}
For a given input distortion $D_X>0$, and the corresponding lowest possible intermediate distortion achievable by an $X$-optimal approximation, $D_Y^{min} = \min \mathbfcal{D}^*_{Y}(D_X;X)$,  the minimal achievable rate of distilling the intermediate layer directly is is upper-bounded by the input rate-distortion (for the corresponding distortion values). Formally:
$$R_X(D_Y^{min};Y)\leq R_X(D_X;X).$$
\end{corollary}

Up to this point we have only shown that the supervised approach achieves no worse rate-distortion performance, we show next that under very reasonable conditions, the supervised approach is strictly better.
\begin{theorem}\label{thm:task-optimal-strict} 
Begin with a set of $X$-optimal distributions $\mathbfcal{R}_X(D_X;X)$, and a corresponding lowest possible task distortion, $D_T^{min} = \min{\mathbfcal{D}}^*_{T}(D_X;X)$. If, for any $p(\hat{x}|x)\in\mathbfcal{R}_X(D_X;X)$, there exist two points $\hat{x}_1 \neq \hat{x}_2$ with non-zero probabilities, $p(\hat{x}_1), p(\hat{x}_2)\neq 0$, for which the task output is identical,\footnote{In fact this only has to hold for $p(\hat{x}|x)\in\mathbfcal{R}_X(D;X)$, which also satisfy $\mathbb{E}\left[d_T\left(f(X),f(\widehat{X})\right)\right] = D_T^{min}$} $f(\hat{x}_1) = f(\hat{x}_2)$, and at least one input $x$ for which $p(x|\hat{x}=x_1)\neq p(x|\hat{x}=x_2)$, then the minimal achievable rate of the supervised approach is strictly lower than the input rate-distortion (for the corresponding distortion values). Formally:
$$ R_X(D_T^{min};T) < R_X(D_X;X) . $$ 
\end{theorem}

Once again, the logic of Theorem~\ref{thm:task-optimal-strict} can be applied to distillation of an intermediate layer, giving the following result.
\begin{corollary}\label{cor:task_optimal_strict} 
    Under equivalent conditions to Theorem~\ref{thm:task-optimal-strict}, for two points $\hat{y}_1,\hat{y_2}$, the minimal achievable rate corresponding to distilling an intermediate layer $Y$ directly, is strictly higher than that of the supervised approach (for corresponding distortion values). Formally: 
    $$R_Y(D_T^{min};T) < R_Y(D_Y;Y).$$
\end{corollary}
\noindent Of course, as in Corollary~\ref{cor:unsupervised-mid-input}, we can consider the intermediate representation as our actual target, which would lead to the following result.   
\begin{corollary}
For two inputs $\hat{x}_1,\hat{x}_2$ with identical intermediate representation $g(\hat{x}_1) = g(\hat{x}_2)$, and under equivalent conditions\footnote{It is important to note, that for practical task deep-models, these conditions are far less likely to hold for an intermediate representation then they are for the task labels.} to Theorem~\ref{thm:task-optimal-strict}, the minimal achievable rate corresponding to distilling an intermediate layer $Y$ directly, is strictly lower than the input-rate-distortion (for corresponding distortion values): 
    $$R_X(D_Y^{min};Y) <  R_X(D_X;X).$$
\end{corollary}

Summarising our theoretic results we can draw some important conclusions, at least \emph{in the limits of optimal encoding}: 
\begin{itemize}
    \item When training a compression method to optimise rate alongside task-distortion, all three considered approaches for coding for machines (full-input, model splitting, and direct) are equivalent.
    \item Optimising the compression model using the desired task labels will achieve better rate-distortion performance than what is achievable by attempting to use any unsupervised proxy, as intuitively expected.
    \item The choice of the cut point does not affect rate-distortion performance, but the choice of the distillation point does. The latter means that we might be inclined to choose the cut point based on practical considerations for a given task, platform, etc.
\end{itemize}
\section{Design Considerations}
\label{sec:design}
Besides the conclusions presented above, a natural question to ask is whether deeper distillation points lead to better rate-distortion? Although the intuitive answer seems to be affirmative, the theory presented above does not yet provide a definitive proof, except in the special case when one of the points is the task output (i.e., the supervised case in Corollaries~\ref{cor:task_optimal_non_strict},~\ref{cor:task_optimal_strict}). The question was tackled in our previous work~\cite{pcs}, where an affirmative answer was proved under certain conditions. Unfortunately, those conditions may be too restrictive in practice, and the search for the less-restrictive proof continues.   
In the meantime, we believe the benefits of using deeper points can be explained by further investigating the distortions used for distillation and their relationship with task labels.
\vspace{-0.3cm}

\subsection{Distillation, Distortion, and Task Performance}\label{subsec:appropriateness}
We explore the relationship above by considering the case of classification, which 
can be generalised to include many popular computer vision (CV) tasks such as object detection and instance/semantic segmentation. In classification, we can consider task labels (and thus an optimal task-model) to be a clustering of the inputs where each cluster contains all the inputs corresponding to that label. Given a classification problem with $C$ classes, we denote the clusters as follows: 
\begin{equation}
    \mathbfcal{X}^{(i)}_T = \{x\in\mathbfcal{X}:f(x) = i\},\;\; i = 1,2,...,C.
\end{equation}
Next, using our input distortion $d_X(x,\hat{x})$, define the expected distortion of cluster $i$ relative to $\bar{x}$ as
\begin{equation}
    d_X^{(i)}(\bar{x}) = \sum_{x \in \mathbfcal{X}^{(i)}_T}p(x|X\in \mathbfcal{X}^{(i)}_T)d_X(x, \bar{x}).
\end{equation}
Then, intra-cluster distortion can be defined as follows:
\begin{equation}
    D^{(i)}_X = d_X^{(i)}(\bar{x}^{(i)}), \quad \text{where} \; \; \bar{x}^{(i)} = \argmin_{\bar{x}\in\mathbfcal{X}^{(i)}_T}d_X^{(i)}(\bar{x}).
\end{equation}
Here, $\bar{x}^{(i)}$ is the centroid of the cluster in terms of the distortion metric $d_X(\cdot,\cdot)$. If, for example, the distortion is mean square error (MSE), a common metric for image compression, then $\bar{x}^{(i)} = \mathbb{E}\left[X|X\in\mathbfcal{X}^{(i)}_T\right]$ is the cluster mean, and $D^{(i)}_X = \text{var}\left(X|X\in\mathbfcal{X}^{(i)}_T\right)$ is the cluster variance. Similarly, we can define the inter-cluster distortion of two clusters using their respective centroids:
\begin{equation}
    D^{(i,j)}_X = d_X(\bar{x}^{(i)}, \bar{x}^{(j)}).
\end{equation}

Finally, to get a measure of how well the task clustering coincides with the distortion on the input itself, we can use a variation of the silhouette score~\cite{rousseeuw1987silhouettes} as follows:

\begin{align}
    \rho^{(i,j)}_X &= \frac{D^{(i,j)}_X}{\sqrt{D^{(i)}_XD^{(j)}_X}}\\
    \rho^{(i)}_X &= \min_{j\in\{1,2,...,C\}}\rho^{(i,j)}_X \\
    \rho_X &= \mathbb{E}\left[\rho^{(i)}_X\right] = \sum_{j=1}^C Pr(X\in\mathbfcal{X}^{(i)}_T)\rho^{(i)}_X
\end{align}
Under this formulation, the higher $\rho^{(i,j)}_X$ is, the easier it is to distinguish between the two clusters $\mathbfcal{X}^{(i)}_T,\mathbfcal{X}^{(j)}_T$ using the input distortion $d_X$. Consequently $\rho^{(i)}_X$ is high if the cluster $\mathbfcal{X}^{(i)}_T$ is easily distinguishable from all other clusters. We name the final expression, $\rho_X$, the \emph{task-appropriateness} of the distortion $d_X(\cdot,\cdot)$. When $\rho_X$ is high, it means that by simply minimizing the distortion on $X$ we can get good clustering with respect to our task $T$. 

Next we 
examine the effect of optimising distortion at the input compared to doing so at an intermediate layer $Y$, and analogously, between different 
layers $Y_1$ and $Y_2$. We can do this by defining an equivalent clustering for the intermediate representations of each input:
\begin{equation}
    \mathbfcal{Y}^{(i)}_T = \{y\in\mathbfcal{Y}:h(y) = i\},\;\; i = 1,2,...,C.
\end{equation}
Using this clustering we can now define the same set of metrics defined above: the centroid $\bar{y}^{(i)}$, the inter and intra cluster distortions $D^{(i)}_Y,D^{(i,j)}_Y$, and the silhouette-like scores $\rho^{(i,j)}_Y, \rho^{(i)}_Y,\rho_Y $. Once again, importantly, the higher the value of $\rho_Y$, the more appropriate the distortion metric $d_Y$ is for clustering according to our task labels.  

We claim that in many deep classification models, it can be shown that using equivalent distortion, such as MSE, the task-appropriateness of intermediate representations improves as we proceed deeper into the model. That is, given our previous notation of $X,Y_1,Y_2$:
\begin{equation}
    \rho_X\leq\rho_{Y_1}\leq\rho_{Y_2}
\end{equation}
In our experiments presented in Section~\ref{sec:empirical_evidence}, we demonstrate that this holds in practice: first for a toy problem, and subsequently for several classification datasets and DNN models. Later, in Section~\ref{sec:applications}, we will show that in a practical system, this allows us to improve rate-distortion by choosing a deeper distillation point.

\textcolor{black}{For other tasks, popular architectures (e.g. U-net or autoencoders) reduce the dimensions of the latent space up to a certain point, often referred to as the bottleneck, and then increase its dimensions back up in subsequent layers. In such models, we hypothesise that the task-appropriateness of a norm based distortion metric (such as MSE or MAE) will be maximised at the bottleneck, for two reasons. First, as the network reduces the dimensions of the latent representation it must learn to discard information. Once the dimensions begin to increase the network is no longer incentivized to lose information (though it cannot add additional information due to the data-processing inequality). Second, as the dimensions increase (but no new information is introduced), the representations inherently must contain redundancies - features that have limited or overlapping effect on the output of the model. Because the norm based distortion metric treats all features equally, the resulting clustering may be distracted by irrelevant features, reducing task appropriateness. Later in Section~\ref{subsec:direct} we see an example where the best distillation point is, in fact, the most compact layer in terms of latent space dimensions.} 

\subsection{Empirical Evidence}
\label{sec:empirical_evidence}
\subsubsection{Toy Example}
We begin in a setting where optimal quantisation can be analytically derived. Consider the classification task of two classes, which we will call ``circles'' and ``squares'', located in a two-dimensional space (which, to distinguish from our previous notation, we label $u,v$). For both classes, $u$ values are uniformly distributed in $[0,10]$, but $v$ values differ: for the ``squares'', $v$ values are uniformly distributed in $[0.8,1.2]$, whereas for the ``circles'', $v$ values are uniformly distributed in $[2.8,3.2]$. In both cases, $u,v$ values are independent. Next, we design a task model with two steps, similar to our formulation: $f = h \circ g$, such that $T =f(X)$ and $Y = g(X)$, where:
\begin{equation*}
\small
\begin{aligned}
        g(u,v) & = \begin{cases}
            (2.5 + \text{sgn}(u\text{--}5)\cdot \sqrt{|0.2^2\text{--}(v\text{--}3)^2|},v), \; v>2, \\
            (7.5 + \text{sgn}(u\text{--}5)\cdot \sqrt{|0.2^2\text{--}(v\text{--}1)^2|},v), \; v\leq2,
        \end{cases}\\
        h(u,v) &= \begin{cases}
            \text{``circle''},\quad v>2, \\
            \text{``square''}, \quad v\leq2.
        \end{cases}
\end{aligned}
\end{equation*}

\begin{figure}[htbp] 
  \centering
  \includegraphics[width=0.95\linewidth]{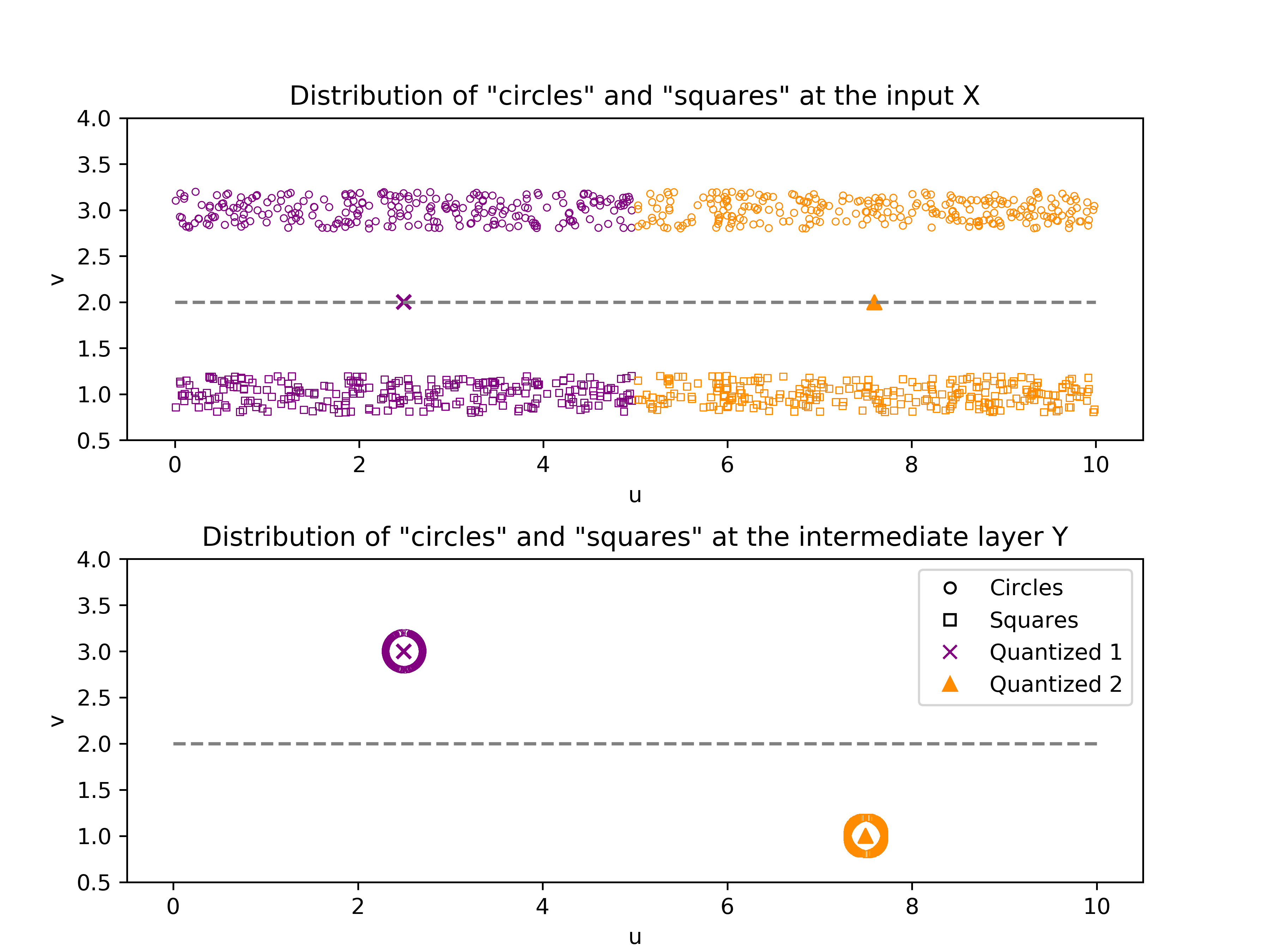}
   \caption{Distribution and quantisation of ``circles'' and ``squares'' at both the input and the intermediate layer. The marker symbol corresponds to the class of each point, whereas the color corresponds to the 1-bit bin to which each point is quantised to minimise MSE.}
   \label{fig:toy}
\end{figure}

\figref{fig:toy} shows the distribution of the two classes at the input $X$ and the intermediate layer $Y$. By using MSE\footnote{MSE here uses the $\ell_2$ norm squared, meaning $d^{MSE}\left((u_1,v_1),(u_2,v_2)\right) = (u_1-u_2)^2 + (v_1-v_2)^2$.} as our distortion for both the input $X$ and the intermediate layer $Y$ we can calculate the task-appropriateness of MSE at each layer. We get $\rho_X = \rho_X^{\text{squares}} = \rho_X^{\text{circles}} \approx 0.48$, and $\rho_Y = \rho_Y^{\text{squares}} = \rho_Y^{\text{circles}} = 725$.
\vspace{-0.2cm}

Next, we consider the optimal 1-bit quantisation, and observe how the choice of distillation point affects task performance under such rate-constrained quantisation.\footnote{We choose to start at the rate, which is equivalent to investigating the distortion-rate, rather than the rate-distortion, for convenience, as the two are equivalent.} It is simple to show that in the input space $X$, the optimal 1-bit quantisation in terms of MSE 
is simply to use the two bins $u\leq5,\,u>5$ and quantised values $\hat{x}_1 = (2.5,2),\, \hat{x}_2 = (7.5,2)$. In intermediate layer $Y$, the points in each class are distributed on the circumference of a circle with different centers, with ``circles'' centered at $(2.5,3)$ and ``squares'' centered at $(7.5,1)$. Here there are many equivalent optimal bin choices (due to the large empty space between the two distributions) but for simplicity, the same bins as in the input space can be used, leading to the quantised values $\hat{y}_1 = (2.5,3),\, \hat{y}_2 = (7.5,1)$. The quantisation bins are indicated in \figref{fig:toy} by the color of each point, with the quantised values described by correspondingly colored marker.

By observing \figref{fig:toy} we can see clearly that using quantisation optimised for MSE at the input leads to a classification error of $50\%$. This is because each bin contains an equal amount of observations from both classes. Since each bin is assigned a single corresponding value, the task-model $f$ can only output one label for it (in fact, both bins are assigned ``square''), 
leading to the aforementioned classification error.  In contrast, when optimising quantisation for MSE at the intermediate layer $Y$, we can see that all observations in each of the two bins are from the same class, and their representative values are such that the task back-end $h$ will assign them the correct label, leading to perfect classification. Thus, under the 1-bit rate constraint and using MSE distortion, a deeper representation ($Y$) offers better performance for the classification task than the input ($X$). 

Although highly contrived, the example above is helpful in understanding the relationship between task-appropriateness of a distillation point, the distortion metric,\footnote{The toy classification problem would have been easy to solve even in the input space if we used a distortion metric other than MSE.} and rate-distortion. In general, however, we do not know the exact distribution of inputs in each class, nor do we have an optimal task model or optimal compression. Instead, we first demonstrate (Section~\ref{sec:deep_classification}) that the behaviour of task-appropriateness remains consistent with our claim in practical deep classification models. Later in Section~\ref{sec:applications}, 
we will show how this translates to improved rate-distortion in the unsupervised training of learned codecs for machines.

\subsubsection{Deep Classification Models}
\label{sec:deep_classification}
\begin{figure}[ht]
    \centering
    \begin{subfigure}[b]{0.49\linewidth}
        \centering
        \includegraphics[width=0.9\textwidth]
        {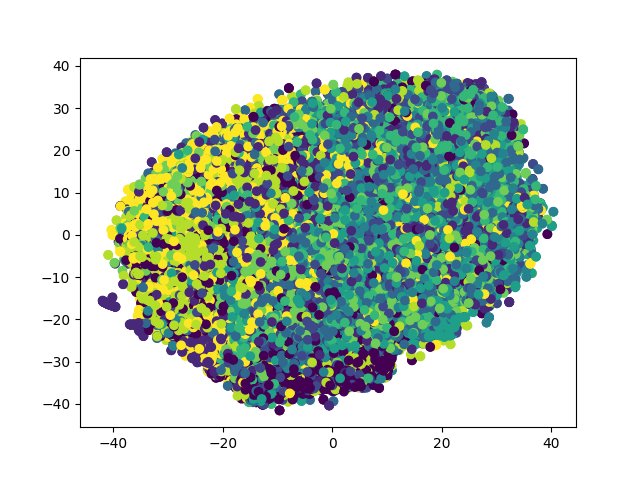}
        \caption{Input $\rho = 0.038$}
    \end{subfigure}
    \hfill
    \begin{subfigure}[b]{0.49\linewidth}
        \centering
        \includegraphics[width=0.9\textwidth]{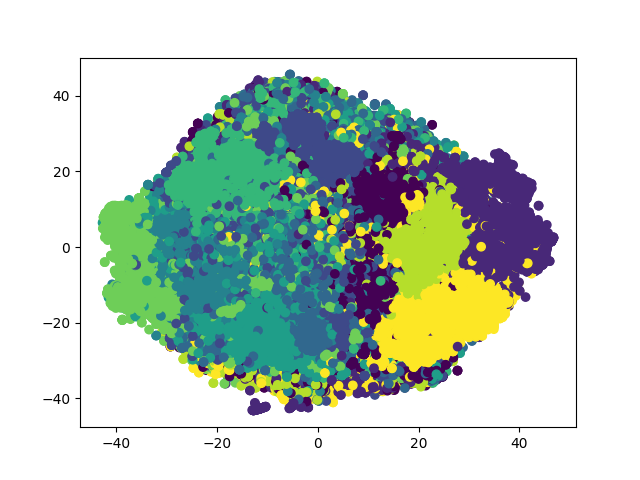}
        \caption{Features.19 $\rho = 0.052$}
     \end{subfigure}

    \begin{subfigure}[b]{0.49\linewidth}
        \centering
        \includegraphics[width=0.9\textwidth]{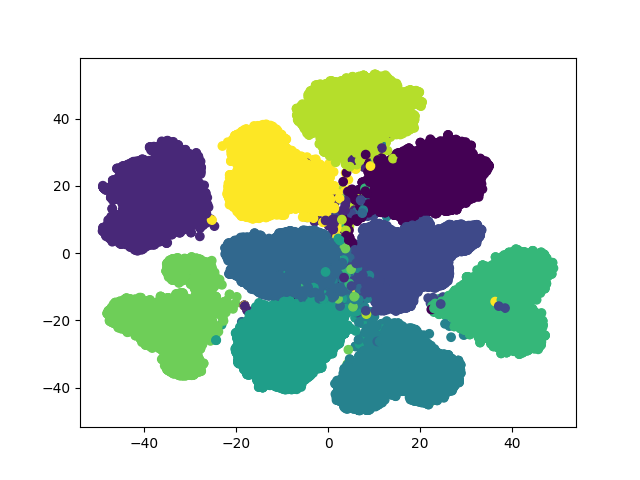}
        \caption{Features.26 $\rho = 0.517$}
     \end{subfigure}
    \hfill
    \begin{subfigure}[b]{0.49\linewidth}
        \centering
        \includegraphics[width=0.9\textwidth]{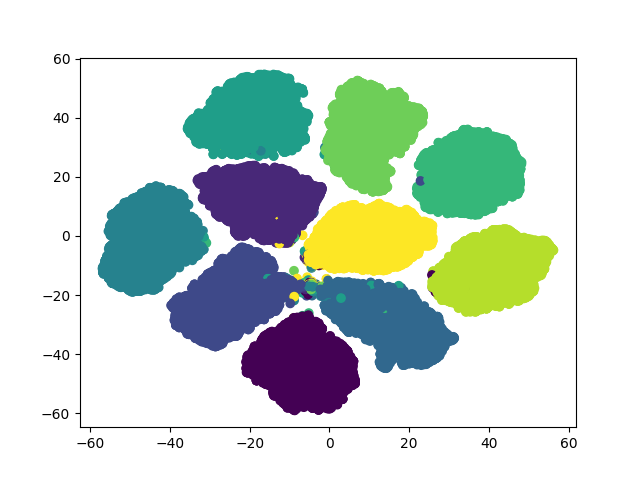}
        \caption{Features.32 $\rho = 1.0$}
       \end{subfigure}

    \caption{Task appropriateness and t-SNE visualisation for various layers in VGG16, using the CIFAR-10 dataset and MSE distortion. The improvement in $\rho$ values suggests that using deeper layers, such as 'Features.32' will lead to better rate-distortion performance for this task. 
    }
    \label{fig:task_app_cifar10_vgg}
\end{figure}

In order to examine the behaviour of task-appropriateness in practical models, we consider two well known classification models - VGG16\cite{vgg} and ResNet50\cite{resnet}, and two well known datasets - CIFAR-10 and CIFAR-100~\cite{krizhevsky2009learning}. We use MSE as our distortion metric, which allows us to use $\bar{x}^{(i)} = \mathbb{E}\left[X|X\in\mathbfcal{X}^{(i)}_T\right]$. We replace the expectation with a sample mean, and calculate the average task-appropriateness score for several layers along each model. Because the representations are very high dimensional, we utilise t-SNE~\cite{van2008visualizing} to help visualise the relationship between the spatial clusters and the class labels. \figref{fig:task_app_cifar10_vgg} shows this relationship for several layers of VGG16, using MSE, for the CIFAR-10 dataset. Additional similar figures are included in Appendix~\ref{app:appropriateness} of the supplemental material. As seen in Table~\ref{tbl:task-app}, where bold type indicates the highest value for a given (model, dataset) pair, in both models and both datasets, our claim regarding the behaviour of the task-appropriateness holds. This result, combined with the toy-example, suggests that there is good reason to expect that distilling deeper layers results in improved rate-distortion for machines using deep task-models.

\vspace{0.2cm}
\begin{table}[h]
\centering
\caption{Task-Appropriateness Using Mean Squared Error for Various Layers in Deep Classification Models}
\label{tbl:task-app}
\resizebox{0.9\linewidth}{!}{%
\begin{tabular}{@{}cccc@{}}
\toprule
Dataset & Model &  Layer & Task-  \\ 
        &       &        & Appropriateness $\rho$ \\
\midrule \\
CIFAR-10  & -                  & Input       & $0.038$ \\ \midrule
          & VGG16               & Features.19 & $0.052$ \\
          &                     & Features.26 & $0.517$ \\ 
          &                     & Features.32 & $\mathbf{1.0}$ \\ 
          & ResNet50              & Layer2      & $0.021$ \\
          &                     & Layer3      & $0.102$ \\
          &                     & Layer4      & $\mathbf{1.0}$ \\\midrule
CIFAR-100 &  -                  & Input       & $0.036$ \\ \midrule
          & VGG16               & Features.19  & $0.046$ \\
          &                     & Features.26 & $0.120$ \\
          &                     & Features.32 & $\mathbf{0.555}$ \\ 
          & ResNet50              & Layer2      & $0.032$ \\
          &                     & Layer3      & $0.050$ \\
          &                     & Layer4      & $\mathbf{0.997}$ \\
          \bottomrule
\end{tabular}
    }
\end{table}

\section{Image Coding for Machines} \label{sec:applications}

The theoretical analysis, alongside our motivating hypotheses and corresponding empirical evidence, can be combined to inform critical design choices in coding for machines. \textcolor{black}{We focus specifically on applications of our theory in image coding for machines, as it is the domain with the largest body of work in CfM. In this section we demonstrate how our theory can be leveraged in several different settings in image coding for machines, including a variety of computer vision tasks.}  We explore the performance of both the model-splitting approach and direct coding for machines on a combination of supervised and unsupervised optimisation settings. In all settings we use learned image codecs, with a loss function introduced earlier in~\eqref{eq:learned}:

\begin{equation*}
    L = R + \lambda D,
\end{equation*}
where $R$ is the rate and $D$ is the task-related distortion. 

\begin{figure}[htbp]
  \centering
  \includegraphics[width=0.7\linewidth]{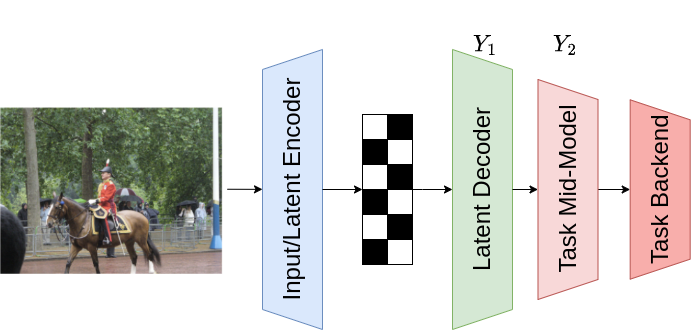}
  
  \caption{Model distillation approach - Note that our latent decoder only recreates the cut-point $Y_1$. To obtain the distillation loss at the point $Y_2$, we make use of pre-trained layers of the original task-model, which we denote the task mid-model.}
  \label{fig:mid-model}
\end{figure}

In our supervised setting experiments, we have access to task labels and thus use task-distortion directly as $D$. On the other hand, in all unsupervised experiments, task-labels are assumed to be unavailable, meaning we must use some feature tensor as a target. In such cases we refer to $D$ as distillation loss, and the training process as model distillation (sometimes known as knowledge-distillation~\cite{gou2021knowledge}). We draw upon the results of Corollaries~\ref{cor:design_considerations}, \ref{cor:direct_design_considerations}, and \ref{cor:distillation}, and decouple the choice of cut point from that of the distillation point. Furthermore, we utilise pretrained layers from the original task-model, which we refer to as the \emph{mid-model}, as part of the decoding process. Model distillation is thus performed as a two-step procedure, which can be seen in \figref{fig:mid-model}. First our codec is used to recreate the output of the chosen cut point; the resulting quantised feature tensor is then passed to the mid-model to obtain the reconstructed distillation point; which is then compared to its uncompressed equivalent derived from the same input. 

In all tasks we present rate-accuracy curves (rather than rate-distortion) and utilise the Bj\o{}ntegaard Delta (BD) metric~\cite{Bjontegaard, vcm_template}, a well-established metric for estimating the average difference between two such curves. The original BD metric is design using peak signal to noise ratio (PSNR), which of course is not available in the case of most computer vision tasks. Instead, we use modified BD metrics replacing the PSNR with an appropriate task metric such as classification accuracy, mean average precision, etc. Considering the significant number of different experimental settings we present, the description of each setup in this section will be limited, with further details regarding task-models, codecs, and training procedures available in Appendix~\ref{app:model_details} of the supplemental material. Furthermore, because of the difference between the CV tasks and compression settings, we present comparable work for each experiment individually.

\subsection{Model Splitting}\label{subsec:model_splitting}

We have seen, in Theorems~\ref{thm:task-optimal-non-strict} and~\ref{thm:task-optimal-strict}, that supervised optimisation of compression models for machines has better theoretic rate-distortion performance. We also showed that the choice of cut point does not, in the optimal case, change rate-distortion performance. Furthermore, in Section~\ref{sec:design} we showed that classification task-appropriateness of MSE distortion on feature tensors increases as we go deeper into the task-model, suggesting that distilling deeper model layers is preferable in unsupervised optimisation. Thus, our first experiment aims to compare the compression performance of various supervised and unsupervised variations of an otherwise equivalent model-splitting coding scheme. 

Our model-splitting coding scheme is used to evaluate rate-distortion performance for an image classification task. To make our approach viable in a real-world setting we impose the following additional constraints:
\begin{enumerate}
    \item The latent encoder and decoder should have low computational complexity compared to the original DNN model being considered (which comprises the task frontend and backend in \figref{fig:approach-split}) so that the overhead introduced by our method during actual deployment is low.
    \item The parameters (weights etc.) of the original DNN model remain frozen throughout the entire process. 
    Thus, whenever the compression model is trained using the loss in \eqnref{eq:learned} -- either for a different cut-point, or for a different $\lambda$ value -- it is only the parameters of the latent encoder and decoder that are updated, not those of the original model.
\end{enumerate}
Such constraints have been considered in our recent work~\cite{datta2022low, ahuja2023neural} and are motivated by conditions that exist in real-world setups. To abide by these constraints, we restrict the latent encoder to be a single depth-wise separable convolutional layer. We do this because this method has the fewest parameters and lowest complexity compared to other topologies~\cite{howard2017mobilenets}
that transform an intermediate tensor of size $H\times W \times C$ into a lower dimensional tensor of size $H_r\times W_r \times C_r$, such that $H_r \leq H,W_r \leq W,C_r \leq C$. $W_r$ and $H_r$ are related to $W$ and $H$ by $W_r=W/S$ and $H_r=H/S$, where $S$ is the stride factor of the convolutional kernel. 

The encoder is followed by quantisation with step size $Q$, and an entropy coder.  The decoder contains an entropy decoder followed by a simple mirror image of the encoder. As explained in \cite{datta2022low}, the rate-distortion performance is influenced by architectural hyper-parameters of the latent encoder such as the stride $S$, and number of output channels $C_r$, and also by compression-related hyperparameters such as the Lagrange multiplier $\lambda$ and the quantisation step-size $Q$. These hyperparameters interact in complex ways to determine the eventual rate-distortion performance. We therefore have to search this hyperparameter space to arrive at an optimal set. We follow the search space procedure described in \cite{datta2022low} to arrive at a set of Pareto-optimal design points for the latent encoder and decoder. More details are provided in Appendix~\ref{appsec:modelsplit} of the supplemental material. 

\textcolor{black}{We test our model-splitting scheme on two tasks -- image classification and semantic segmentation} --  using both supervised and unsupervised approaches. The models are trained with a rate-distortion loss objective from \eqnref{eq:learned}. In the unsupervised approach, the distortion term $D$ is essentially an MSE-based distillation loss, while for the supervised approach, \textcolor{black}{the distortion loss are the usual task losses: cross-entropy loss for classification, and sum of per-output-pixel cross-entropy losses for segmentation}. For the rate-loss $R$, we use a neural rate estimator from our previous work~\cite{ahuja2023neural} to estimate the rate at the output of the latent encoder. Briefly, the lower-dimensional latent encoder outputs $Z$ ($y$ in \cite{ahuja2023neural}) are interpreted as reduced-dimension latent-representation of the feature tensor at the split point $Y$ ($x$ in~\cite{ahuja2023neural}). Inspired by \cite{balle2018variational}, we derived a set of hyper-latents $Z_h$ from $Z$ by using an additional variational auto-encoder model for the hyperprior. We use a Gaussian scale-hyperprior, meaning $Z|Z_h\sim\mathcal{N}(0,\sigma(Z_H))$ where $\sigma$ is the output of the hyperprior decoder. The overall rate is estimated as the sum of the rates of the $Z$ and $Z_h$ respectively (More details are provided in Appendix~\ref{appsec:modelsplit} of the supplemental material):
\begin{equation}
\label{eq:variational_rate}
    \mathcal{R} = \mathbb{E}\left[ -\log_2 p(Z_h) - \log_2 p(Z|Z_h)\right]
\end{equation}

\begin{figure}[htbp]
  \centering
  \includegraphics[width=0.85\linewidth]{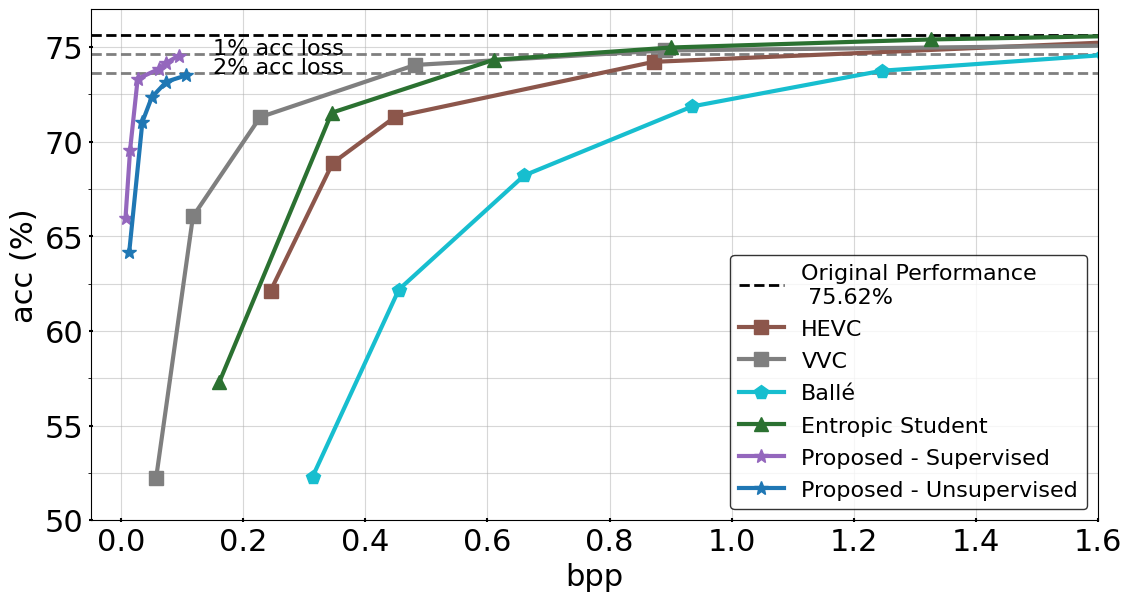}
  \caption{Benchmark comparison for image classification in a model-splitting setup using ResNet50, on the Imagenet validation set.}
  \label{fig:classification_benchmark}
\end{figure}

\begin{figure}[htbp]
  \centering
  \includegraphics[width=0.85\linewidth]{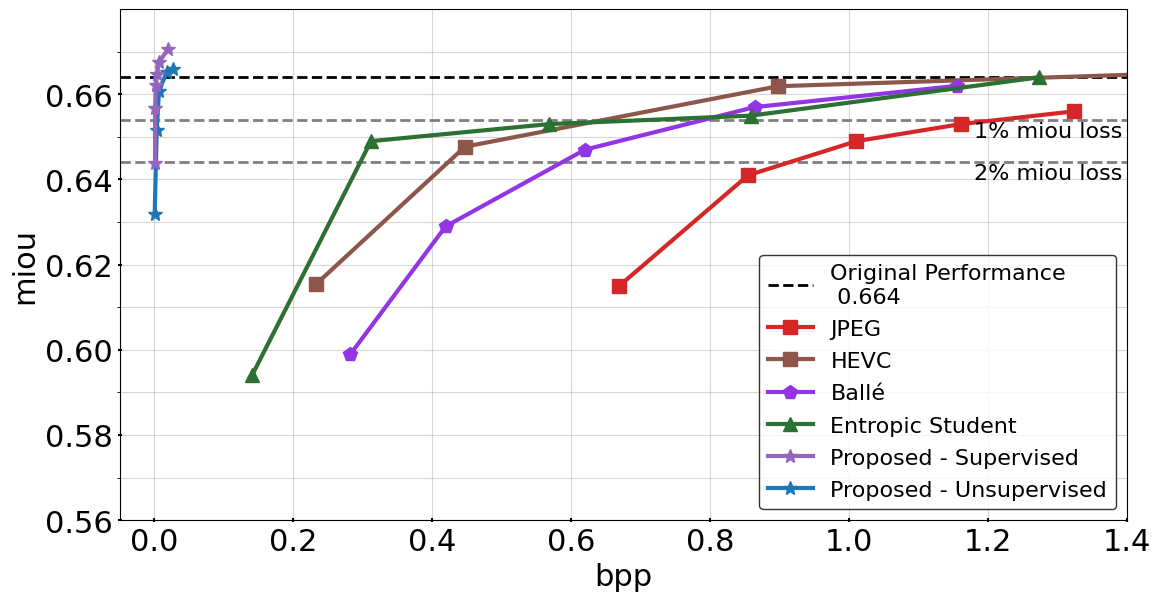}
  \caption{Benchmark comparison for semantic segmentation in a model-splitting setup using Deeplab-v3 with a Resnet50 backbone, on COCO 2017 validation set.}
  \label{fig:segmentation_benchmark}
\end{figure}

\textcolor{black}{For the classification task, we select a ResNet50~\cite{resnet} network trained on the ImageNet dataset~\cite{imagenet2015} as our task model; for the segmentation task, we select a Deeplab-v3 \cite{deeplabv3} network with a ResNet50 backbone trained on the COCO 2017 dataset~\cite{COCO}. In both cases, the evaluation is performed as follows. First, we present results from \cite{ahuja2023neural} in which we compare our approach against state-of-the-art benchmarks which include two machine-learning-based compression methods -- variational image compression \cite{balle2018variational} and Entropic Student \cite{entropic_student} -- and two standards based compression -- the previously reported HEVC (and newly included VVC for the classification task). We see from both \figref{fig:classification_benchmark} and \figref{fig:segmentation_benchmark}}
that our method easily outperforms these benchmarks by achieving lower bit-rates for any given accuracy or mIOU level. Note that because our method achieves significantly lower bit-rates than the benchmark curves, there is insufficient overlap to calculate BD-metrics.

Next, to validate the theory developed in Section \ref{sec:RD_theory}, we evaluate in greater depth the impact of the choice of distillation layer for different cut-points for both classification and \textcolor{black}{semantic segmentation} tasks (as opposed to selecting only the last layer as the distillation layer for unsupervised training as was done in \cite{ahuja2023neural}). \textcolor{black}{Here, we select two different cut-points for each task ($C_1$, $C_2$ for classification, and $C_3$, $C_4$ for segmentation). For classification, we choose two different distillation points ($M_1$, $M_2$), with $M_2$ being the output of the last layer of the network (but before the softmax operation is applied), while for segmentation we choose three ($M_3$, $M_4$, and $M_5$) with $M_4$ being the output of the Resnet50 backbone in Deeplab-v3. See Appendices \ref{appsec:resnet50} and \ref{appsec:deeplab_v3} for a visualisation of the various cut-points and split-points.}

\begin{figure}[htbp]
  \centering
  \includegraphics[width=0.8\linewidth]{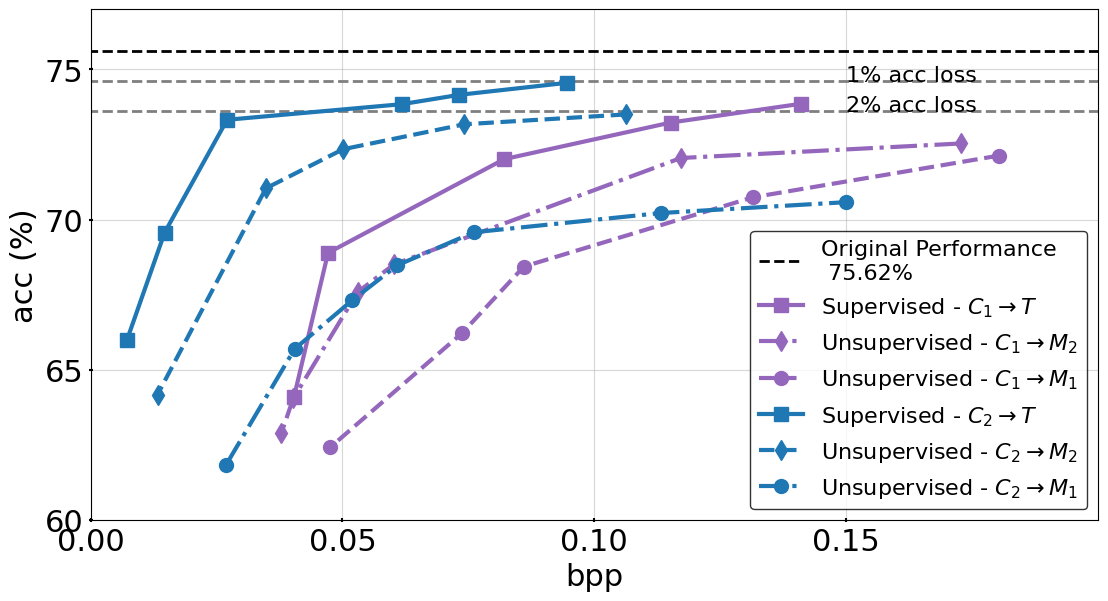}
  \caption{Impact of choice of distillation points for classification using Resnet50 on the Imagenet validation set. Note that as predicted by our theoretical analysis, deeper distillation points yield better performance curves.}
  \label{fig:classification_distillation}
\end{figure}

\begin{figure}[htbp]
  \centering
  \includegraphics[width=0.85\linewidth]{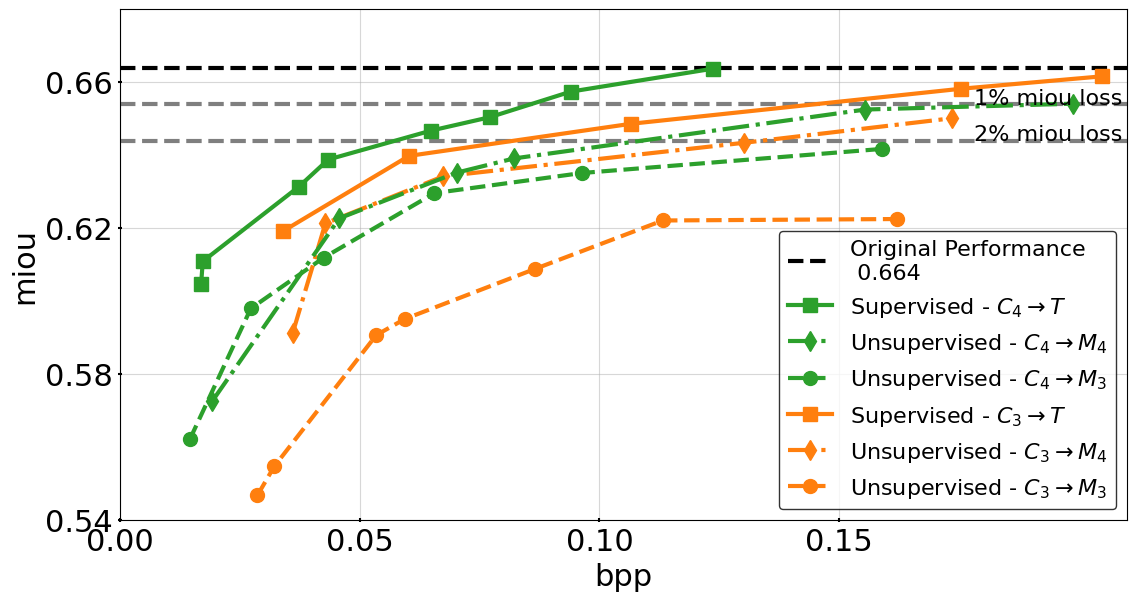}
  \caption{\textcolor{black}{Impact of choice of distillation points for semantic segmentation using Deeplab-v3 on the COCO 2017 validation set. Note that as predicted by our theoretical analysis,the deeper distillation point $M_4$ yields better performance curves than $M_3$.}}
  \label{fig:segmentation_distillation}
\end{figure}

\begin{figure}[htbp]
  \centering
  \includegraphics[width=0.85\linewidth]{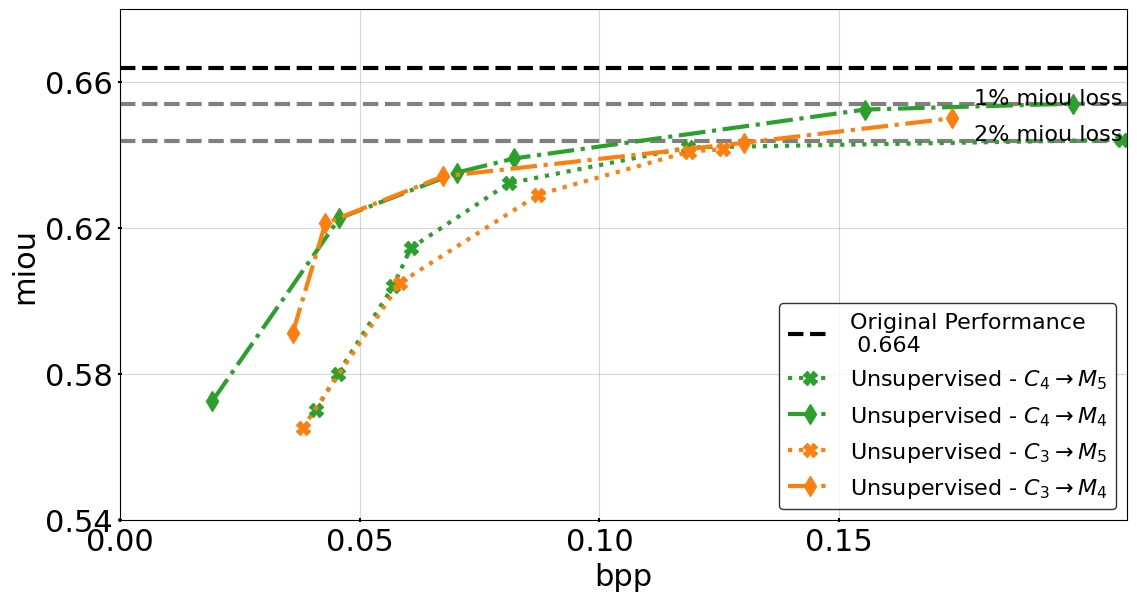}
  \caption{\textcolor{black}{Choosing distillation layers ($M_5$) downstream of the Resnet50 backbone output ($M_4$), which is the most compact in the Deeplab-v3 model, results in degraded performance.}}
  \label{fig:segmentation_distillation_downstream}
\end{figure}

We also reproduce some of the curves from \cite{ahuja2023neural} for the supervised approach which uses a supervised training loss (cross-entropy) for the task $T$. For each cut-point, therefore, we generate three curves: two using an unsupervised distillation loss, and one using a supervised task loss. These results are shown in \figref{fig:classification_distillation} \textcolor{black}{for classification and \figref{fig:segmentation_distillation} for semantic segmentation. In general,} we observe that as predicted by our theoretical analysis, choosing deeper distillation points leads to improved rate-distortion performance for both cut-points, and that using a supervised loss outperforms the unsupervised alternatives. 
\textcolor{black}{For segmentation, we observe that distilling the backbone output $M_4$ provides the best performance and that using the deeper $M_5$ actually results in somewhat degraded performance (see \figref{fig:segmentation_distillation_downstream}). This coincides with our analysis from section~\ref{subsec:appropriateness} as $M_4$ is the most compact representation within the network; subsequent layers in Deeplab-v3 are responsible for reconstructing a high-resolution map from this representation and offer no benefit from a compression perspective.}

Interestingly, though, we see that under this experimental setup the cut point does, in fact, change rate-distortion performance, with deeper cut points achieving better performance. Clearly this is a result of the design and computational constraints we have chosen for our coding scheme. To ensure this is the case, we increase the computational capacity and complexity of the compression model in our next set of experiments, and re-evaluate the effect of the different cut points in that setting.

\color{black}
\subsection{Direct Coding for Machines} \label{subsec:direct}

Having established that our theory regarding the choice of distillation point lends itself well to design considerations in a model-splitting coding scheme, we move on to the direct coding approach. As in the model-splitting experiments, we compare the rate-distortion performance of multiple possible choices of either cut point or distillation point in an otherwise identical model.  Additionally, as explained above, we remove the computational constraints on the coding scheme, to more closely approach the optimal limit of our theoretic formulation, especially regarding the effects of the choice of cut point. \textcolor{black}{For our first three experiments we focus on the more difficult unsupervised setting using three CV tasks-models which are featured in the standardisation efforts of coding for machines~\cite{vcm_call_for_evidence}} - object detection using Faster R-CNN~\cite{ren2015faster} as well as YOLOv3~\cite{Redmon2018_yolov3}, and image segmentation using Mask R-CNN~\cite{he2017mask}. \textcolor{black}{Finally, we add one additional experiment, in which we focus on outright RD performance, and efficiency. We do this by replacing the task model with the more modern SWIN-Transformer~\cite{Liu_2021_ICCV}, while also utilising a more modern learned codec, ELIC~\cite{he2022elic}.}

It is important to note that more complex CV tasks, such as the ones in question, are often best performed by multi-scale models, which process an input image at several resolutions at once.  Often, doing this involves multi-stream processing in which shallower feature tensors are still needed for the task backend, even in the presence of deeper ones. This means that when we perform model distillation on a multi-stream model, we must take special care in choosing distillation points that adequately cover all processing streams. For example, we may choose an early layer, before the computation has split into multiple streams, but this greatly limits the depth of our distillation point. Alternatively, we may choose several feature tensors which together ensure all processing streams are accounted for, allowing us to effectively select a deeper distillation point.

In \textcolor{black}{the first 3 task} experiments, we use an identical compression model, similar to the "base-layer" of the scalable codec in~\cite{hyomin}, which in turn is largely based on~\cite{cheng2020image}. First, a synthesis transform is used to produce a latent representation. This transform is comprised of downsampling blocks, as well as residual convolutional blocks, all using generalised divisive normalisation (GDN~\cite{gdn}) activations. Next the latent representation is quantised and encoded using an autoregressive hyperprior entropy model~\cite{minnen2018joint}, followed by arithmetic encoding. After decoding the resulting bitstream, the recreated latent representation is processed by a latent decoder\footnote{This was referred to as the latent space transform (LST) in~\cite{hyomin}.} comprised of residual blocks and inverse GDN activations, as well as upsampling blocks. The output of the latent decoder is used as our recreated cut point, which can then be fed to the mid-model to obtain the distillation point during training, or to the full task backend during inference. 

\textcolor{black}{For our final experiment in this direct-coding approach, we utilise a variant of the more modern and efficient ELIC~\cite{he2022elic} learned codec. In ELIC, efficiency is improved by replacing the autoregressive context model of~\cite{cheng2020image} with a space-channel context model (labeled SCCTX). The SCCTX is composed of a spatial checkerboard~\cite{he2021checkerboard} as well as an unevenly sized channel-group context model (building on ~\cite{minnen2020channel}) and a parameter aggregation model connecting the two. Overall, ELIC achieves improved performance compared with~\cite{cheng2020image} while also greatly reducing latency computational cost. As such an efficient codec, it is highly beneficial for the setting of coding for machines.} Further details on the architecture of our direct-coding models can be seen in Appendix~\ref{appsec:compression_model} of the supplemental material. 

\subsubsection{Object Detection Using Faster R-CNN}\label{subsubsec:Faster}

Our first task-model used for direct-coding is Faster R-CNN~\cite{ren2015faster}, a well established benchmark in object detection. The Faster R-CNN architecture is comprised of a feature proposal network (FPN) based on a backbone model, followed by region proposal network (RPN), and finally Region of Interest (RoI) pooling to provide bounding boxes and class labels. More details regarding the full architecture of Faster R-CNN, and specifically the FPN, can be seen in Appendix~\ref{appsec:rcnn} of the supplemental material. Notably, this model is an example of a multi-stream model, which means our distillation points must be carefully chosen as explained above. On the other hand, Corollary~\ref{cor:direct_design_considerations} suggests that the choice of cut point does not have a strong impact on our rate-distortion performance. Thus we can greatly simplify our approach by choosing our cut points from the shallower, single-stream portion of the model.

For our experiments we use on ResNet50~\cite{resnet} as the backbone model and compare 3 possible distillation points and 2 cut points, all taken from the FPN portion of Faster R-CNN. Our cut points, in order of depth, are labelled $C_5$, and $C_6$ (to distinguish from those used in the \textcolor{black}{previous experiments}), and our distillation points include the two cut points, as well as the deeper $M_6$. We use MSE as our distillation loss metric, and compare the rate-distortion performance of the various configurations. For obvious reasons, we cannot use a deeper cut point than the corresponding distillation point, leading to 5 possible combinations. Importantly, because the point $M_6$ consists of  5 tensors of different dimensions, we choose to use the unweighted average of the 5 MSE losses as our distillation loss when using $M_6$:
\begin{equation} \label{eq:multi_distillation}
    MSE(M_6,\widehat{M}_6) = \frac{1}{5}\sum_{i=1}^5 MSE(Y_i,\widehat{Y}_i),
\end{equation}
where $Y_i$ are the tensors which comprise $M_6$. For details on the location of $C_5, C_6, M_6$ and the architecture of Faster R-CNN see Appendix~\ref{appsec:rcnn} of the supplemental material.

Training is performed in two stages using a combination of CLIC~\cite{clic_dataset}, JPEG-AI~\cite{jpeg_ai}, and VIMEO-90K~\cite{xue2019video_vimeo} datasets. For full details, including hyper parameters, see Appendix~\ref{appsec:rcnn} of the supplemental material.
The Faster R-CNN task model in all experiments was pretrained with the weights taken from the DetectronV2~\cite{wu2019detectron2} implementation. Since benchmark performance in DetectronV2 is reported using the COCO2017~\cite{COCO} validation set, we evaluate our models using the same dataset. For our accuracy metric, we choose the commonly used mean average precision, averaged over a range of intersection of union (IoU) thresholds between $50-95\%$, which we denote mAP for brevity.

\begin{figure}[htbp]
  \centering
  \includegraphics[width=0.85\linewidth]{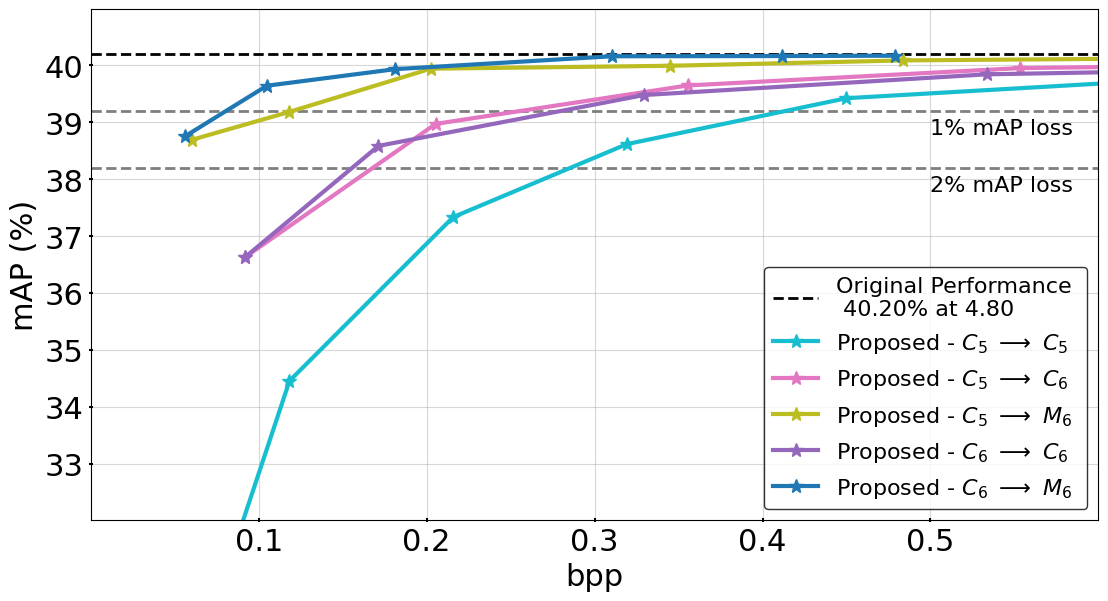}
  \caption{Comparison of multiple choices of cut and distillation points for object detection using Faster R-CNN on the COCO2017 validation set. Note that as predicted by our theoretical analysis, the effects of the choice of cut point on rate-distortion are far smaller than those of the choice of distillation point. The baseline bit-rate is calculated using the original format of the dataset (JPEG).}
  \label{fig:faster_ablation}
\end{figure}


As mentioned earlier, we choose 2 cut points and 3 distillation points for Faster R-CNN model as the machine task. After training, we evaluated the model's performance on the COCO2017 validation set, which contains 5000 RGB images. The results, shown in Figure \ref{fig:faster_ablation}, demonstrate that the distillation points have a significant impact on the model's performance. The BD-rates corresponding to the curves in \figref{fig:faster_ablation}, calculated using our $C_5\rightarrow C_5$ model as the anchor, are reported in Table \ref{tbl:RCNN}. Analysing these results demonstrates, that as our theory suggests, using deeper distillation points results in BD-rate and BD-mAP improvements, while changing the cut point does not change RD performance much. Changing the distillation point from $C_5$ to $C_6$ results in a 36.57\% BD-rate saving, while also changing the cut point to $C_6$ adds a mere 2\% more savings in BD-rate with nearly identical BD-mAP. Lastly, using our deepest distillation point,  $M_6$ (with $C_6$ as the cut point), we observe an 82.34\% BD-rate savings and 4.38\% BD-mAP improvement, compared with the shallowest. This strengthens our hypothesis that deepening the distillation point contributes significantly to improving the RD performance.

\begin{table}[h]
\centering
\captionsetup{font = small}
\parbox{1\linewidth}{\centering\caption{Rate-Distortion Performance of Various Cut and Distillation Points for Faster R-CNN and Mask R-CNN}\label{tbl:RCNN}}
\resizebox{1\linewidth}{!}{%
\begin{tabularx}{0.95\linewidth}{@{}>{\centering\arraybackslash}X>{\centering\arraybackslash}X>{\centering\arraybackslash}X>{\centering\arraybackslash}X>{\centering\arraybackslash}X@{}} 
\toprule
Model  &  \multicolumn{2}{c}{Faster R-CNN} & \multicolumn{2}{c}{Mask R-CNN}   \\       
        &  BD-Rate     & BD-mAP  &  BD-Rate    & BD-mAP \\
\midrule
$C_5\rightarrow C_5$ & 0 & 0 & 0 & 0 \\
$C_5\rightarrow C_6$ & -36.57 & 1.46 & -44.02 & 1.47 \\
$C_5\rightarrow M_6$ & -77.43 & 3.53 & -65.41 & 2.67 \\
$C_6\rightarrow C_6$ & -38.50 & 1.35 & -45.58 & 2.040 \\
$C_6\rightarrow M_6$ & -82.34 & 4.38 & -78.45 & 3.28 \\
\midrule
Choi2022 & 0 & 0 & 0 & 0 \\
$C_6\rightarrow M_6$ & -36.58 & 0.27  & -18.95 & 0.14  \\
VVC                  & 239.37 & -2.33 & 241.85 & -2.33 \\
HEVC                 & 268.01 & -3.07 & 278.26 & -2.98 \\
Cheng2020            & 305.83 & -3.62 & 266.13 & -3.49   \\

\end{tabularx}
    }
\end{table}

\begin{figure}[htbp]
  \centering
  \includegraphics[width=0.85\linewidth]{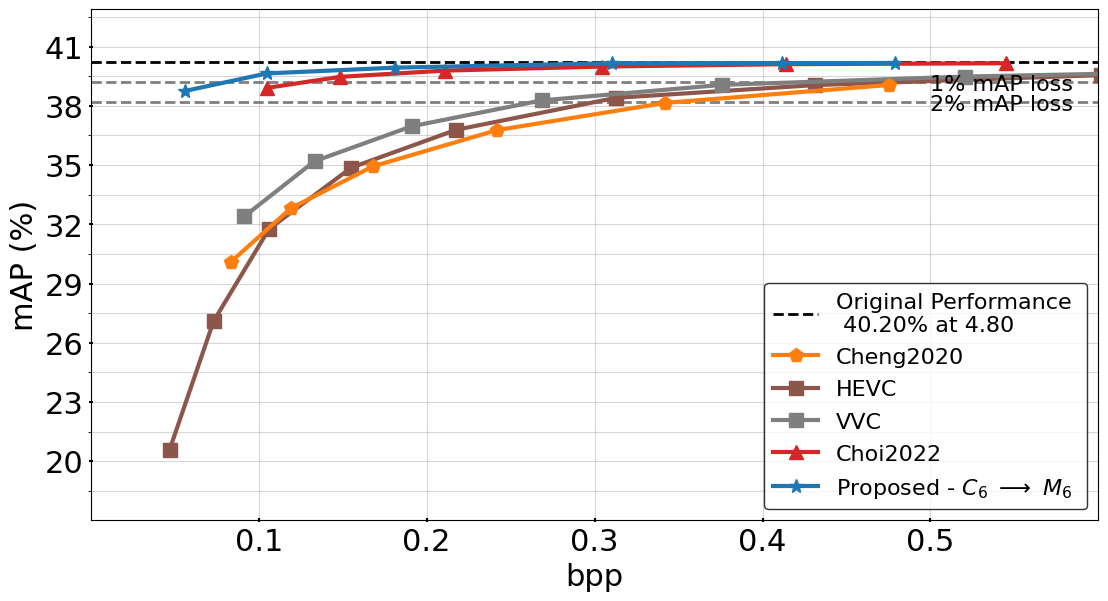}
  
  \caption{Benchmark comparison for object detection using Faster R-CNN, on the COCO2017 validation set.}\label{fig:faster_benchmark}
\end{figure}

After observing the results of our study regarding the effects of the various cut and distillation points, we proceed to compare our best performing configuration ($C_6\rightarrow M_6$) with three traditional compression benchmarks: VVC~\cite{vvc_std} using the VTM 12.3~\cite{VTM12.3} reference software, HEVC~\cite{hevc_std_2019} using the HM16.20 reference software~\cite{HM16.20}, as well as the learned codec of~\cite{cheng2020image} (which we refer to as Cheng2020), as implemented in CompressAI~\cite{begaint2020compressai}. 
Additionally, we also include the base layer from scalable human and machine codec presented in~\cite{hyomin} (to which we refer as Choi2022) as one of our benchmarks.

Observing the results, shown in \figref{fig:faster_benchmark} and the corresponding BD-metrics in \tabref{tbl:RCNN}, we see that our proposed method represents  significant improvement over the various benchmarks. For example our model remains within $1\%$ mAP at rates of lower than 0.1 bits per pixel, where traditional compression methods already suffer over $6\%$ of degradation. Even when compared with the previous SOTA, the proposed method achieves BD-rate savings of over $36\%$, representing a significant improvement. 

As its title suggests~\cite{hyomin}, the previous state of the art in compression for YOLOv3, Faster R-CNN, and Mask R-CNN was established by a scalable codec. Choi \etal use a single synthesis transform to produce two latent representations: the base layer which is used for both machine analysis and image reconstruction, and an enhancement layer which is used alongside base for human vision only. 
The dual use of the base layer means that during optimisation, its features must support both the CV task and image reconstruction, likely causing suboptimal RD performance for the base. Interestingly, the base layer distillation loss in~\cite{hyomin} is calculated equivalently to our $C_6\rightarrow M_6$ model, and still yields better RD performance compared with task-only models using shallower distillation. For example, when compared with our $C_6\rightarrow C_6$ model, ~\cite{hyomin} achieves BD-rate savings of close to $55\%$. This means that the effects of a deeper distillation point are strong enough to overcome a secondary task. 


\subsubsection{Instance Segmentation Using Mask R-CNN}\label{subsubsec:Mask}

Our second task-model for the direct coding for machines approach is Mask R-CNN~\cite{he2017mask}, a well established benchmark for instance segmentation. This model shares many characteristics with Faster R-CNN, of which the most important for our experiments is the FPN. Because the FPN for Mask R-CNN is identical in architecture to that of  Faster R-CNN, we are able to select the same cut and distillation points, $C_5,C_6, M_6$, as well as the same distortion loss for $M_6$ as in \eqnref{eq:multi_distillation}. For further details on the architecture of Mask R-CNN see Appendix~\ref{appsec:rcnn} of the supplemental material. 

Training for our Mask R-CNN models was also performed in an identical manner to that of Faster R-CNN, with the pretrained task-model taken from DetectronV2. Once again, we begin with the ablation study observing the effects of our choice of cut and distillation point before picking the most successful configuration to compare with the benchmarks. Although several different metrics are commonly used in instance segmentation literature such as mean intersection over union (mIOU), we choose to use the same metric as in object detection for convenience in comparing the two experiments.

\begin{figure}[htbp]
  \centering
  \includegraphics[width=0.85\linewidth]{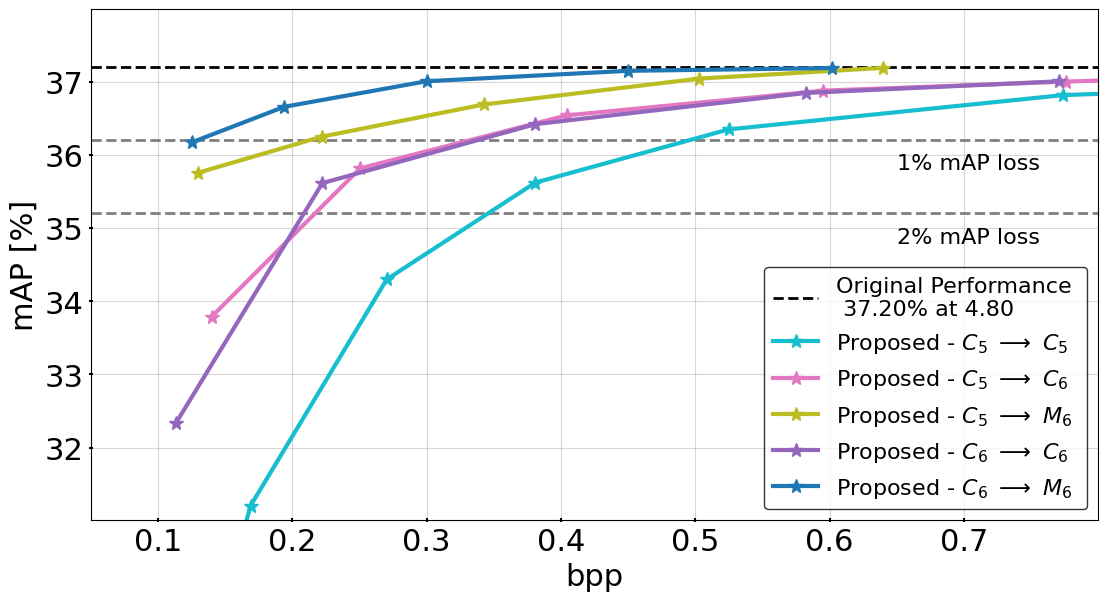}
  \caption{Comparison of multiple choices of cut and distillation points for instance segmentation using Mask R-CNN on the COCO2017 validation set. Once again we see the increased importance of the choice of distillation point compared to the choice of cut point.}
  \label{fig:mask_ablation}
\end{figure}

Figure \ref{fig:mask_ablation} and Table \ref{tbl:RCNN} demonstrate that the RD-performance trend seen with Faster R-CNN and suggested by our theoretic analysis holds for Mask R-CNN. Here too we see a significant boost in RD-performance by using deeper distillation point, and very little change due to cut point. Specifically, calculating the distortion based on $C_6$ (with $C_5$ as the cut point) leads to $44\%$ savings in BD-rate and $1.47\%$ better mAP on average, while using deepest distillation point $M_6$ results in $78.45\%$ BD-rate savings and a $3.28\%$ improvement in BD-mAP when the model is split at $C_6$. Here too we compare our deepest, best performing configuration $C_6\rightarrow M_6$ to the same benchmarks used in the Faster R-CNN experiment. Once again, we see that the proposed method outperforms traditional methods by a large margin, and shows BD-rate savings of close to $20\%$ over the previous SOTA of~\cite{hyomin}.

\begin{figure}[htbp]
  \centering
  \includegraphics[width=0.85\linewidth]{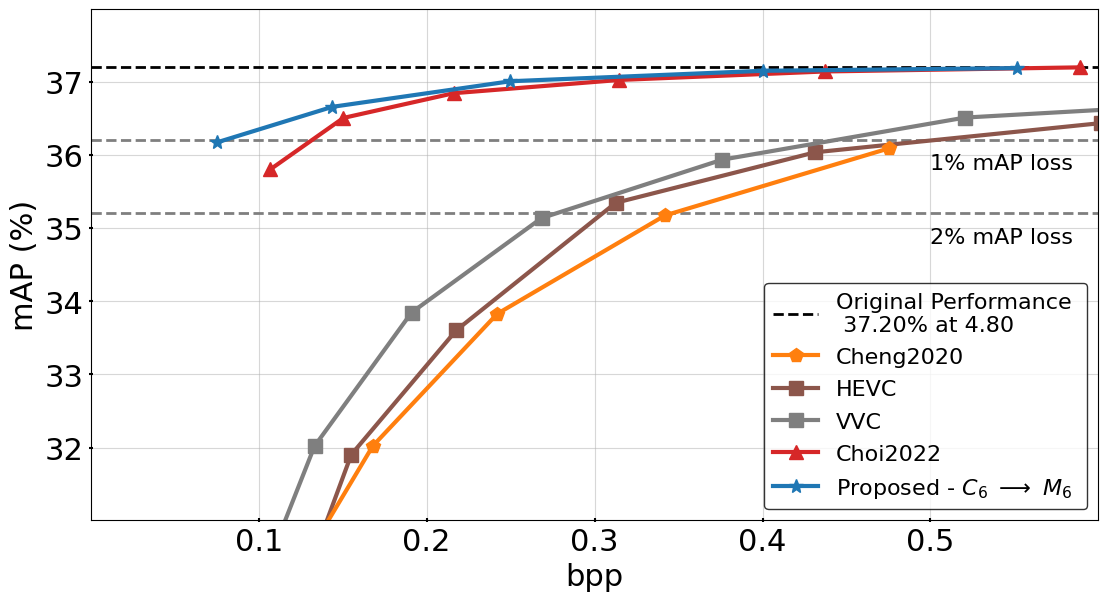}
  \label{fig:mask_benchmarks}
  \caption{Benchmark comparison for instance segmentation using Mask R-CNN, on the COCO2017 validation set.}
\end{figure}

\subsubsection{Object Detection Using YOLOv3}\label{subsubsec:yolo}

Our third task-model is YOLOv3~\cite{Redmon2018_yolov3}, another well established method for object detection. Having established the relative insignificance of the choice of the cut point in the Faster R-CNN and Mask R-CNN experiments (as expected from Corollary~\ref{cor:direct_design_considerations}), we pick a single cut point and focus on 4 different distillation points. Our different distillation points are labeled $C_7,M_7,M_8,O$, representing the cut point, 2 different choices of the mid-model, and the final multi-scale output of YOLOv3 (not to be confused with task labels $T$). For more details on the exact location of the different model layers see Appendix~\ref{appsec:yolo} of the supplemental material. Similarly to the previous two experiments, distillation points $M_7, M_8$ and $O$ are comprised of multiple tensors. For the case of YOLOv3, we found it beneficial to average the MSE of the multiple tensors by weighting each tensor by the number of elements it contains, leading to:
\begin{equation} \label{eq:multi_distillation_flatten}
    MSE(Y,\widehat{Y}) = \frac{1}{\sum_{i=1}^KE_i} \sum_{i=1}^K ||Y_i-\widehat{Y}_i||_2^2,
\end{equation}
where $Y_i,i=1,...,K$ are the tensor in the distillation point, $E_i$ is the number of elements in the tensor $Y_i$, and $||\cdot||_2^2$ is the squared $l_2$ norm\footnote{In practice this is implemented by flattening and concatenating the tensors, followed by a standard element-wise MSE}.

Training is performed using the same two stage approach used in the R-CNN experiments (including datasets and learning rate strategy), using the same loss from \eqnref{eq:learned}. Nonetheless there were slight differences in hyper-parameters, which are reported in Appendix~\ref{appsec:yolo} of the supplemental material. Here too, the task model is maintained fixed throughout training and uses pretrained weights, this time from the Darknet implementation~\cite{AlexeyAB_darknet, darknet_weights}, which was also used for performing inference in all configurations and benchmarks. All models were evaluated using 5000 images from the COCO2014~\cite{COCO} dataset, using the mean average precision at $50\%$ IOU, which we denote mAP@50, as was done in the previous SOTA~\cite{hyomin}. 

 \begin{figure}[htbp]
  \centering
  \includegraphics[width=0.85\linewidth]{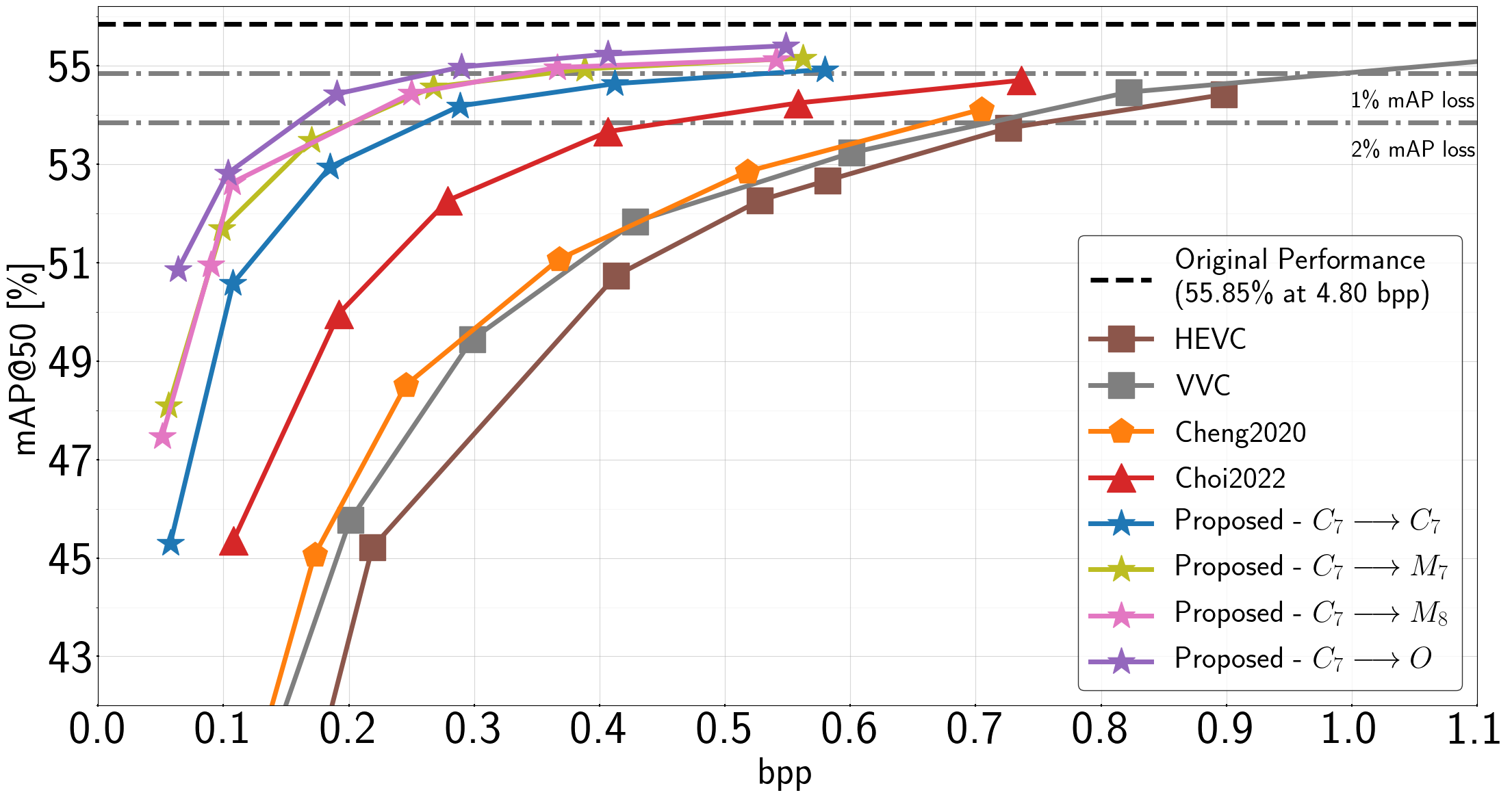}
  
  \caption{Benchmark comparison for Object detection using YOLOv3, on 5000 images from the COCO2014 validation set. The baseline bit-rate is calculated using the original format of the dataset (JPEG).}\label{fig:yolo_benchmark}
\end{figure}

Our results, as shown in \figref{fig:yolo_benchmark} and summarised in \tabref{tbl:yolo_only} demonstrate that in the case of YOLOv3, as in the previous experiments, using deeper distillation points leads to improved rate-distortion, with BD-Rate savings of $43.1\%$ for distillation point $O$ when compared with distillation point $C_7$, and BD-Rate saving of $67.5\%$ when compared with the previously SOTA base layer from~\cite{hyomin}. Interestingly, even our models with earlier distillation points achieve better RD performance than the previous SOTA (which is equivalent to our $C_7\rightarrow C_7$ setup). This likely means that our improvements result from a combination of the deeper distillation point and our models not having to balance the performance of the machine vision task with the quality of image reconstruction (the enhancement layer in ~\cite{hyomin}). 
In our previous work~\cite{pcs}, we isolated the effects of the distillation point by comparing otherwise identical scalable coding models, only separated by their choice of distillation point. For completeness, a full detailing of this experiment alongside the results can be found in Appendix~\ref{app:scalable} of the supplemental material.

\begin{table}[h]
\centering
\captionsetup{font = small}
\caption{Rate-Distortion Performance for YOLOv3}\label{tbl:yolo_only}
\begin{tabularx}{0.95\linewidth}{@{}>{\centering\arraybackslash}X>{\centering\arraybackslash}X>{\centering\arraybackslash}X@{}}
\toprule
Model   &  BD-Rate$[\%]$   & BD-mAP$[\%]$\\
\midrule
$C_7\rightarrow C_7$ & -44.2 & 2.44 \\
$C_7\rightarrow M_7$ & -58.1 & 3.14 \\
$C_7\rightarrow M_8$ & -59.7 & 3.32 \\
$C_7\rightarrow O$ & \textbf{-67.4} & \textbf{3.65} \\
\midrule
Choi2022  & 0  & 0  \\
VVC       & 66.4 & -3.90  \\
HEVC      & 89.3 & -5.86  \\
Cheng2020 & 54.5 & -3.47  \\

\end{tabularx}
\end{table}

\color{black}

\subsubsection{Object Detection and Instance Segmentation Using SWIN-Transformer}\label{subsubsec:swin}

In this experiment we aim to fully showcase the potential of the direct-coding for machines approach. We do this by utilising a more modern task model - SWIN-Transformer~\cite{Liu_2021_ICCV} (SWIN) to perform both object detection and instance segmentation using a single model. SWIN builds on the classic vision transformer~\cite{vit} by introducing sliding, overlapping windows instead of non-overlapping image patches. As is the case with many transformer based architectures, SWIN is pretrained on a large corpus of data which makes it useful as a backbone for a variety of CV architectures, which are often referred to as "heads". In our case, we use the RepPointsV2~\cite{reppointsv2} head, which performs both object detection as well as instance segmentation simultaneously.  

The SWIN architecture is comprised of several blocks, known as stages, each resulting in a decreased resolution representation of the input image. Simliarly to the case of Mask RCNN and Faster RCNN, the RepPointsV2 head is a multistream model, utilising the output of each of the stages to perform inference. Drawing from our theory and previous experimental results, we know that the choice of optimisation target is more consequential than the choice of cutting point, and thus we choose to use the output of the first stage, which we label $C_8$ as our cutting point. Drawing upon the results from section~\ref{subsec:model_splitting} as well as Theorem~\ref{thm:task-optimal-strict}, we choose a supervised approach using the task labels $T$ and the original training loss from~\cite{Liu_2021_ICCV}. We use the official, publicly available implementation of SWIN\footnote{\url{https://github.com/microsoft/Swin-Transformer}}, which utilises the MMDetection~\cite{mmdetection} framework for this experiment.

For our compression model we use the CompressAI~\cite{begaint2020compressai} implementation of ELIC as a basis of our modified ELIC-CfM model. We modifiy ELIC by replacing the synthesis transform with a latent space transform similar to~\cite{hyomin}, followed by a patch embedding layer (taken from SWIN) to match the shape of the output of stage 1. Since we are performing supervised training we can no longer utilise the VIMEO-90k~\cite{xue2019video_vimeo} or CLIC~\cite{clic_dataset} datasets, and instead train our model on the COCO2017~\cite{COCO} training set (from which we take a $10\%$ subset to be used for validation). For further details on SWIN, the cut-point, and training hyper-parameters see Appendix~\ref{app:swin} of the supplemental material. 

\begin{figure}[htbp]
  \centering
  \includegraphics[width=0.85\linewidth]{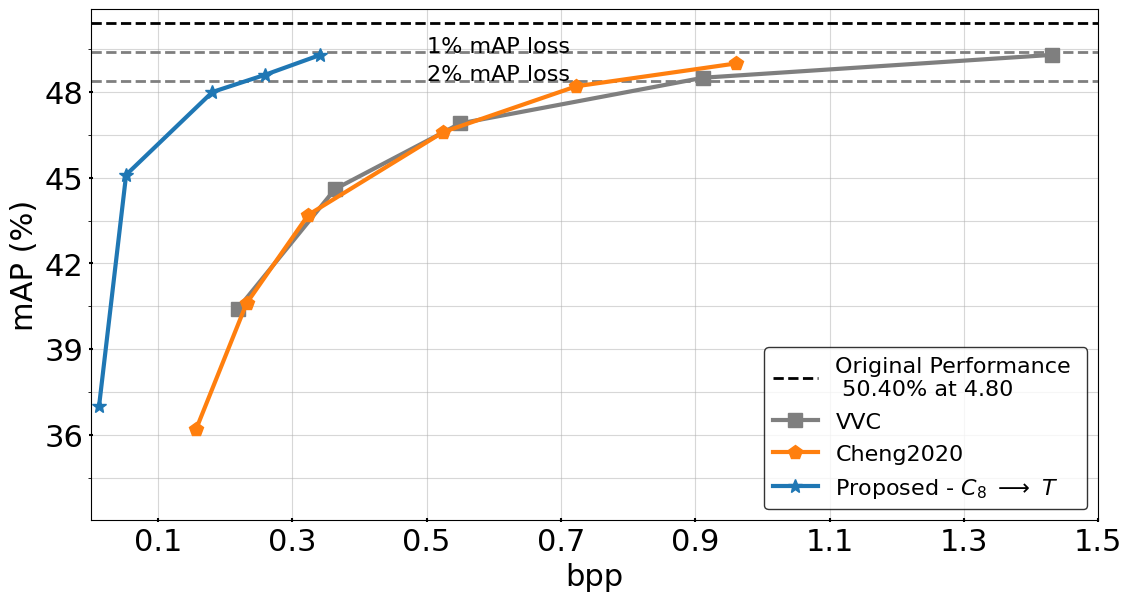}
  
  \caption{\textcolor{black}{Benchmark comparison for object detection using SWIN, on the COCO2017 validation set.}}\label{fig:swin-obj}
\end{figure}

\begin{figure}[htbp]
  \centering
  \includegraphics[width=0.85\linewidth]{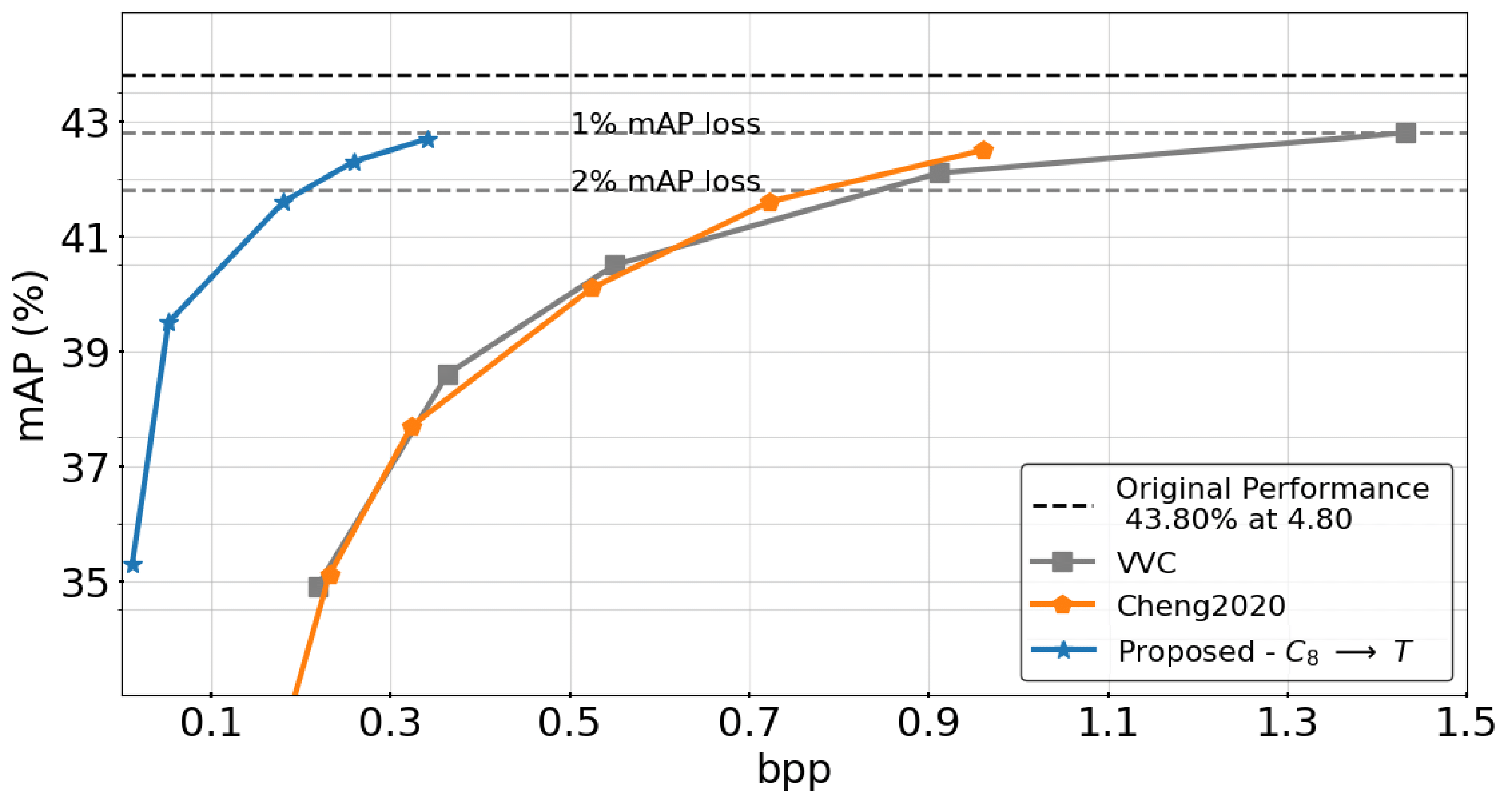}
  
  \caption{\textcolor{black}{Benchmark comparison for instance  segmentation using SWIN, on the COCO2017 validation set.}}\label{fig:swin-seg}
\end{figure}

We evaluate our model using the same commonly used average of mAP values used in Sections~\ref{subsubsec:Faster} and \ref{subsubsec:Mask}, averaged over a range of intersection of union (IoU) thresholds between 50-95\%, which we once again denote mAP for brevity. To the best of our knowledge no previously published work has utilised the SWIN, RepPointsV2 combination and thus no direct comparison can be made with other CfM approaches. Instead we compare our method with VVC~\cite{vvc_std} as implemented by the VVEnc and VVDec repositories~\cite{vvenc_vvdec} as well as Cheng2020~\cite{cheng2020image}. As can be seen from \figref{fig:swin-obj} our proposed method greatly outperforms both VVC and \cite{cheng2020image}   on object detection. For example, to achieve $45\%$ mAP, our approach requires 0.053 bits per pixel whereas VVC and \cite{cheng2020image} require almost eight-times as much with 0.4 bits per pixel. Overall, our model achieves BD-Rate savings of 87.4\% when compared with VVC and 89.7\% when compared with \cite{cheng2020image}. Similar results can be seen for instance segmentation  in \figref{fig:swin-seg}, where our model performs better with 0.012 bits per pixel than VVC does with 0.22 bpp and \cite{cheng2020image} does with 0.23 bpp. Overall, our model achieves BD-rate savings of 89.3\% and 89.5\% compared with VVC and Cheng2020 respectively.

\color{black}

\section{Summary and Conclusion}
\label{sec:conclusion}

The field of coding for machines is rapidly evolving with promising developments and a growing number of potential applications. In this work we have presented a formulation of rate-distortion theory as it pertains to coding for machines, with specific attention to coding for deep models. We have proven, that in the optimal case, three of the most commonly used approaches today are essentially equivalent in terms of their optimal RD performance. 

Furthermore, we have shown both theoretically and empirically, that using a supervised approach leads to superior RD performance and were able to achieve SOTA compression for image classification \textcolor{black}{and object detection} using this insight. In the unsupervised case, where one does not have access to task labels, we argue that distilling deeper layers is preferable. While our theory does not provide definitive proof of this claim, we provide strong empirically-based hypotheses for the cause at the root of our claim. Furthermore, by selecting deeper distillation points for our model we are able to achieve SOTA rate-distortion performance for several CV tasks, all trained in an unsupervised manner.

\textcolor{black}{Although our own applications focus on image coding for machines, our theory remains agnostic to both the input signal modality as well as the nature of the analysis task and task-model. Existing work in other settings of coding for machines serves as evidence to this claim. For example, in point-cloud CfM, PCHM-Net~\cite{pchm_net}, SPCGC~\cite{spcgc}, and~\cite{mateen_point_cloud} all utilise a distillation loss or supervised labels and achieve superior RD performance to codecs optimised only for geometric similarities, as predicted by Theorems~\ref{thm:task-optimal-non-strict},~\ref{thm:task-optimal-strict} and their corresponding corollaries. In image CfM with traditional codecs, the addition of learned pre-processing layers optimised for feature similarity or task performance, which once again corresponds to Theorems~\ref{thm:task-optimal-non-strict} and~\ref{thm:task-optimal-strict}, has been shown to improve RD performance compared with the same codecs alone in~\cite{pre_transform_icm, multitask_preprocess_ICM}. More generally, the continued prevalence of all three major coding approaches in CfM serves, in and of its own, as empirical evidence for Theorems~\ref{thm:split-direct} and~\ref{thm:distortion-only}.}

We believe that grounding the research in coding for machines with relevant, sound theoretical background is crucial to ensure the longevity of resulting methods. At the same time, using theoretical insights enables the creation of stronger, more efficient codecs, as evident by the state-of-the-art empirical performance of our proposed methods.


%

\ifCLASSOPTIONcaptionsoff
  \newpage
\fi



\bibliographystyle{IEEEtran}
%
\bibliography{main.bib}

%
\vspace{-1.1cm}
\clearpage
\begin{IEEEbiography}[{\includegraphics[width=1in,height=1.25in,clip,keepaspectratio]{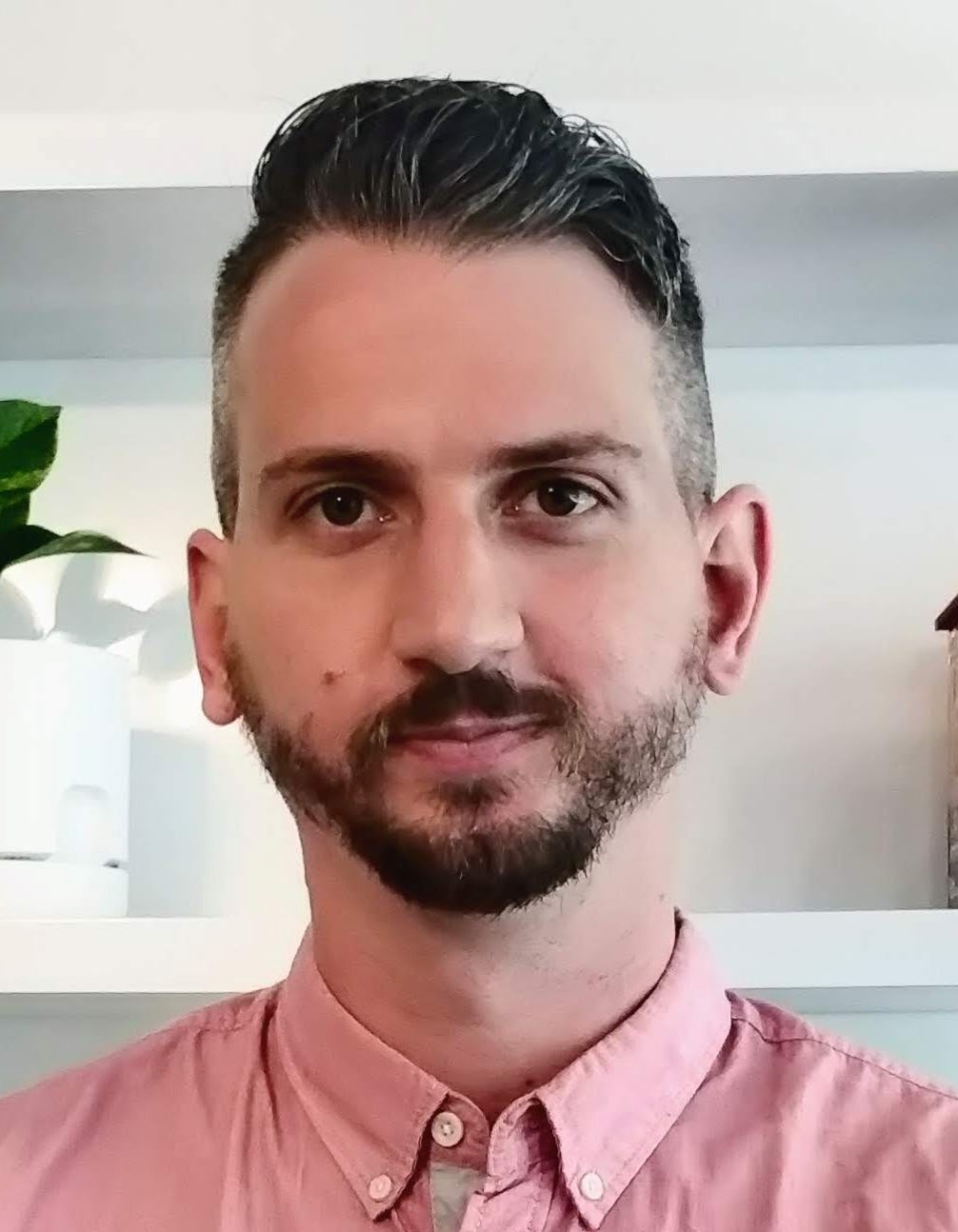}}]{Alon Harell}(S'19) 
received the M.A.Sc. degree in electrical engineering from Simon Fraser University, Burnaby, BC, Canada  in 2020 focusing on deep learning applications for non-intrusive load monitoring. 
Since 2020 Alon has been pursuing his PhD in engineering science at Simon Fraser University. His research interests include information theory as it applies to deep learning, coding for machines, and sports analytics. He has published at major conferences including ICASSP, ICM Multimedia, and AAAI, and has been awarded both NSERC CGS-M and PGS-D scholarships.)
\end{IEEEbiography}
\vspace{-0.7cm}
\begin{IEEEbiography}
[{\includegraphics[width=1in,height=1.25in,clip,keepaspectratio]
{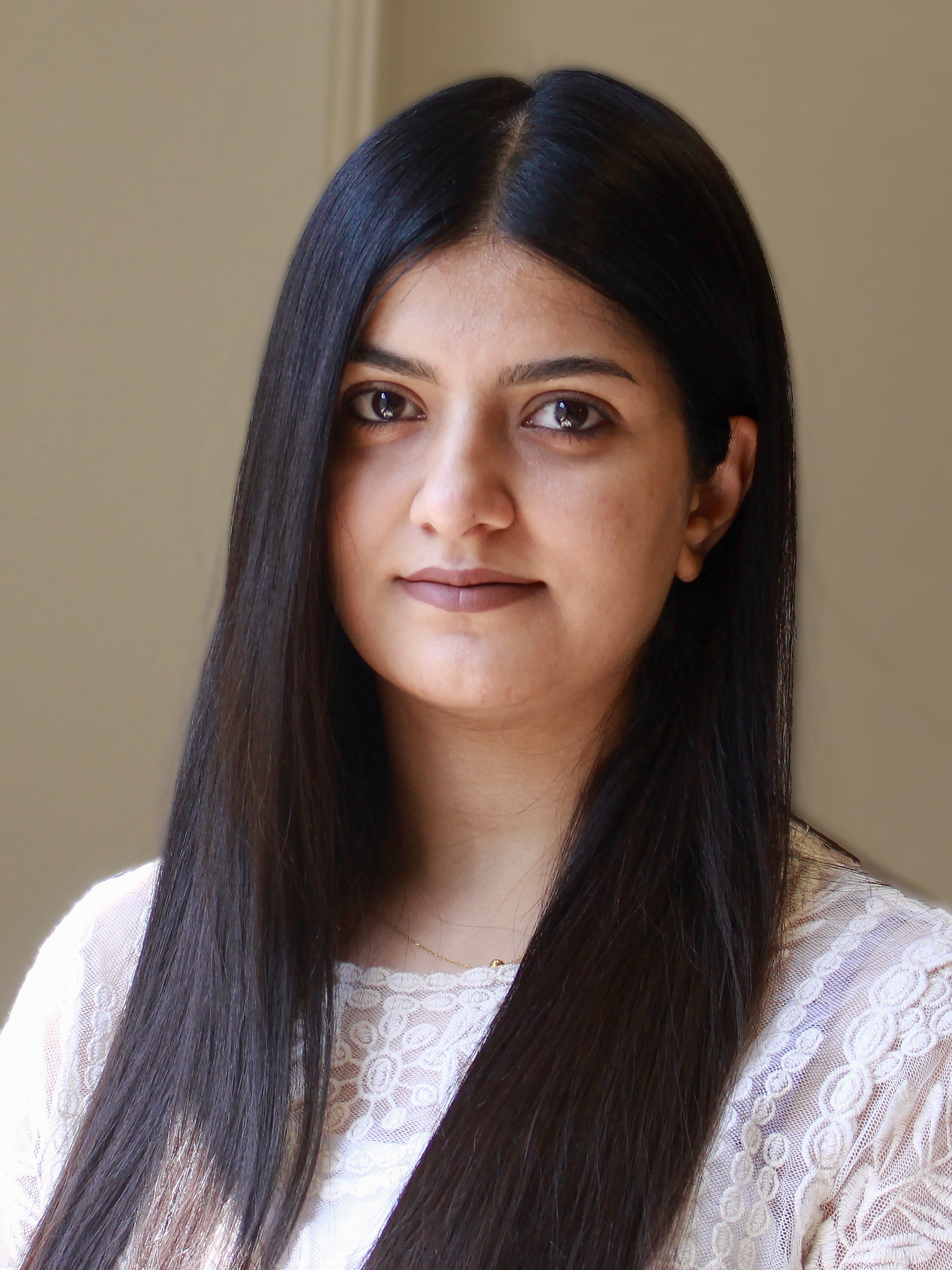}}]{Yalda Foroutan}(S'23)
received her Bachelor's degree in Electrical Engineering from Amirkabir University of Tehran in 2012. She then pursued a Master's degree in Electrical Engineering at Tehran University, specializing in deep learning from 2017 to 2020. During this time, she focused on object detection and hand gesture recognition models to control computer mouse. In 2021, Yalda started her Ph.D. in Engineering Science at Simon Fraser University, where she is currently exploring learned human-machine coding. Her research interests include deep learning and computer vision.
\end{IEEEbiography}
\vspace{-0.6cm}
\begin{IEEEbiography}[{\includegraphics[width=1in,height=1.25in,clip,keepaspectratio]{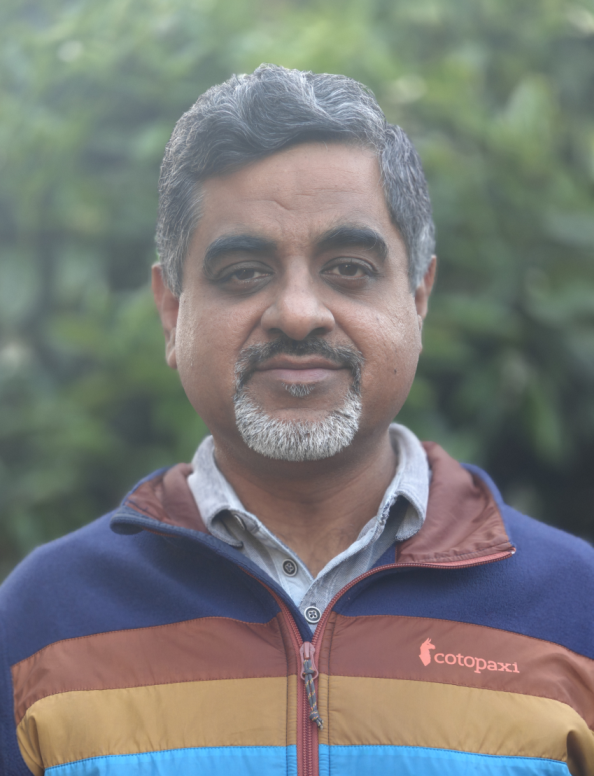}}]{Nilesh A. Ahuja} is an AI Research Scientist in Intel Labs.  His current research focus is in the area of adaptive AI systems for the Edge. This includes development of efficient and reliable methods for uncertainty estimation in AI systems; its applications to real-world problems such as out-of-distribution detection for industrial anomaly detection and novelty detection for continual learning systems; and efficient and adaptive deployments on Edge systems via split computing. His other research interests include 3D computer-vision; odometry and SLAM; super-resolution; image, and video processing; and AI methods for video compression. He received his Ph.D. degree in Electrical Engineering from Pennsylvania State University in 2008. His work has been published in top-tier journals and conferences, and he has over 20 issued or pending US and international patents.
\end{IEEEbiography}
\vspace{-0.7cm}

\begin{IEEEbiography}[{\includegraphics[width=1in,height=1.25in,clip,keepaspectratio]{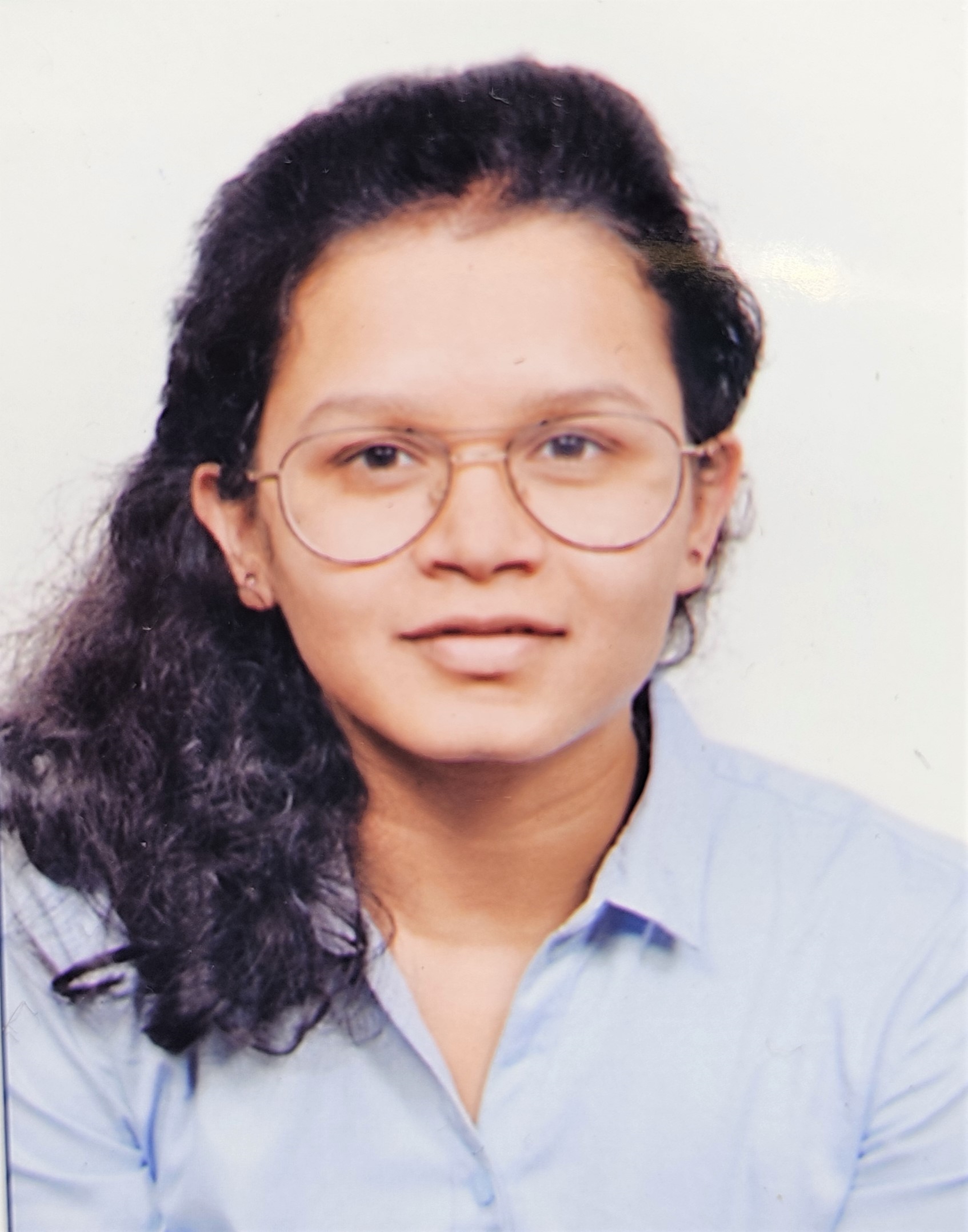}}]{Parual Datta} is an AI research scientist at Intel Labs in India. She received a Master's in Communication and Signal Processing in 2017 for her thesis in Multi-Person Pose Estimation at the Indian Institute of Technology, Indore. Her research focuses on image and video processing, particularly machine-learning-based image compression. She has published in top-tier conferences and holds several US patents. 
\end{IEEEbiography}

\begin{IEEEbiography}[{\includegraphics[width=1in,height=1.25in,clip,keepaspectratio]{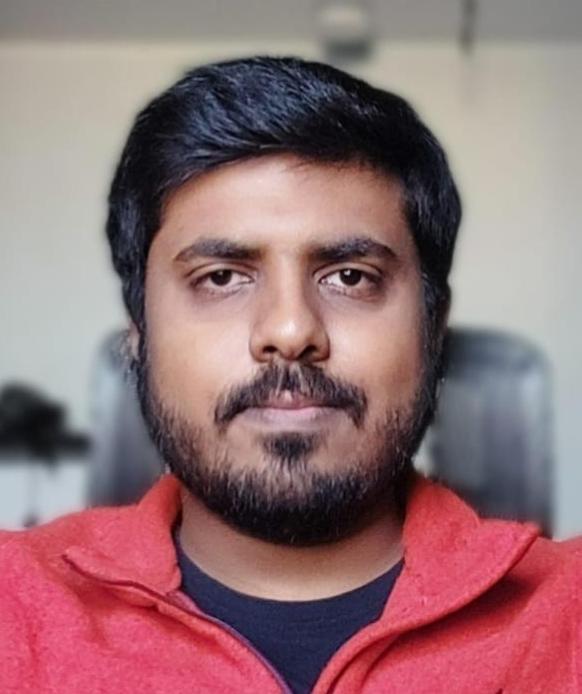}}]{Bhavya Kanzariya} worked as a Graduate Intern at Intel Labs, Bangalore. His interests are Computer Vision, Edge computing, Probabilistic computing, and robust ML. He has co-published an article in CVPR. He holds an MTech. in Artificial Intelligence from the Indian Institute of Technology, Hyderabad during which he has earned the award for academic excellence.
\end{IEEEbiography}
\vspace{-0.5cm}

\begin{IEEEbiography}[{\includegraphics[width=1in,height=1.25in,clip,keepaspectratio]
{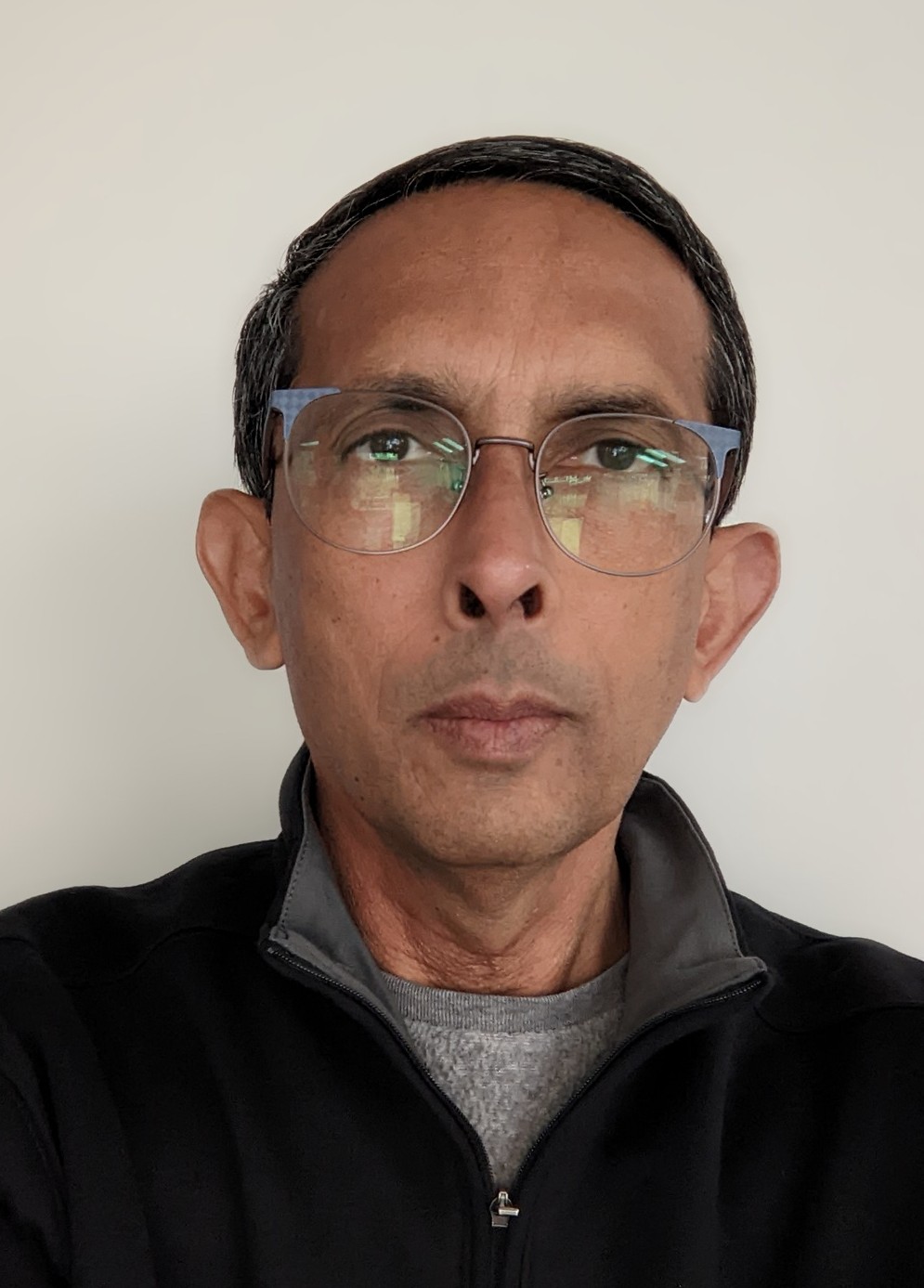}}]{V. Srinivasa Somayazulu}
is a researcher at Intel Labs, with expertise in wireless communications, networking, and multimedia signal processing.  His current research interests include AI in video processing, compression, communications, and machine vision.  He has co-authored over 30 publications and over 60 patents in these and related fields.  He received his Ph.D. degree from the U. of California, Santa Barbara in Electrical Engineering.
\end{IEEEbiography}
\vspace{-0.5cm}

\begin{IEEEbiography}[{\includegraphics[width=1in,height=1.25in,clip,keepaspectratio]
{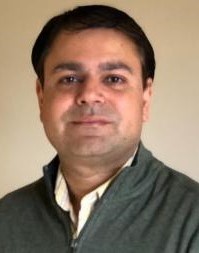}}]{Omesh Tickoo}
is a Principal AI Engineer and Research Manager at Intel Labs, Hillsboro. His current interests include probabilistic computing, interactive multi-modal scene understanding and contextual learning. Omesh received his PhD from Rensselaer Polytechnic Institute for his thesis on Analysis and Improvement of Multimedia Transmission over Wireless Networks. Omesh has authored more than 40 papers in premier international Journals and Conferences and holds more than 25 patents. Omesh has served as chair of multiple committees for IEEE conferences and regularly serves as a Technical Program Committee member and reviewer for international conferences and journals.
\end{IEEEbiography}
\vspace{-0.5cm}

\begin{IEEEbiography}[{\includegraphics[width=1in,height=1.25in,clip,keepaspectratio]
{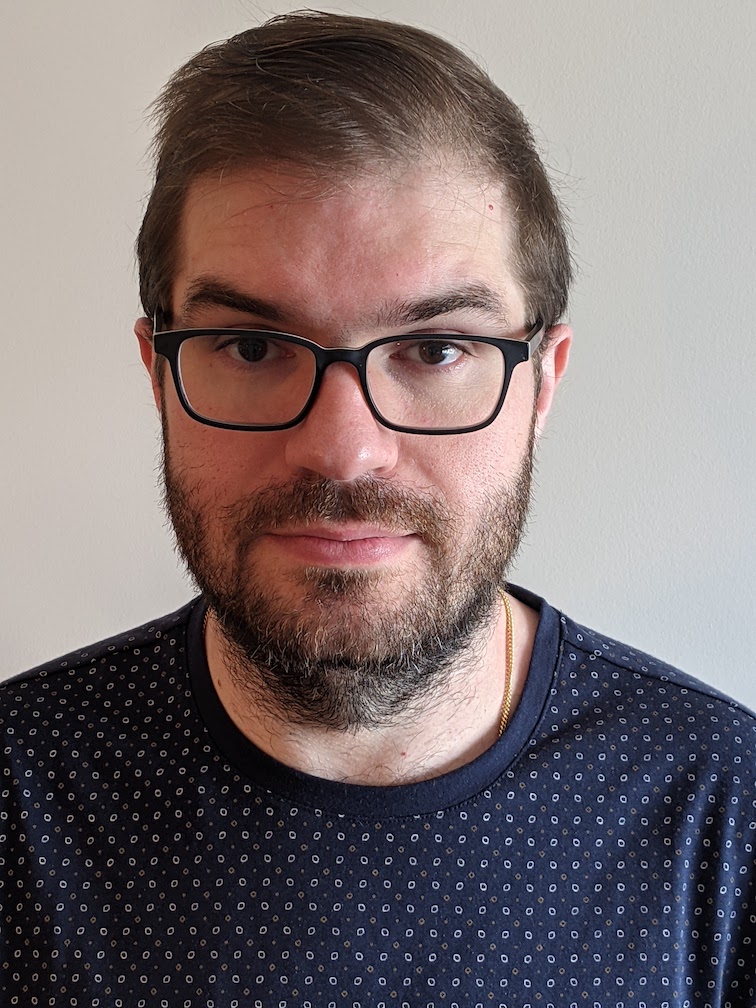}}]{Anderson de Andrade}
(S'22) received his M.Sc. in Applied Computing from the University of Toronto in 2015 and obtained a B.Eng. degree in Networks and Communications in 2007 from Universidad Tecnológica del Centro, Venezuela. He is currently an Engineering Science Ph.D. student at Simon Fraser University. His research interests include learned compression for humans and machines, representation learning, information theory, and collaborative intelligence.
\end{IEEEbiography}
\vspace{-0.5cm}

\begin{IEEEbiography}[{\includegraphics[width=1in,height=1.25in,clip,keepaspectratio]{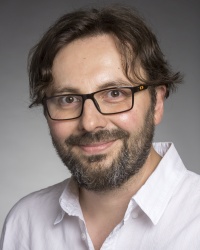}}]{Ivan V. Baji\'{c}}(S'99--M’04--SM’11) is a Professor of Engineering Science and co-director of the Multimedia Lab at Simon Fraser University, Canada. His research interests include signal processing and machine learning with applications to multimedia signal processing, compression, and collaborative intelligence. His group’s work has received the 2023 IEEE TCSVT Best Paper Award, conference paper awards at ICME 2012, ICIP 2019,  MMSP 2022, and ISCAS 2023, and other recognitions (e.g., paper award finalist, top n\%) at Asilomar, ICIP, ICME, and CVPR. He is the Past Chair of the IEEE Multimedia Signal Processing Technical Committee and currently serves as a Senior Area Editor of IEEE Signal Processing Letters.
\end{IEEEbiography}



\newpage
\input\onecolumn
\begin{center}
\bf
SUPPLEMENT TO

Rate-Distortion Theory in Coding for Machines and its Applications
\end{center}
\appendices
\section{Theorem Proofs}\label{app:proofs}
\begin{reptheorem}{thm:split-direct}
The minimal achievable rates for direct coding for machines and model splitting are identical, that is, $$R_{XY}(D;T) = R_Y(D;T)$$.
\end{reptheorem}
\begin{proof}
We begin by proving that $R_{XY}(D;T) \leq R_Y(D;T)$. Let $p^*(\tilde{y}|y)\in  \mathbfcal{P}_{\widehat{Y}}(D;T)$ be a conditional distribution of $\tilde{y}$ on $y$ which achieves the rate-distortion function of the model splitting approach $R_Y(D;T)$. By definition, $y=g(x)$ for some $x$, so we can re-write $p^*(\tilde{y}|y) = p^*(\tilde{y}|g(x)) = q(\tilde{y}|x)$. We know that $q(\tilde{y}|x) \in \mathbfcal{Q}_{X\widehat{Y}}(D;T)$ because 
$$\mathbb{E}\left[d_T\left(f({X}),h(\widetilde{{Y}})\right)\right] =  \mathbb{E}\left[d_T\left(h({Y}),h(\widetilde{{Y}})\right)\right]\leq D.$$
The equality comes from $f = h\circ g$ and the definition of $Y$, and the inequality is true because  $p^*(\tilde{y}|y)\in  \mathbfcal{P}_{\widehat{Y}}(D;T)$.
Next we note that $X \rightarrow Y \rightarrow \widetilde{Y}$ is a Markov chain and thus we can apply the data processing inequality to learn that $I(X;\widetilde{Y})\leq I(Y;\widetilde{Y}) = R_Y(D;T)$. However, because $R_{XY}(D;T)$ is the minimum of $I(X;\widehat{Y})$ for all $q(\hat{y}|x) \in \mathbfcal{Q}_{X\widehat{Y}}(D;T)$ we know that $R_{XY}(D;T)\leq I(X;\widetilde{Y}) \leq R_Y(D;T)$. 

To finish our proof we show that $R_{Y}(D;T) \leq R_XY(D;T)$. Analogously to the first part of our proof, let $q^*(\tilde{y}|x)\in  \mathbfcal{Q}_{X\widehat{Y}}(D;T)$ be a conditional distribution of $\tilde{y}$ on $x$ which achieves the rate-distortion function of direct coding for machines $R_{XY}(D;T)$. For a given observation of $x$ we get an exact value of $y = g(x)$ by definition, which induces a distribution $q(y) = \sum_{x\in g^{-1}(y)}p(x)$ (here $g^{-1}(y) = \{x: y =g(x)\}$ is the set inverse of $g$, and $p(x)$ is the density of $x$). This, alongside $q^*(\tilde{y}|x)$ induces a conditional distribution $p(\tilde{y}|y)$. We know that $p(\tilde{y}|y)\in  \mathbfcal{P}_{\widehat{Y}}(D;T)$ because:
\begin{flalign}
\begin{aligned}
\mathbb{E}\left[d_T\left(h({Y}),h(\widetilde{{Y}})\right)\right] & =  \sum_{y,\tilde{y}}p(\tilde{y}|y)q(y)d_T(y,\tilde{y})   \\
& \equaltext{(a)} \sum_{x,y,\tilde{y}}d_T(y,\tilde{y}) q(y)q(\tilde{y}|x,y) p(x|y)\\
& \equaltext{(b)} \sum_{x,y,\tilde{y}}d_T(y,\tilde{y}) q(y)q^*(\tilde{y}|x) p(y|x) \frac{p(x)}{q(y)} \\
& = \sum_{x,\tilde{y}}q^*(\tilde{y}|x) p(x) \sum_y d_T(y,\tilde{y}) p(y|x) \\
& \equaltext{(c)} \sum_{x,\tilde{y}}q^*(\tilde{y}|x) p(x)d_T(g(x),\tilde{y}) \\
& = \mathbb{E}\left[d_T\left(f({X}),h(\widetilde{{Y}})\right)\right]\leq D.  
\end{aligned}
\end{flalign}
(a) comes from the law of total probability; (b) includes Bayes' law alongside the fact that $q(\tilde{y}|x,y) = q^*(\tilde{y}|x)$ because $y$ is completely determined by $x$; (c) is simply a result of $p(y|x) = 1$ whenever $y=g(x)$ and zero otherwise. The final inequality is true because  $q^*(\tilde{y}|x)\in  \mathbfcal{Q}_{X\widehat{Y}}(D;T)$. Trivially, $X\rightarrow \widetilde{Y}$ is a Markov chain and thus its inverse $\widetilde{Y}\rightarrow X$ is also one. We can now add the processing step $Y = G(X)$ to get the Markov chain $\widetilde{Y}\rightarrow X \rightarrow{Y}$. We apply the DPI to the last chain to get $I(\widetilde{Y};Y) \leq I(\widetilde{Y};X) = R_{XY}(D;T)$. Finally, once again we note that as the minimum over all distributions $p(\hat{y}|y) \in \mathbfcal{P}_Y(D_T)$, we have $R_Y(D;T) \leq I(\widetilde{Y};Y) \leq R_{XY}(D;T)$ concluding our proof. 
\end{proof}

\begin{reptheorem}{thm:distortion-only}
Let $\mathbfcal{I}_f(\mathbfcal{X})\subseteq \mathbfcal{T}$ be the image set of a task model on all possible inputs, and let $\widehat{\mathbfcal{Y}}\subseteq \mathbfcal{Y}$ be the set of all possible approximations of $Y$.
If $h(\widehat{\mathbfcal{Y}})\subseteq \mathbfcal{I}_f(\mathbfcal{X})$ then, for any given distortion $D>0$, the minimal achievable rate for model splitting is equal to the minimal achievable rate for the classical approach: 
\begin{equation*}
    R_X(D;T) = R_Y(D;T)
\end{equation*}
\end{reptheorem}
\begin{proof}
From~\cite{hyomin} we already know that $R_{{Y}}(D;T)\leq R_{{X}}(D;T)$, which means it is enough to show that $R_{{X}}(D;T)\leq R_{{Y}}(D;T)$ to show equality. Let $p^*(\hat{y}|y)\in \mathbfcal{P}_Y(D;T)$ be a distribution which achieves the corresponding rate-distortion $R_Y(D;T)$. Next, define the indirect inverse of $g$ through $h$ as: $g^{-1}_h(y) =f^{-1}\left(h(y)\right)$, where $f^{-1}(t) = \left\{ x: f(x) = t \right\}$ is the set inverse of $f$. We can now define an approximation of $X$ as the median of the indirect inverse of $g$ through $h$ applied to $\hat{y}$, that is $\hat{x} = median\left(g^{-1}_h(\hat{y})\right)$. We use a convention for the median such that it is always a member of the set. Additionally, the choice of the median here is arbitrary, any member of the set $(f^{-1}\left(h(\hat{y})\right)$ is a suitable choice.

First, by the conditions of the theorem, we know that the set $g^{-1}_h(\hat{y})$ is not empty for any value of $\hat{y}$, and thus $\hat{x}$ is well defined. Next, we note that the following is a Markov chain: $X \rightarrow Y\rightarrow \widehat{Y}\rightarrow \widehat{X}$. Applying the DPI, we see that $I(X;\widehat{X})\leq I(Y;\widehat{Y}) = R_Y(D;T)$. The definition above (along with the Markov chain) induces a conditional distribution $p(\hat{x}|x)$, which satisfies $p(\hat{x}|x)\in \mathbfcal{P}_X(D;T)$ because:
\begin{equation}
\begin{aligned}
    \mathbb{E}\left[d_T\left(f(X), f(\widehat{X})\right)\right] & = \mathbb{E}\left[d_T\left(h\big(g(X)\big),f\left(g_h^{-1}(\widehat{Y})\right)\right)\right] \\
    & = \mathbb{E}\left[d_T\left(h(Y), h(\widehat{Y})\right)\right] \leq D
\end{aligned}
\end{equation}
Where the last inequality is because $p^*(\hat{y}|y)\in\mathbfcal{P}_Y(D;T)$.
Finally, because of its definition as the minimum over all approximations in $\mathbfcal{P}_X(D;T)$ we know that $R_X(D;T)\leq I(\widehat{X};X) \leq R_Y(D;T)$ concluding the proof.
\end{proof}

\begin{reptheorem}{subthm:distortion_only}
    Let $\hat{y}: h(\hat{y})\notin \mathbfcal{I}_f(\mathbfcal{X})$ be an approximation of a subset $\mathbfcal{Y}_{\hat{y}} \subseteq~\mathbfcal{Y}$ of values of $Y$, for which the output of the task backend is not contained in the image set of $f$ (the task model). If, for any such $\hat{y}$, there exists an alternative approximation, $\tilde{y}$ such that $h(\tilde{y}) \in \mathbfcal{I}_f(\mathbfcal{X})$ and $d_T\left(h(y),h(\tilde{y})\right)\leq d_T\left((h(y),h(\hat{y})\right) \forall y \in\mathbfcal{Y}_{\hat{y}} $ then \eqnref{eq:equality} still holds.
\end{reptheorem}
\begin{proof}
    First, define the restriction of $\mathbfcal{Y}$ to values for which the output of the task backend is in the image of the task model, $\tilde{\mathbfcal{Y}} = \left\{ y \in \mathbfcal{Y}: h(y)\in \mathbfcal{I}_f(\mathbfcal{X}) \right\}$. Any approximation for which the output of the task backend is not in the image of the original task model satisfies $\hat{y} \in \mathbfcal{Y} \setminus \tilde{\mathbfcal{Y}}$.
    Next, denote $\tilde{y}(\hat{y}) \in \tilde{\mathbfcal{Y}}$ to be the alternative approximation corresponding a specific value of $\hat{y}$. Now we can define the following mapping $\tilde{g}:\mathbfcal{Y}\rightarrow \tilde{\mathbfcal{Y}}$ so that: 
    $$\tilde{Y} = \tilde{g}(\widehat{Y}) = \begin{cases} 
    \tilde{y}(\hat{y}) \;\;,\;\; Y=\hat{y} \notin \widetilde{\mathbfcal{Y}}\\
    \hat{y} \;\; ,\;\; \text{otherwise.}
    \end{cases}
    $$
    By the conditions of this corollary such a mapping exists, and the following is a Markov chain: $Y \rightarrow \widehat{Y} \rightarrow \tilde{Y}$, which allows us to apply the DPI and get $I(Y;\tilde{Y})\leq I(Y;\widehat{Y})$. Additionally, because every value of $\tilde{Y}$ achieves no worse distortion than its equivalent value of $\widehat{Y}$ (regardless of the value $y$ currently being encoded), we also know $\mathbb{E}\left[d_T(h(Y),h(\tilde{Y})\right] \leq \mathbb{E}\left[d_T(h(Y),h(\widehat{Y})\right]$. Thus, for any value of $D>0$, if $p(\hat{y}|y)$ achieves $R_Y(D;T)$ then so does $p(\tilde{y}|y)$. 
    Finally, since $\tilde{Y}\subseteq \tilde{\mathbfcal{Y}}$, for any $D>0$ there exists a distribution $p(\tilde{y}|y) = p(\tilde{g}(\hat{y})|y)$ which achieves the rate-distortion function $R_Y(D;T)$ while still meeting the conditions of the proof of Theorem~\ref{thm:distortion-only}.
\end{proof}

\begin{reptheorem}{thm:task-optimal-non-strict} 
For some input distortion, $D_X>0$, and the corresponding lowest possible task distortion achievable by an $X$-optimal approximation, $D_T^{min} = \min {D}^*_{T}(D_X;X)$,  the minimal achievable rate of the supervised approach is upper bound by the input rate-distortion (for the corresponding distortion values). Formally:
$$R_X(D_T^{min};T)\leq R_X(D_X;X)$$
\end{reptheorem}
\begin{proof}
Let $p^*(\tilde{x}|x) \in \mathbfcal{R}_X(D_X;X)$ be the best possible $X$-optimal distribution in terms of task distortion, meaning $\mathbb{E}\left[d_T\left(f(X),f(\widetilde{X})\right)\right] = D_T^{min}$. By definition this means that $p^*(\tilde{x}|x) \in \mathbfcal{P}_X(D_T^{min};T)$, and recall that because $p^*(\tilde{x}|x) \in \mathbfcal{R}_X(D_X;X)$, we know that $I(X;\tilde{X}) = R_X(D_X;X)$. Finally, also by definition: $$R_X(D_T^{min};T) = \min_{p(\hat{{x}}|{x})~\in~\mathbfcal{P}_{{X}}(D_T^{min};T)} I({X};\widehat{{X}}),$$ which gives us $R_X(D_T^{min};T) \leq I(X;\tilde{X}) = R_X(D_X;X)$, concluding the proof.
\end{proof}

\begin{reptheorem}{thm:task-optimal-strict} 
Begin with a set of $X$-optimal distributions $\mathbfcal{R}_X(D_X;X)$, and a corresponding lowest possible task distortion, $D_T^{min} = \min {\mathbfcal{D}}^*_{T}(D_X;X)$. If, for any $p(\hat{x}|x)\in\mathbfcal{R}_X(D_X;X)$ there exist two points $\hat{x}_1 \neq \hat{x}_2$ with non-zero probabilities, $p(\hat{x_1}), p(\hat{x_2})\neq 0$, for which the task output is identical\footnote{In fact this only has to hold for $p(\hat{x}|x)\in\mathbfcal{R}_X(D_X;X)$, which also satisfy $\mathbb{E}\left[d_T\left(f(X),f(\widehat{X})\right)\right] = D_T^{min}$}, $f(\hat{x}_1) = f(\hat{x}_2)$, and at least one input $x$ for which $p(x|\hat{x}=x_1)\neq p(x|\hat{x}=x_2)$, then the minimal achievable rate of the supervised approach is strictly lower than the input rate-distortion(for the corresponding distortion values):
$$ R_X(D_T^{min};T) < R_X(D_X;X) . $$ 
\end{reptheorem}
\begin{proof}
Let $p^*(\hat{x}|x) \in \mathbfcal{R}_X(D;X)$ be the best possible $X$-optimal distribution in terms of task distortion, meaning $\mathbb{E}\left[d_T\left(f(X),f(\widehat{X})\right)\right] = D_T^{min}$. Next, consider $\widetilde{X}$ which is defined by applying the following transformation to $\widehat{X}$:
\begin{equation*}
    \tilde{X} = \begin{cases} \widehat{X}\;\;,\;\; \widehat{X}\neq \hat{x}_2 \\
                             \hat{x}_1\;\;,\;\; \widehat{X} = \hat{x}_2
    \end{cases}
\end{equation*}
First, note that the task-distortion for $\tilde{X}$ is unchanged because $f(\tilde{X}) = f(\widehat{X})$, which means $p(\tilde{x}|x)\in\mathbfcal{P}_X(D_T^{min};T)$. Next note the minimum rate needed to encode $\tilde{X}$, equal to $I(X;\tilde{X})$ is strictly lower than the equivalent for $\widehat{X}$:
\begin{equation}
    \begin{aligned}
    I(X;\widehat{X})-I(X;\tilde{X}) & \equaltext{(a)} H(X|\tilde{X}) - H(X|\widehat{X}) \\ 
    & \equaltext{(b)} 
    \sum_x \sum_{\hat{x}}{p(x,\hat{x})\log\left(p(x|\hat{x})\right)} - \sum_x\sum_{\hat{x}}{p(x,\tilde{x})\log\left(p(x|\tilde{x})\right)}\\
    & \equaltext{(c)}   \sum_x\bigg[{p(x,\hat{x}=\hat{x}_1)\log\left(\frac{p(x,\hat{x} =\hat{x}_1)}{p(\hat{x}=\hat{x}_1)}\right) + p(x,\hat{x}=\hat{x}_2)\log{\frac{p(x,\hat{x} =\hat{x}_2)}{p(\hat{x}=\hat{x}_2)}}}\bigg] - \\    & \quad \; \sum_x{\bigg[\big(p(x,\hat{x}=\hat{x}_1)+p(x,\hat{x}=\hat{x}_2)\big)\log\left(\frac{p(x,\hat{x} =\hat{x}_1) + p(x,\hat{x}=\hat{x}_2)}{p(\hat{x}=\hat{x}_1) + p(\hat{x}=\hat{x}_2)}\right)\bigg]}\\
    & \largertext{(d)}{0}, 
    \end{aligned}
\end{equation}
where $(a)$ is a result of $I(U;V) = H(U)-H(U|V)$, and $(b)$ is simply the definition of conditional entropy. The step $(c)$ involves a few parts: first, we use the definition of conditional probability $p(u|v) = \frac{p(u,v)}{p(v)}$; next we note that for any $x^*\notin\{\hat{x}_1,\hat{x}_2\}$ we have $p(x,\hat{x} = x^*)=p(x,\tilde{x} = x^*)$ as well as $p(x|\hat{x} = x^*)=p(x|\tilde{x} = x^*)$ ; and finally, we know that $p(x,\tilde{x} = \hat{x}_1) = p(x,\hat{x} = \hat{x}_1)+p(x,\hat{x} = \hat{x}_2)$, which also means that $p(\tilde{x} = \hat{x}_1) = p(\hat{x} = \hat{x}_1)+p(\hat{x} = \hat{x}_2)$. Step $(d)$ is true because of the \emph{log-sum inequality} - Theorem 2.7.1 in~\cite{Cover_Thomas_2006}, where the condition that there exists at least one input $x$ for which $p(x|\hat{x}=x_1)\neq p(x|\hat{x}=x_2)$, makes the log-sum inequality strict.
Finally, recall that $R_X(D_T^{min};T)$  is the minimum of mutual information over all potential approximations of X in $\mathbfcal{P}_X(D_T^{min};T)$ and thus specifically is no greater that $I(X;\tilde{X})$ giving us:
$R_X(D_T^{min};T)\leq I(X;\tilde{X}) < I(X,\widehat{X}) = R_X(D;X)$ concluding our proof.
\end{proof}

\section{Model Architectures and configurations}\label{app:model_details}

The following appendix contains more detailed description of the model architectures used in our experiments, including the task models. Of course, since these task-models are taken from previously published work, we limit the presentation here to an overview, and the reader to explore the original publication for any additional detail.

\subsection{Model splitting}\label{appsec:modelsplit}

\begin{figure}[htbp]
  \centering
  \includegraphics[width=\textwidth]{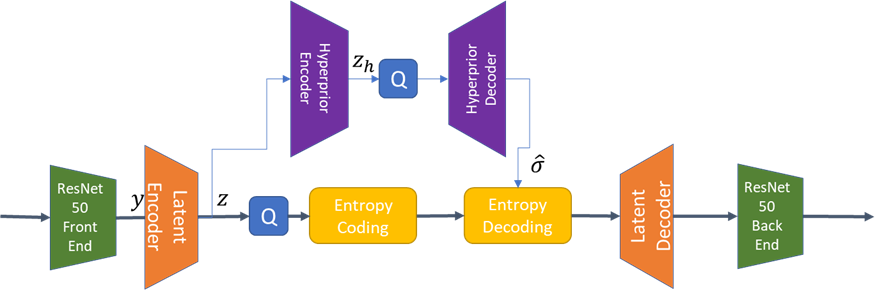}
  \caption{Overall block diagram of our model splitting approach showing the latent encoder and decoder and the hyperprior network.}
  \label{fig:nre_block_diagram}
\end{figure}

\paragraph*{\textbf{Training of the compression unit}}
We follow the approach of Datta \etal \cite{datta2022low} for the design and training of the latent encoder and decoder. The procedure involves exploring a joint space of hyperparameters related to the architecture of the latent encoder, and those related to compression and encoding (recall that the latent decoder is the mirror image of the latent encoder and hence shares the same architectural hyperparameters). The architectural hyper-parameters we tune for are the number of channels $C_r$ at the output of the latent encoder and the stride $S$ of the convolutional kernels used therein. The compression hyper-parameters include the Lagrange multiplier $\lambda$ in Eq. (\ref{eq:learned}), and the quantisation step size, $Q$, used to discretise the output of the latent encoder. 

As we did in our previous work~\cite{ahuja2023neural, datta2022low}, we perform the hyperparameter search as follows: first, a random sample is generated from the joint hyperparameter space which fixes the topology of the latent encoder (and decoder) and the compression related parameters. A training run is then performed to train the compression unit and the resultant classification accuracy is measured along with the average bit-rate (measured, as explained earlier in bits-per-pixel). This process is then repeated multiple times -- each time with a different sample from the joint hyperparameter space -- to generate sufficient number of such candidates. From this set, Pareto optimal set of points are determined. The points lying on the Pareto frontier correspond to the set of trained bottleneck layers that yield the optimal accuracy vs compression performance. 

\begin{figure}[htbp]
  \centering
  \includegraphics[width=\textwidth]{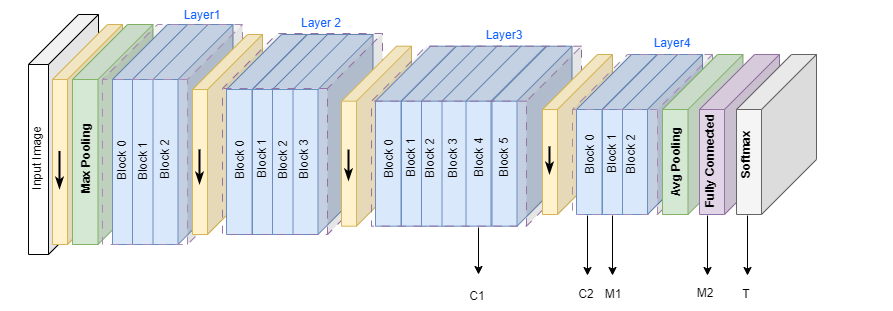}
  \caption{ResNet50 architecture. The potential cut and distillation (middle) points are indicated with $C_i$'s and $M_i$'s respectively. The blocks marked with a $\downarrow$ represent down-sampling blocks, which are implemented with strided convolutions. Each block is comprised of several convolutional layers with residual connections, not pictured here.}
  \label{fig:resnet50}
\end{figure}

\subsection{ResNet50}\label{appsec:resnet50}
For the classification task using a model-splitting approach, we use the standard ResNet50 architecture \cite{resnet} that has been trained on the ImageNet dataset. The main body of the ResNet50 network comprises four units (Layer1 to Layer4), each comprising multiple residual blocks as shown in \figref{fig:resnet50} (refer to~\cite{resnet} for additional details). The two cut-points $C_1,C_2$, and distillation points $M_1,M_2$, which are used in our evaluation, are at the outputs of various blocks as shown in the figure. Importantly, when training in a supervised manner, the ground truth labels $T$, are used, and not the model output (though $T$ is shown in \figref{fig:resnet50} at the output for simplicity).

\begin{figure}[htbp]
  \centering
  \includegraphics[width=\textwidth]{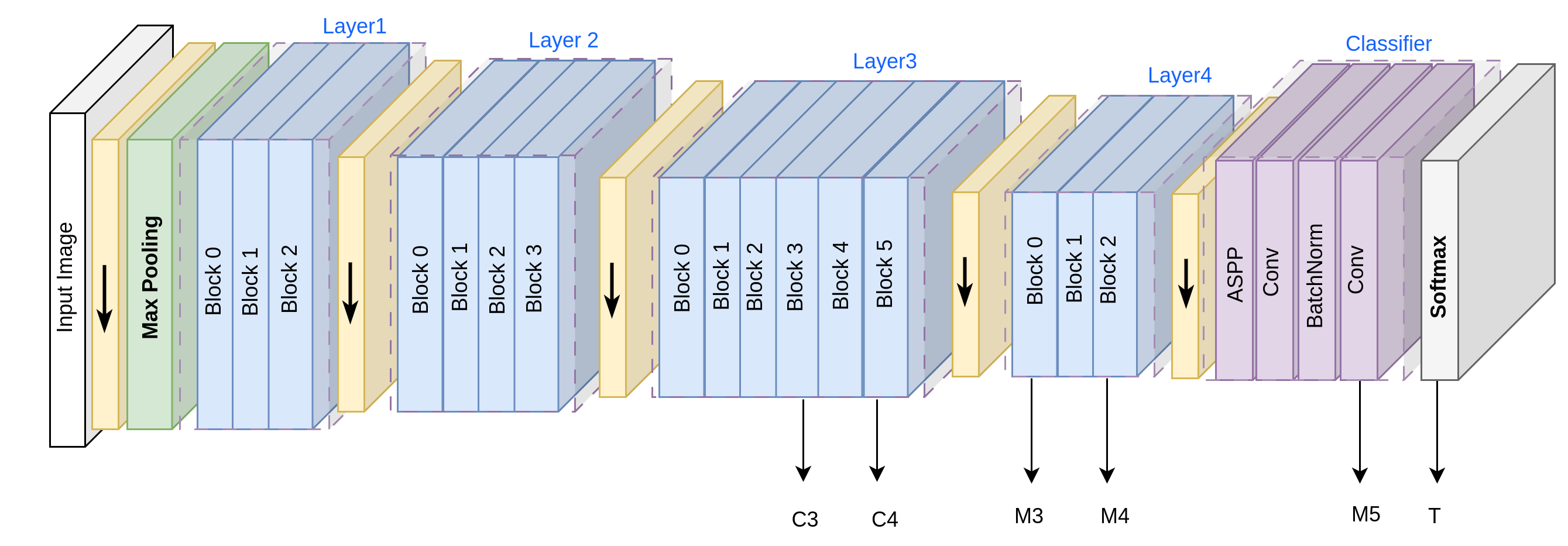}
  \caption{\textcolor{black}{Deeplab-V3 architecture with potential cut and distillation (middle) points indicated, respectively, as $C_i$'s and $M_i$'s.}}
  \label{fig:deeplab_v3}
\end{figure}

\color{black}

\subsection{Deeplab V3}\label{appsec:deeplab_v3}
For the segmentation task using a model-splitting approach, we use the Deeplab-v3 architecture \cite{deeplabv3} that has been trained on the COCO 2017 dataset. It comprises a Resnet50 backbone followed by a classification network. The architecture of the backbone is identical to that of the standard Resnet50 described in the previous subsection. The classification network primarily comprises ASPP layers (Atrous Spatial Pyramid Pooling) followed by some convolutional layers. The classification network is responsible for reconstructing a high-resolution segmentation map from the compact representation learnt by the backbone. The two cut-points $C_3,C_4$, and distillation points $M_3, M_4, M_5$, which are used in our evaluation, are at the outputs of various blocks as shown in the figure. Note that the most compact representation is found at $M_4$ and beyond that the resolution increases again, until the full image resolution is available at the output. Similar to the classification case in the previous subsection, when training in a supervised manner, the ground truth labels $T$, are used, and not the model output (though $T$ is shown at the output for simplicity).
\color{black}

\subsection{Direct Coding For Machines Model}\label{appsec:compression_model}

\textcolor{black}{In direct coding for machines, the encoder takes in the input $X$, while the decoder's goal is to decode a certain set of features from the task model. The distinction from model splitting is that the model-splitting approach takes in features from the task model (e.g., from cut-points $C_1$ or $C_2$ above), rather than taking the input directly. In other words, model-splitting employs the initial layers of the task model, while direct coding need not do that. Both approaches might target the same set of features (e.g., $M_1$, $M_2$, or $T$ above).}

\subsubsection{Autoregressive model based on Cheng2020}

\begin{figure}[htbp]
  \centering
  \includegraphics[width=\textwidth]{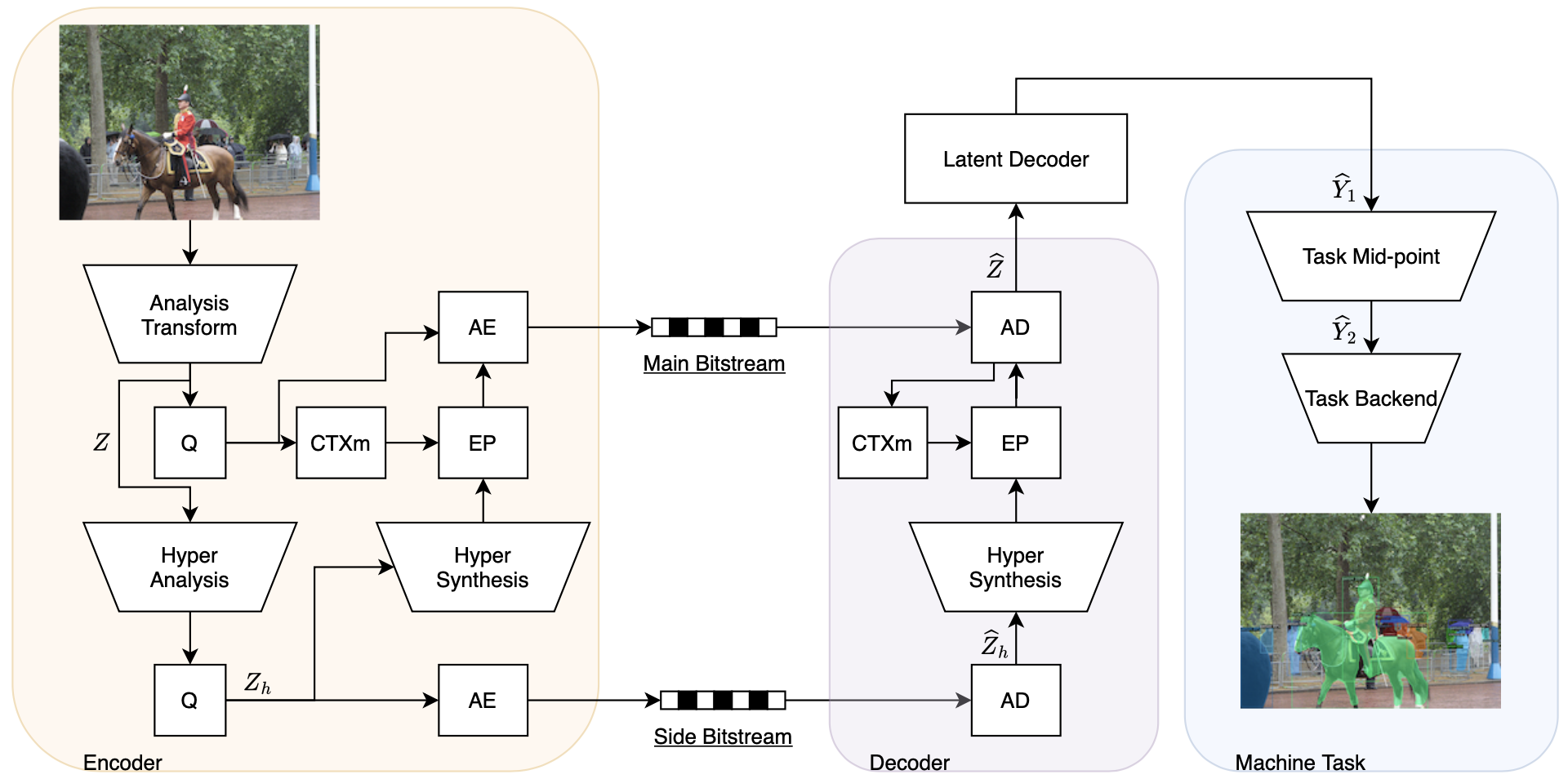}
  \caption{Architecture of our codec model for direct coding for machines. CTX and EP denote the context model and entropy parameters, respectively. AE/AD stands for arithmetic encoder/decoder.
}
  \label{fig:compression_model}
\end{figure}

\textcolor{black}{In the first three experiments utilising} direct coding for machines we use the following learned compression model, based on the "base-layer" of the scalable codec in~\cite{hyomin}, which in turn is largely based on~\cite{cheng2020image}, shown in \figref{fig:compression_model}. On the encoder side, we begin by creating a latent representation $Z$ using a synthesis transform $g_a$, which is comprised of downsampling blocks as well as  convlutional layers, all with a fixed amount of channels (192 in~\cite{hyomin}), and generalised divisive normalisation (GDN) activations~\cite{gdn}. We use a slightly modified $g_a$ because although we need a relatively smaller of channels for $Z$ than 192, we found reducing the number of channels immediately leads to inferior performance. Instead, we keep the width of synthesis transform layers at 192, reducing dimensionality at the final layer to our desired size, which depends on the task-model and thus detailed in the following sections.

In the next step, we model the elements of the latent representation $Z$ as conditionally independent Gaussian similarly to~\cite{minnen2018joint, balle2017end}, conditioned on their mean and scale, which are calculated as follows. The latent representation $Z$ is input into a second analysis transform $h_a$, known as the hyperprior model to produce the side information $Z_h$. This side information is then quantised and encoded using an entropy bottleneck~\cite{balle2017end} followed by an arithmetic encoder, producing the side bitstream. The quantised side information is then used to produce estimates for mean and scale of each element in the quantised latent representation $Z$. In parallel to the side information, a second estimate of the mean and scale is produced by the autoregressive context model of~\cite{minnen2018joint}. The two representations are then merged using the entropy parameter (EP) estimation block, before being used in an arithmetic encoder  to produce the main bitstream.

After decoding the resulting bitstreams, the reconstructed side information $\widehat{Z}_h$ is used alongside previously decoded elements of $\widehat{Z}$ to recreate the necessary means and scaled to support the arithmetic decoding of $\widehat{Z}$. The recreated latent representation is then processed by a latent decoder, referred to as the latent space transform (LST) in~\cite{hyomin}, comprised of residual blocks and inverse GDN activations, as well as upsampling blocks. The output of the latent decoder is used as our recreated cut point, $\widehat{Y}_1$ which can then be fed to the mid-model to obtain the distillation point $\widehat{Y}_2$ during training, or to the full task backend during inference. Importantly, when training our codec, we replace all quantisation with uniform random noise of magnitude 1, as is commonly done for learned compression. 

\subsubsection{Efficient compression model based on ELIC}
\color{black}
\begin{figure}[htbp]
  \centering
  \includegraphics[width=\textwidth]{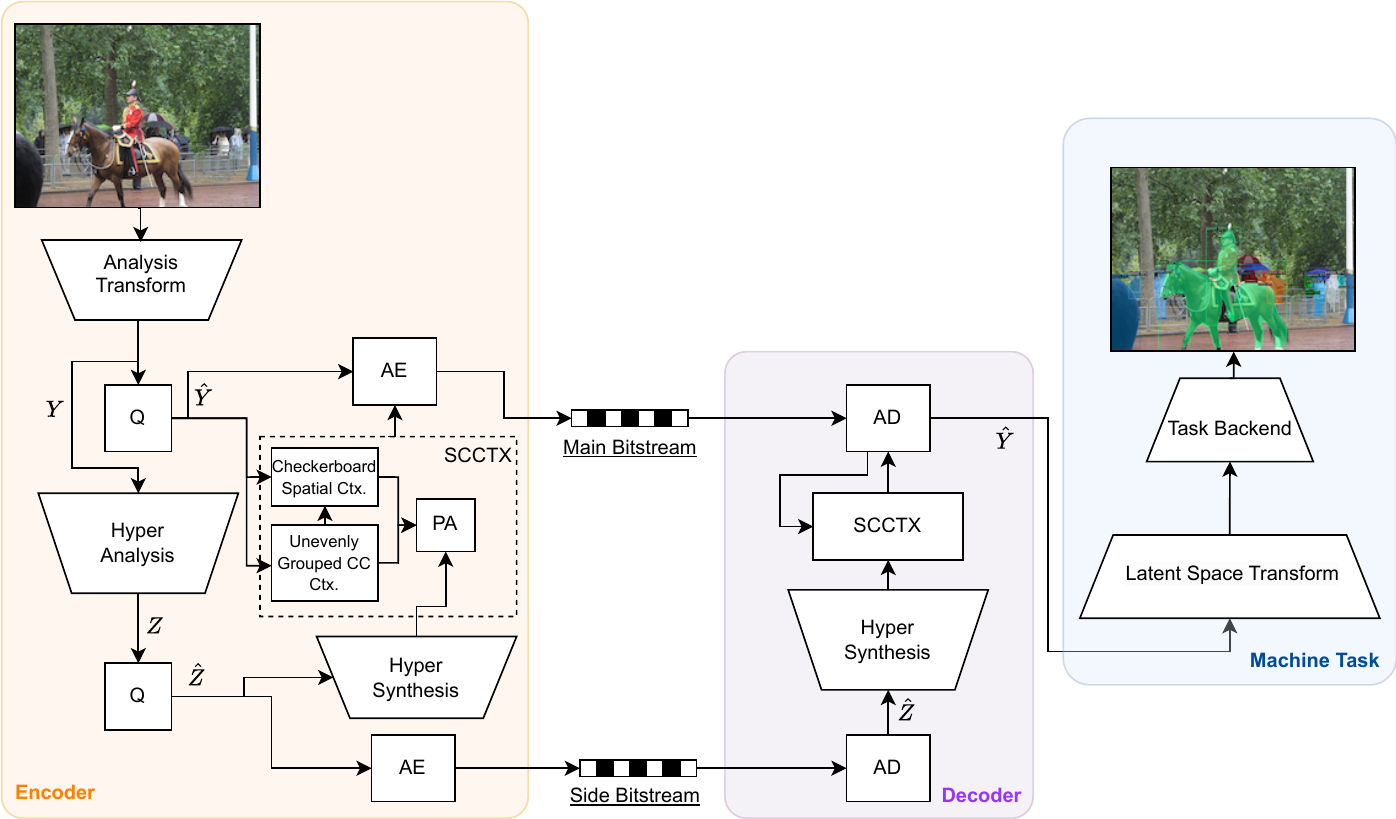}
  \caption{\textcolor{black}{Diagram of the proposed ELIC-based codec architecture for computer vision tasks. The Space-Channel Context model (SCCTX) of ELIC is comprised of 3 modules: a checkerboard spatial context model, an unevenly grouped channel-conditional context model, and a parameter aggregation (PA) module.}}
  \label{fig:compression_model-elic}
\end{figure}

For our final direct-compression we choose to adapt a more modern learned codec in ELIC.
The ELIC\cite{he2022elic} architecture proposes an efficient entropy model in which the spatial dimensions are not processed auto-regressively but in two steps, following a checkerboard pattern as in \cite{he2021checkerboard}. In addition to this, the channels dimension of the feature space is grouped and the channels within each group are processed together using previous groups as context. These groupings are done unevenly in increasing group size, motivated by an observed information compaction property. In the first step, half of the spatial dimensions of the current group are inferred using the previous groups as context. The values to be inferred in this step follow a checkerboard pattern. In the second step, these values are used as contextual \textit{anchors} to infer the rest of the values within the channel group. As additional context, a synthesis of the hyper-prior is used. The hyper-prior, channel, and spatial anchors produce individual contextual representations that are concatenated and aggregated to infer the parameters of a multivariate independent Gaussian distribution placed on the latent representation.

In addition to this entropy model, ELIC proposes different analysis and synthesis transformations of the latent space. It replaces the previously commonly used GDN activations~\cite{gdn} with residual bottleneck blocks and attention modules. These attention modules have been used previously in \cite{cheng2020image}. Fig.~\ref{fig:compression_model-elic} shows the proposed codec architecture for computer vision tasks using ELIC. The hyper-prior analysis and synthesis transforms are identical to the ones from \cite{balle2017end}. The decoded latent representation is processed by a latent-space transform to match the target task backend input. Whereas the latent representation in ELIC has at least 128 and as many as 320 channels and 5 groups, we set the dimensionality of the latent space to 64 channels and propose the consecutive group sizes of [3, 3, 6, 12, 40]. 

The latent space transform seen here is slightly modified from the version used in our Cheng2020 based codec in that it uses similar components to the ELIC synthesis transform. We carefully control the channel sizes and upscaling strides to obtain the correct dimensional shape prior to utilising a patch embedding layer taken from SWIN~\cite{Liu_2021_ICCV}. This final layer changes the latent space into 2D representations, commonly used in vision-transformer-based models, rather than the 3D ones commonly used in convolutional networks.

\color{black}
\subsection{Faster R-CNN \& Mask R-CNN}\label{appsec:rcnn}

\begin{figure}[htbp]
  \centering
  \includegraphics[width=\textwidth]{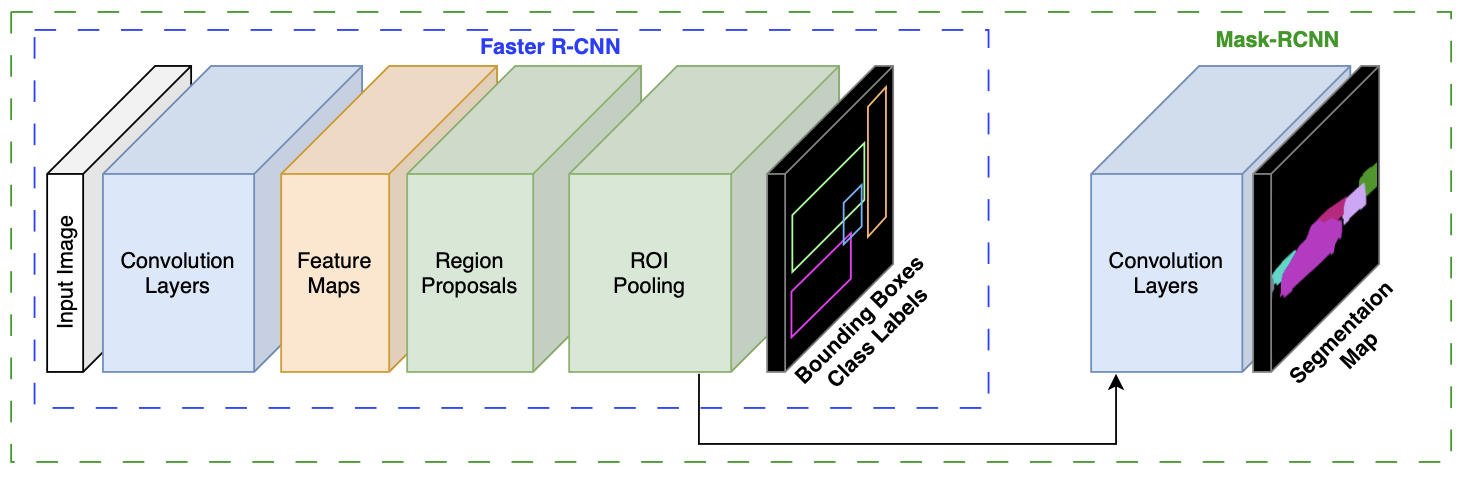}
  \caption{Faster R-CNN \protect\cite{ren2015faster} and Mask R-CNN\protect\cite{he2017mask} architectures. Mask R-CNN extends Faster R-CNN with an additional branch for generating the segmentation map.}
  \label{fig:faster_vs_mask}
\end{figure}

Faster R-CNN is a region-based convolutional neural network that identifies objects within an image, providing their corresponding bounding boxes, class labels. The architecture of Faster R-CNN comprises a backbone that generates feature maps used by a region proposal network to create region proposals. These proposals and feature maps are then passed through a Region of Interest (RoI) pooling layer to predict bounding boxes and class labels, as shown in Figure \ref{fig:faster_vs_mask}. Mask R-CNN is an extension of Faster R-CNN that benefits from an additional branch comprising a Fully Convolutional Network, which is used to predict segmentation masks on the RoIs.

\begin{figure}[htbp]
  \centering
  \includegraphics[width=\textwidth]{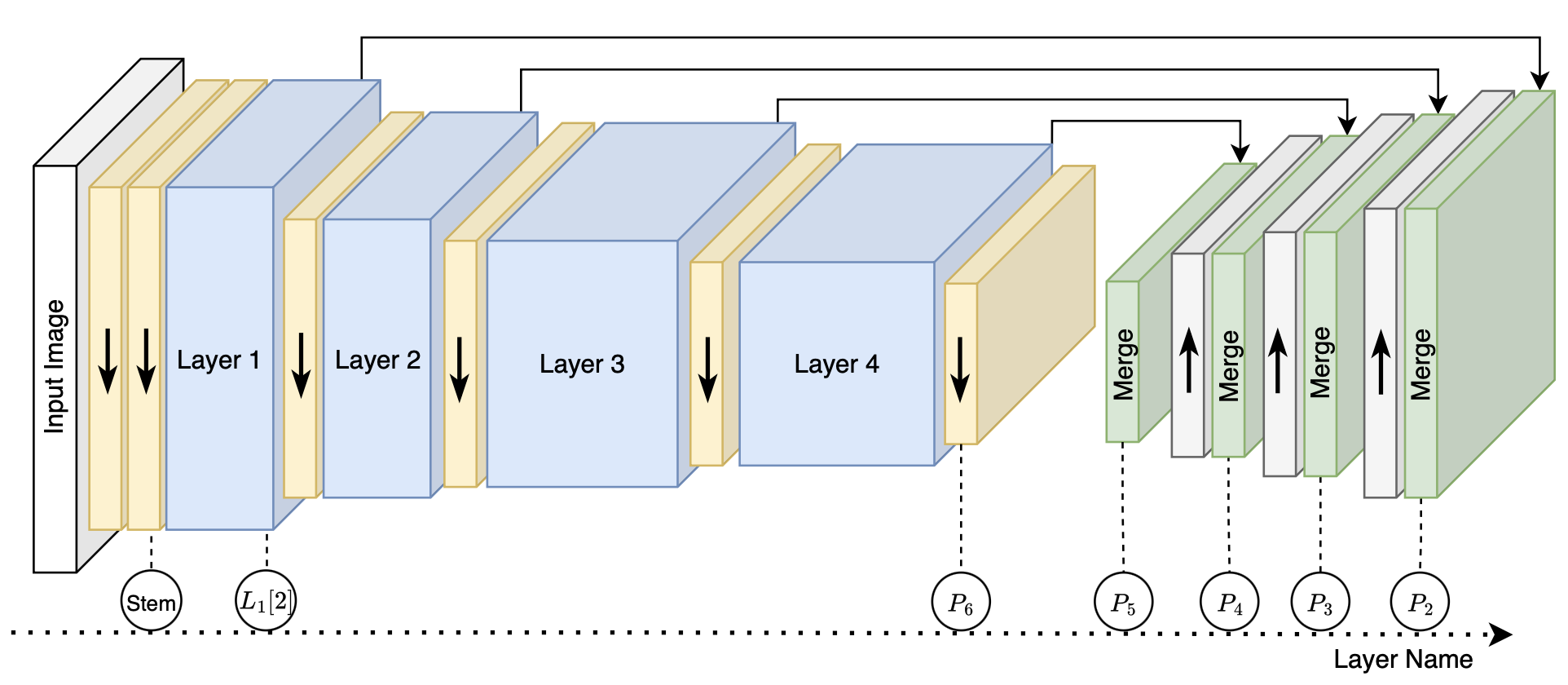}
  \caption{Faster and Mask R-CNN backbone using ResNet50~\protect\cite{resnet} and FPN~\protect\cite{FPN}. The potential cut and distillation points are indicated by circles. The blocks marked with a $\downarrow$ and $\uparrow$ represent down-sampling and up-sampling blocks, respectively.}
  \label{fig:rcnn}
\end{figure}

As seen in Figure \ref{fig:faster_vs_mask}, both Faster R-CNN and Mask R-CNN networks share the same backbone architecture, for example utilising from ResNet50 and FPN. ResNet50 is composed of residual blocks and down-sampling blocks, while the Feature Pyramid Network (FPN) includes convolutional layers and up-sampling blocks to generate multi-scale feature maps, (see Figure \ref{fig:rcnn}).

As the first cut and distillation point $C_5$, we select the output of ``Stem'' or the input of the residual block 2. For the next cut and distillation point, $C_6$, we choose the output of the same residual block, referred to as ``Layer 4'', which is also an input to the FPN. Finally, we select the outputs of ``P2-P6'' as the last distillation point, $M_6$, although it should be noted that since this includes five tensors, so for simplicity, we only utilise it as a distillation point. Note that the notation for the various layers of ResNet50 here is slightly different than in Appendix~\ref{appsec:modelsplit}, due to differences in implementation.


Thus in our various configurations for both object detection with Faster R-CNN and instance segmentation with Mask R-CNN, models are split from the cut points, while the distortion is calculated based on the distillation points, and \eqnref{eq:multi_distillation} is used whenever using $M_6$ for distillation. For instance, in one experiment, we split the Faster R-CNN model at the point ``Stem``, and the distortion is then calculated based on the MSE between pre-trained and estimated values of the Layer 4 tensor. 

Training is performed in two stages, in an unsupervised manner (without the object detection labels) using the loss from \eqnref{eq:learned}. In the first stage we use randomly cropped image patches of size $256 \times 256$ from a combination of the CLIC~\cite{clic_dataset} and JPEG-AI~\cite{jpeg_ai} datasets. In the second stage we use the same size patches taken from the VIMEO-90K~\cite{xue2019video_vimeo} dataset. In both stages we use a batch size of 16, and the ADAM~\cite{kingma2014adam} optimiser. In the first stage of training, we use a fixed learning rate of $10^{-4}$, while in the second stage we employ a polynomial decay for the learning rate after every 10 epochs. The number of channels in the latent representation $Z$ used by the compression model differs slightly between the two models, with Faster R-CNN and Mask R-requiring 96 and 128 channels, respectively. For exact values of these and other hyper-parameters see \tabref{tab:app_hyper}. 

\begin{table}[htbp]
    \caption{Hyper parameter values for direct compression models}
    \vspace{-0.3cm}
    \normalsize
    \centering
    \setlength{\tabcolsep}{4pt} 
    \begin{tabular}{c| cccccc | ccc}
    \bottomrule
     Task model & \multicolumn{6}{c}{$\lambda$} &$Z$ Channels & First stage epochs & Second stage epochs\\
    \hline
    Faster R-CNN  & 1.28e--5 & 3.2e--5 & 8e--5   & 2e--4  & 4e--4  & 5.5e--4 & 96 & 400 & 500 \\
    Mask R-CNN    & 1.28e--5 & 3.2e--5 & 8e-5   & 2e--4  & 4e--4  & 5.5e--4 & 128 & 400 & 500\\
    YOLOv3 $C_7->O$        & 1e--5 &  2.5e--5 & 5e--5   & 1e--4  & 2e--4  & 4e--4 & 64 & 300 & 350\\
    YOLOv3 - Others       & 2.5e--5 &  5e--5   & 1e--4  & 2e--4  & 4e--4 & 1e--3 & 64 & 300 & 350\\
    \toprule
    \end{tabular}
    \label{tab:app_hyper}
\end{table}


\subsection{YOLOv3}\label{appsec:yolo}

\begin{figure}[htbp]
  \centering
  \includegraphics[width=\textwidth]{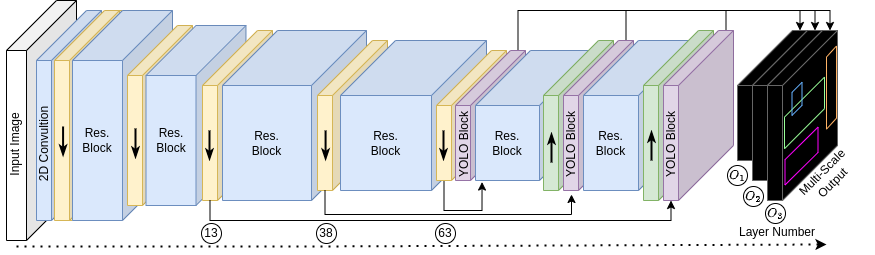}
    \caption{YOLOv3~\protect\cite{Redmon2018_yolov3} architecture. The potential distillation points are marked with numbered circles. Blocks marked with a $\downarrow$ and $\uparrow$ are downsampling and upsampling blocks.}
    \label{fig:YOLOv3}
\end{figure}

YOLOv3 is a popular model of object detection as it combines high accuracy with efficient computation. Its architecture is comprised of residual blocks, downsampling and upsampling blocks, as well as detection blocks, which we refer to as YOLO blocks. Residual blocks are comprised of multiple sub-blocks of convolutions followed by skip connection and additions. The width of residual blocks in the \figref{fig:YOLOv3} correlates roughly to the amount of convolutional layers, though it is not to scale. YOLO blocks, shown in purple in \figref{fig:YOLOv3} are comprised of more convolutional blocks followed by detection layers and each output a 3-D tensor containing a bounding box, an objectness score (an estimate of the likelihood that the box contains any object), and probability estimates for each of the target classes. The final output of YOLOv3 is comprised of three such detection outputs, which we denote $O_1,O_2,O_3$ , as seen in \figref{fig:YOLOv3}.

As detailed in~\cite{Redmon2018_yolov3}, and explained in section~\ref{sec:applications}, YOLOv3 is a multi-scale, multi-stream model. This means that output of certain layers is used as input in multiple downstream computational branches, without being subsequently merged by some addition or concatenation. We refer to such tensors, which are used by downstream layers as branching points, and and they are denoted with numbered circles in \figref{fig:YOLOv3}, corresponding to their PyTorch~\cite{pytorch} implementation~\cite{yolopytorch}. In practice, the multi-stream nature of YOLOv3 means that whenever choosing a distillation point for YOLOv3 that is deeper than layer 13 (the earliest branching point), we need to also include the branching point itself, or at least one tensor from each downstream branch.

In all experiments we use layer 13 as the cut point, with distillation points taken as deep as possible from each branch leading to the following choice of tensors layers: $L_{13}$, $L_{38}$, $L_{63}$, $O_1$, $O_2$ and $O_3$. As explained above, we must keep earlier branching point as part of distillation point whenever going deeper, which gives the following final distillation points (labeled to distinguish from points used in previous models):  $C_7 = \{L_{13}\}, M_7 = \{L_{13}, L_{38}\}, M_8 = \{L_{13},L_{38},L_{63}\}, O = \{O_1,O_2,O_3\}$. In order to calculate a scalar loss value, we choose to flatten and concatenate the relevant tensors before simply using an element-wise MSE loss, which is equivalent to the calculation shown in \eqnref{eq:multi_distillation_flatten}. Training is performed identically to the R-CNN experiments other than the difference in hyperparameters which is summarised in \tabref{tab:app_hyper} above.



\color{black}

\subsection{SWIN-Transformer}\label{app:swin}

Our last task model is the more modern and powerful SWIN-Transformer\cite{Liu_2021_ICCV}. This model builds upon the foundational concept of vision transformers (ViT)~\cite{vit} which first utilised transformers~\cite{transformer}, originally developed for natural language processing, in computer vision. Unlike the original ViT which divided the image into non-overlapping patches of size $16 \times 16$, SWIN utilises a sliding window to create overlapping patch representations. This avoids edge artificats near the boundaries of the patches and allows for improved performance on downstream tasks. 

As in many transformer based arcitechtures, one of the main advantages of SWIN is in its pretraining on large datasets. The majority of well established variations of SWIN were trained on Imagenet-1K~\cite{imagenet2015}, or Imagenet-22K~\cite{ridnik2021imagenet}. Furthermore, the authors present several sizes of SWIN, with a growing number of parameters from 28 (SWIN-T) to 197 million (Swin-L). As expected the larger models achieve better performance on many computer vision tasks, but nonetheless, the small SWIN-T still performs better than many other models with similar parameter size or floating point operations.  Due to our own computational constraints as well as those of a realistic coding for machines scenario we choose the smallest of the SWIN architectures SWIN-T. Note that due to the use of a direct-coding for machines approach, the computational cost of the model frontend would not actually be needed at the edge device, instead only the encoder would be run, with the decoder creating the required latent representation on the server side. 

We train our SWIN-T based models using the same RD loss formulation as \eqnref{eq:learned}, with the distortion replaced by the loss function of the SWIN object detection and instance segmentation model of the official implementation. This loss function combines 4 losses - a class cross-entropy loss, a regression loss for the box locations, a mask loss for the more detailed instance mask locations, and finally a score denoting whether an object exists at all (sometimes referred to as an objectness loss). We use the default parameters given by the official implementation, and select the following lambda values - $\lambda = 0.5,1,2,5,10$ to achieve the various points shown in the RD curves using this task model. All models are trained for 100 epochs on the COCO2017 training set (with a $10\%$ validation set taken out of the training set at random), with the original augmentations from the official implementation with an additional random cropping to $256 \times 256$ patches.

\begin{figure}[h]
  \centering
  \includegraphics[width=0.9\textwidth]{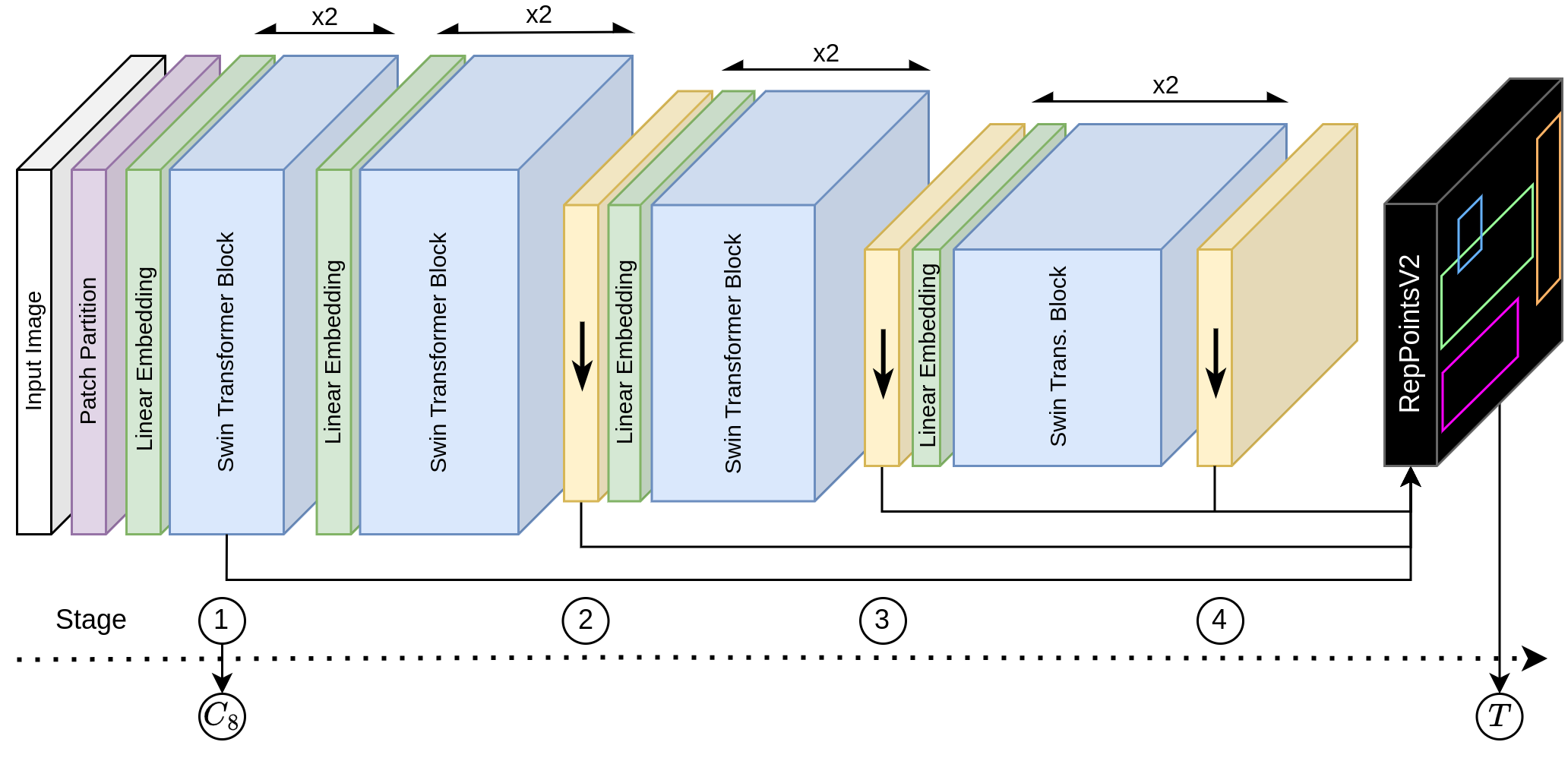}
    \caption{SWIN-Transformer~\protect\cite{Liu_2021_ICCV} architecture with a RepPointsV2~\protect\cite{reppointsv2} head. The cut point and supervised target are marked with numbered circles.}
    \label{fig:Liu_2021_ICCV}
\end{figure}

\color{black}
\section{Additional Experimental Results}

\subsection{Scalable Coding for Humans and Machines}\label{app:scalable}

In order to showcase the advantages of deeper distillation points more directly, we explore their effects in a scalable setting identical to that presented in~\cite{hyomin}. Because our compression now performs optimisation for both human vision and machine analysis the loss terms must change slightly:
\begin{equation}\label{eq:app_scalable}
    L = R + \lambda\cdot(D_{enh} + w\cdot D_{base}).
\end{equation}
Here, $D_{enh} = MSE(X,\widehat{X})$ is the distortion for the input reconstruction task, while $D_{base}$ is a distillation loss identical to \eqnref{eq:multi_distillation_flatten}. $\lambda$ determines the balance between rate and distortion as before, and $w$ controls the balance between machine analysis and human vision tasks. 

\begin{figure*}[htbp]
\centerline{\includegraphics[width = 0.95\linewidth]{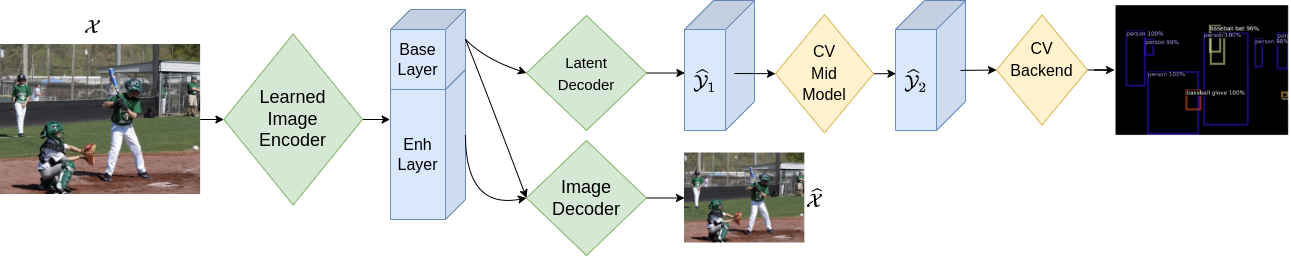}}
\caption{Block diagram of the scalable setting. Note that in \protect\cite{hyomin}, the mid model is not used and the cut and distillation points are identical.}
\label{fig:app_block}
\end{figure*}

Training here follows the same procedure as our direct compression models, though we use a subset of approximately $30\%$ of the VIMEO-90K dataset in our second stage of training due to the larger size of models to be trained. We use the same cut and distillation points as in our direct-compression experiment with YOLOv3. In order to maintain a fair comparison, we also retrain the models of Choi \etal using an identical procedure. Furthermore, because changing the distillation point may produce a similar effect to changing the balance between the base and enhancement task, we train two instances of the Choi22 benchmark with $w=0.06,0.12$.

Evaluation in for the base task is performed identically to Section~\ref{subsubsec:yolo}. For the enhancement task however, we use PSNR as our evaluation metric, and use the Kodak~\cite{kodak_dataset} dataset which consists of 24 high quality uncompressed images. To summarise rate-distortion performance we once again use the BD metrics, using Choi2022 with $w=0.06$ as the anchor for BD-rate and BD-mAP for the machine task. While human vision is not the focus of our work, we include the results for input reconstruction to allow a more comprehensive analysis. Here too we use BD-rate, and BD-PSNR (since that is our accuracy metric) with VVC used as the anchor. We compare our models with the same benchmarks as before, with the important exception that we now have two versions of Choi22.

\begin{figure}[htbp]
\centering
\begin{subfigure}[b]{0.49\linewidth}
     \centering
      \includegraphics[width=\linewidth]{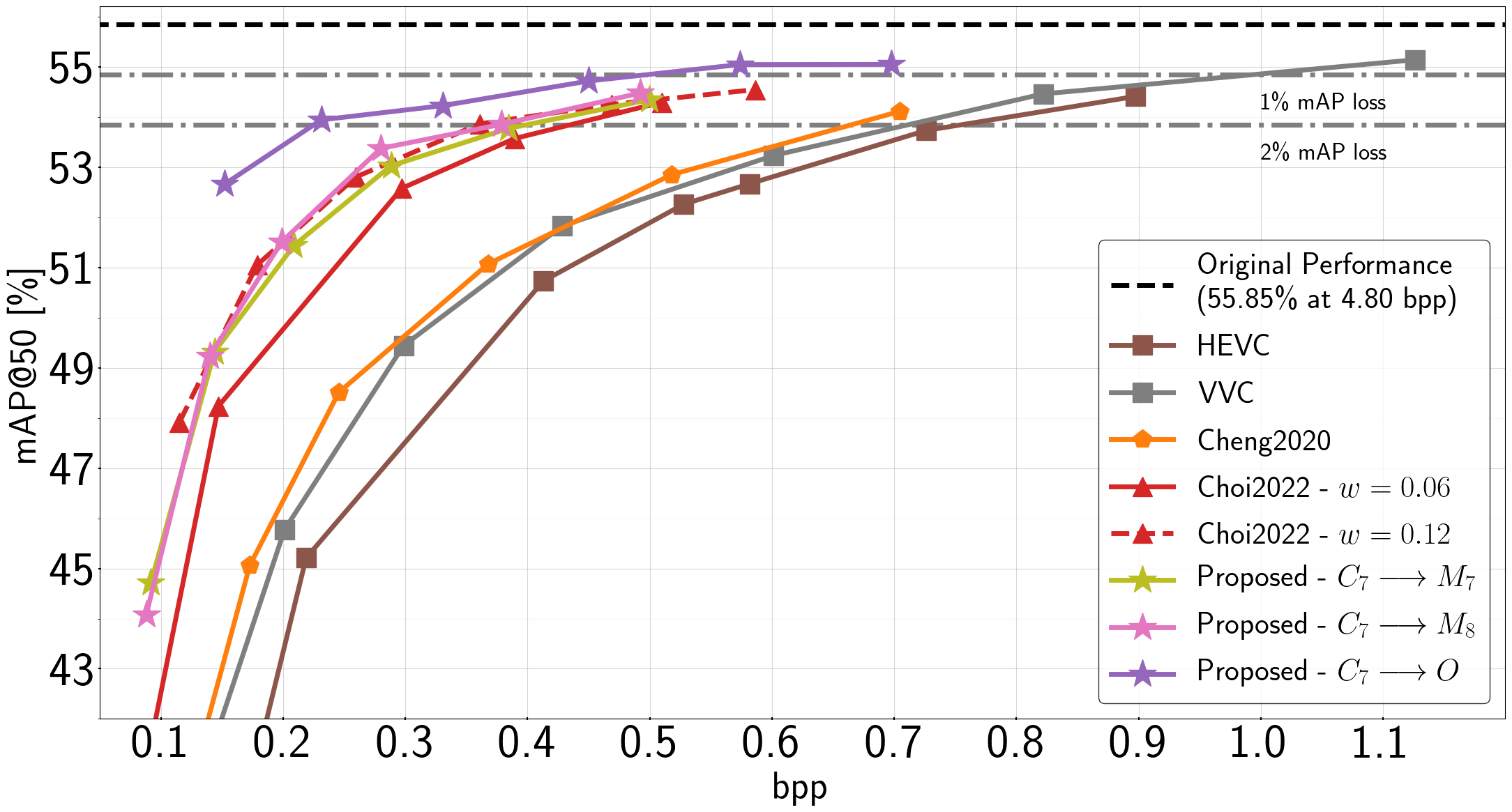}
      
      \caption{Object detection using YOLOv3, on 5000 images from the COCO2014 validation set.}\label{fig:app_obj_benchmark_scalable}
    \end{subfigure}
    \hfill
    \begin{subfigure}[b]{0.49\linewidth}
      \centering
      \includegraphics[width=\linewidth]{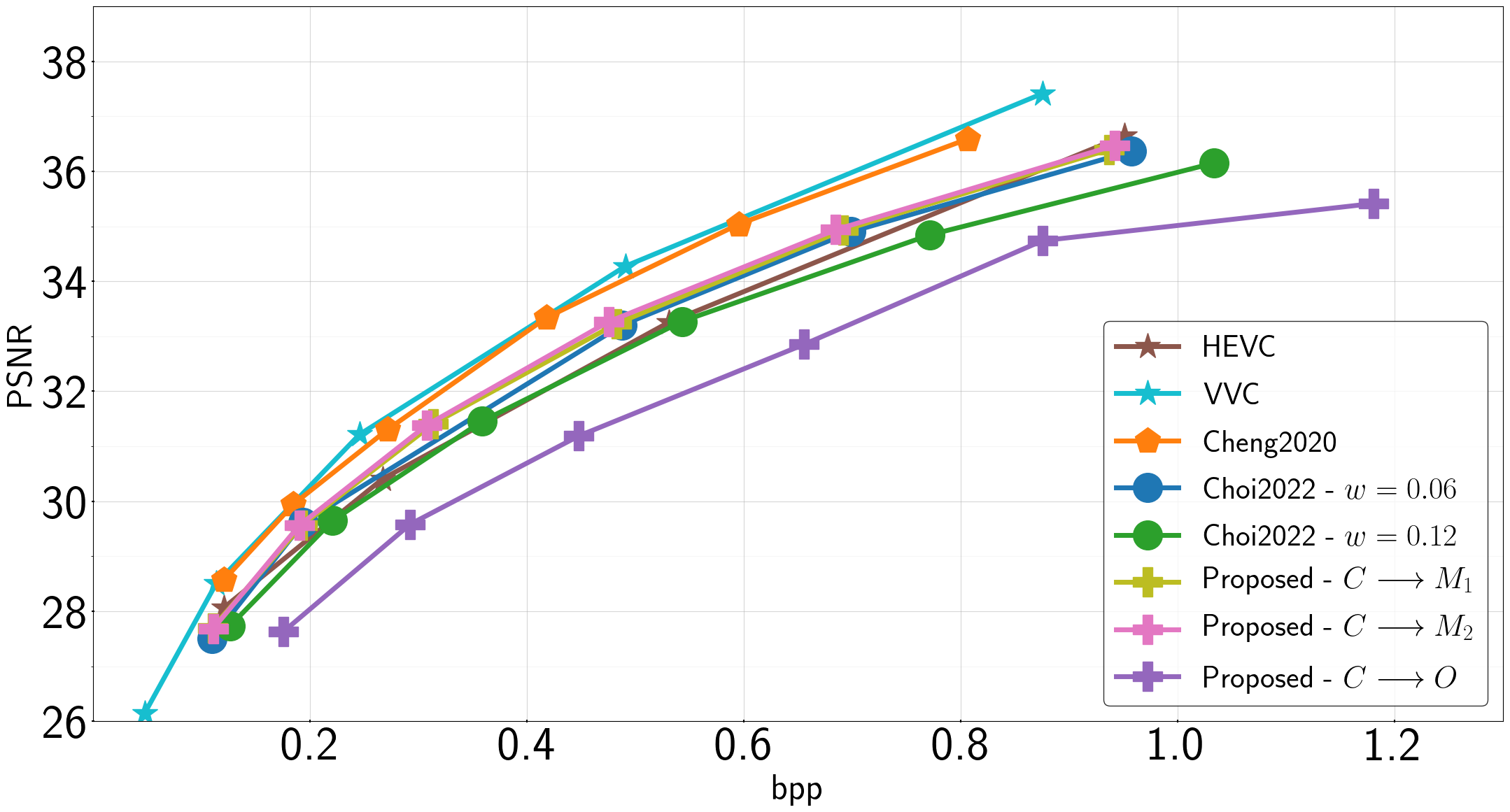}
      
      \caption{Input reconstruction on the Kodak dataset. \newline}\label{fig:app_kodak_benchmark_scalable}
    \end{subfigure}
    \caption{Benchmark comparison for the scalable codec}\label{fig:app_yolo_benchmarks_scalable}
\end{figure}

\begin{table}[h]
\centering
\setlength{\tabcolsep}{4pt}
\caption{Rate-Distortion Performance for the Scalable Compression scenario}\label{tbl:app_yolo_scalable}
\normalsize
\begin{tabular}{@{}ccccc@{}}
\toprule
Model  &  \multicolumn{2}{c}{YOLOv3} & \multicolumn{2}{c}{Input Reconstruction}   \\  
& BD-Rate$[\%]$    & BD-mAP$[\%]$   &  BD-Rate$[\%]$   & BD-PSNR[dB]  \\
\midrule
$C\rightarrow M_5$ & -12.11 & 0.92 & 23.4 & -0.88 \\
$C\rightarrow M_6$ & -13.87 & 1.16 & 21.9 & -0.83\\
$C\rightarrow O$ & \textbf{-50.44} & \textbf{1.94} & 83.5 & -2.56\\
\midrule
Choi2022 $w=0.06$ & 0  & 0 & 24.9 & -0.92 \\
Choi2022 $w=0.12$ & -13.89  & 0.96 & 39.1 & -1.37 \\
\midrule
VVC       & 66.4 & -3.90 & \textbf{0} & \textbf{0} \\
HEVC      & 89.3 & -5.86 & 29.8 & -1.04 \\
Cheng & 54.5 & -3.47 & 5.28 & -0.22 \\

\end{tabular}
\end{table}

Observing the results for the object-detection task, shown in \figref{fig:app_obj_benchmark_scalable} and \tabref{tbl:app_yolo_scalable}, we see once more that the use of the deepest distillation point $O$ has resulted in the best RD performance, achieving a BD-rate improvement of over $50\%$ over previous SOTA. Interestingly, the effect of using distillation points $M_5,M_6$ was nearly identical, with very small difference between the two in terms of BD-mAP. Furthermore, we notice that putting a stronger emphasis on the task in the model of ~\cite{hyomin} using $w=0.12$, gives comparable RD performance to using distillation points $M_5,M_6$. Thus, in order to observe the benefits of the deeper distillation point we can turn to the enhancement task, seen in \figref{fig:app_kodak_benchmark_scalable}, as well as \tabref{tbl:app_yolo_scalable}. There, we see that our proposed models, using deeper distillation points, achieve better reconstruction RD performance for identical object detection RD (or vice-versa). For example, our model using $M_5$ achieves approximately $13\%$ BD-rate savings for input reconstruction when compared directly with the model from~\cite{hyomin} trained with $w=0.12$. Conversely, our model using $M_5$ achieves slightly better input reconstruction RD compared to~\cite{hyomin} trained with $w=0.06$, while achieving $12\%$ BD-rate improvement in object detection RD. \newpage

\subsection{Additional Task-Appropriateness visualisation}\label{app:appropriateness}

\begin{figure}[ht]
    \centering
    \begin{subfigure}[b]{0.245\linewidth}
        \centering
        \includegraphics[width=\textwidth]{Images/CIFAR10/CIFAR10_input_vgg.png}
        \caption{Input, $\rho = 0.038$}
    \end{subfigure}
    \hfill
    \begin{subfigure}[b]{0.245\linewidth}
        \centering
        \includegraphics[width=\textwidth]{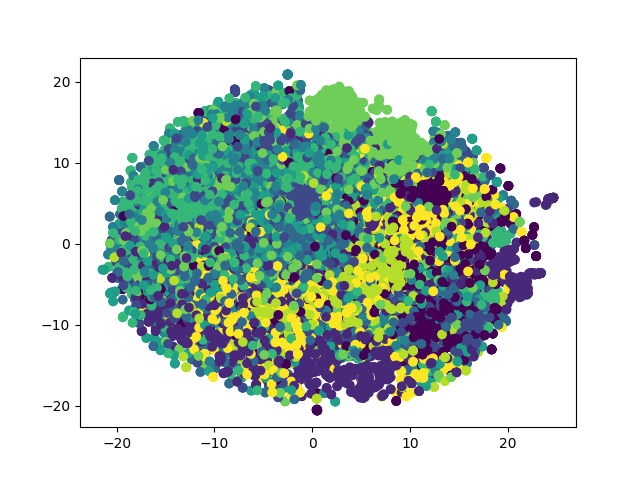}
        \caption{Layer2, $\rho = 0.023$}
     \end{subfigure}
    \begin{subfigure}[b]{0.245\linewidth}
        \centering
        \includegraphics[width=\textwidth]{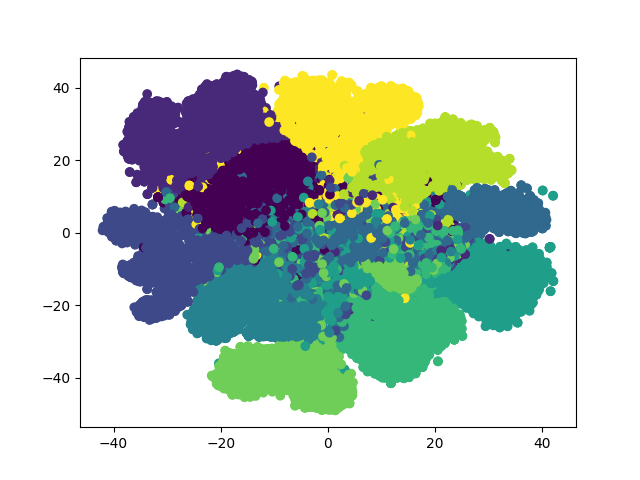}
        \caption{Layer3, $\rho = 0.104$}
     \end{subfigure}
    \hfill
    \begin{subfigure}[b]{0.245\linewidth}
        \centering
        \includegraphics[width=\textwidth]{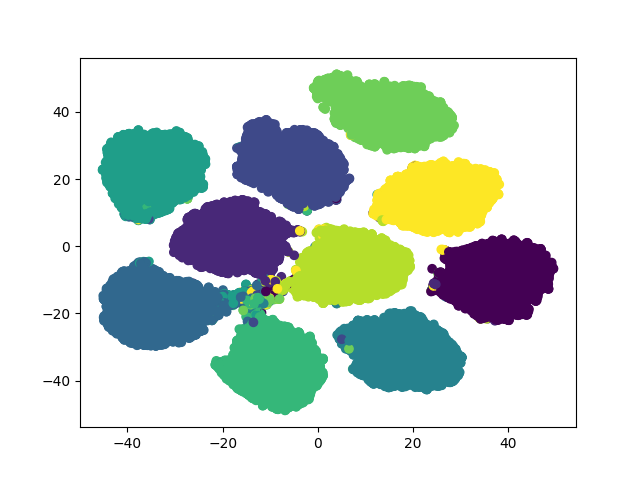}
        \caption{Layer4, $\rho = 1.0$}
       \end{subfigure}

    \caption{Task appropriateness and t-SNE visualisation for various layers in ResNet50, using the CIFAR-10 dataset and MSE distortion. }
    \label{fig:task_app_cifar10_resnet}
\end{figure}

\begin{figure}[ht]
    \centering
    \begin{subfigure}[b]{0.245\linewidth}
        \centering
        \includegraphics[width=\textwidth]{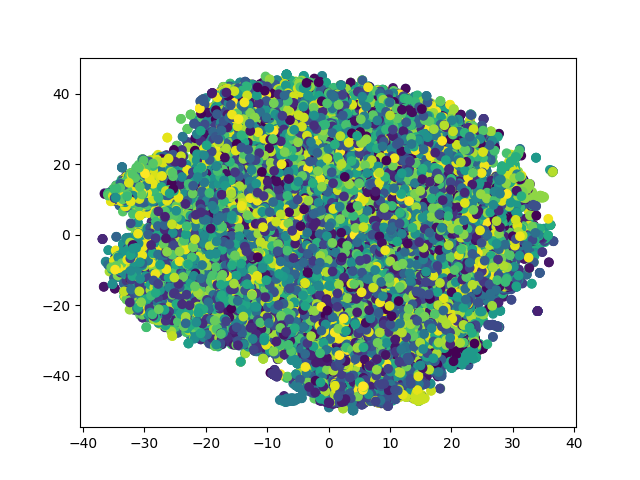}
        \caption{Input, $\rho = 0.036$}
    \end{subfigure}
    \hfill
    \begin{subfigure}[b]{0.245\linewidth}
        \centering
        \includegraphics[width=\textwidth]{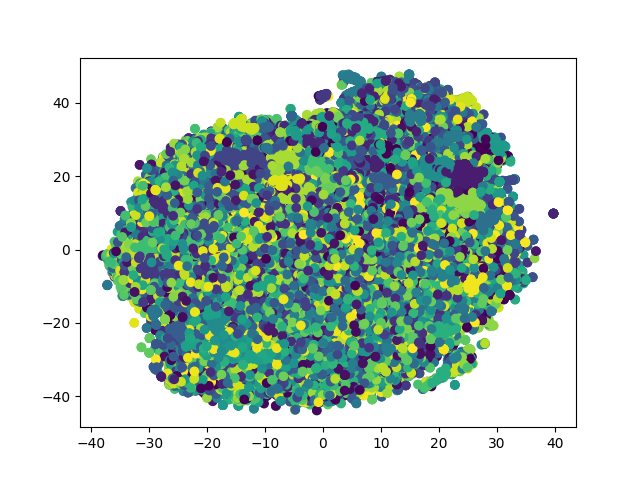}
        \caption{Features.19, $\rho = 0.046$}
     \end{subfigure}
    \begin{subfigure}[b]{0.245\linewidth}
        \centering
        \includegraphics[width=\textwidth]{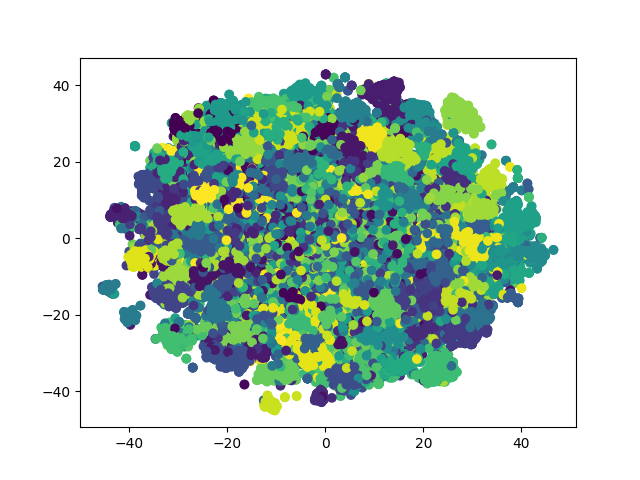}
        \caption{Features.26, $\rho = 0.120$}
     \end{subfigure}
    \hfill
    \begin{subfigure}[b]{0.245\linewidth}
        \centering
        \includegraphics[width=\textwidth]{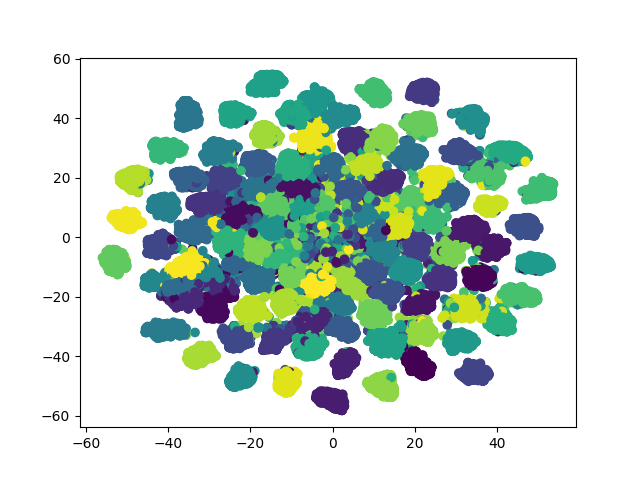}
        \caption{Features.32, $\rho = 0.555$}
       \end{subfigure}
       
    \caption{Task appropriateness and t-SNE visualisation for various layers in VGG16, using the CIFAR-100 dataset and MSE distortion. }
    \label{fig:task_app_cifar100_vgg}
\end{figure}

\begin{figure}[ht]
    \centering
    \begin{subfigure}[b]{0.245\linewidth}
        \centering
        \includegraphics[width=\textwidth]{Images/CIFAR100/CIFAR100_input_resnet.png}
        \caption{Input, $\rho = 0.036$}
    \end{subfigure}
    \hfill
    \begin{subfigure}[b]{0.245\linewidth}
        \centering
        \includegraphics[width=\textwidth]{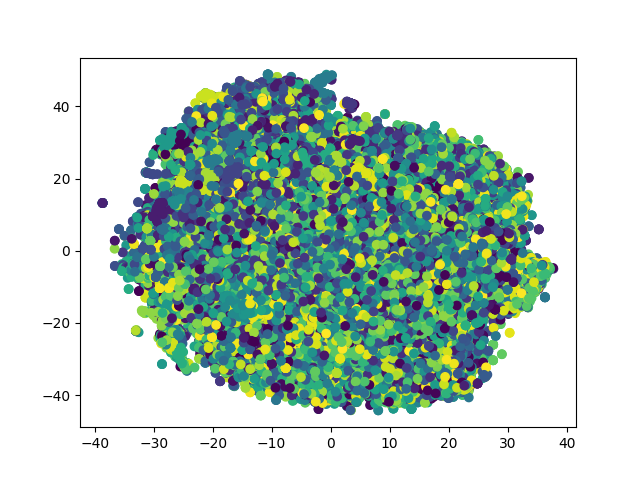}
        \caption{Layer2, $\rho = 0.031$}
     \end{subfigure}
    \begin{subfigure}[b]{0.245\linewidth}
        \centering
        \includegraphics[width=\textwidth]{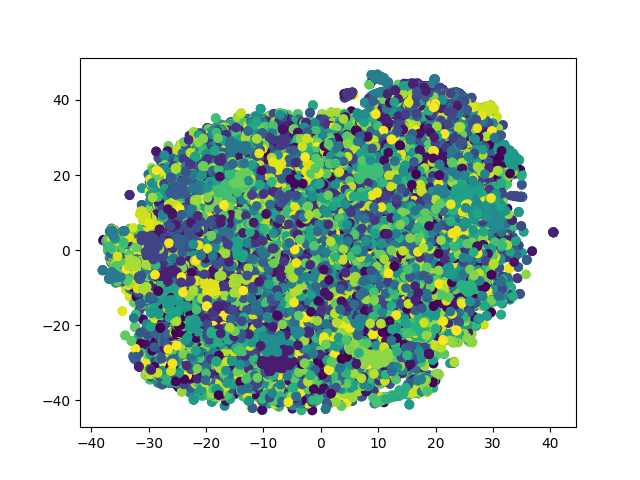}
        \caption{Layer3, $\rho = 0.050$}
     \end{subfigure}
    \hfill
    \begin{subfigure}[b]{0.245\linewidth}
        \centering
        \includegraphics[width=\textwidth]{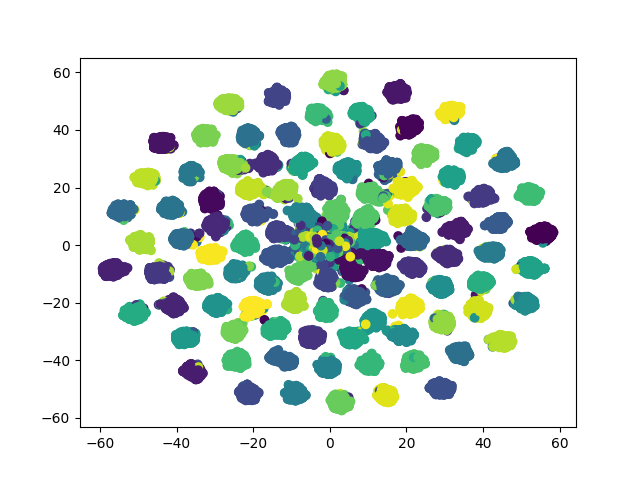}
        \caption{Layer4, $\rho = 0.997$}
       \end{subfigure}
       
    \caption{Task appropriateness and t-SNE visualisation for various layers in ResNet50, using the CIFAR-100 dataset and MSE distortion. }
    \label{fig:task_app_cifar100_resnet}
\end{figure}
\vfill

%

\end{document}